\documentclass[letterpaper,showpacs,preprintnumbers,amsmath,amssymb,nofootinbib,onecolumn,superscriptaddress]{revtex4-1}
\usepackage{amssymb}
\usepackage{graphicx}
\usepackage{amsmath}
\usepackage{multirow}
\usepackage{verbatim}
\usepackage{rotating}
\usepackage[usenames,dvipsnames]{color}
\usepackage[bookmarks=true, bookmarksnumbered=true, colorlinks=true,
  urlcolor=darkblue, citecolor=darkgreen, linkcolor=darkred,
  breaklinks=true]{hyperref} 
\usepackage{xspace}

\definecolor{darkblue}{rgb}{0,0,0.5}
\definecolor{darkgreen}{rgb}{0.1,0,0.3}
\definecolor{darkred}{rgb}{0.6,0,0}

\newcommand{\nc}{\newcommand}
\nc{\like}{\mathcal{L}}
\nc{\Emin}{{E_{\rm min}}}
\nc{\Emax}{{E_{\rm max}}}
\nc{\pdf}{\mathcal{P}}
\begin{document}

\preprint{IFIC/15-06, IPPP/15/04, DCPT/15/08}


\title{Spectral analysis of the high-energy IceCube neutrinos } 

\author{Sergio Palomares-Ruiz}
\email{sergiopr@ific.uv.es}
\affiliation{Instituto de F\'{\i}sica Corpuscular (IFIC)$,$
 CSIC-Universitat de Val\`encia$,$ \\  
 Apartado de Correos 22085$,$ E-46071 Valencia$,$ Spain}
\author{Aaron C. Vincent}
\email{aaron.vincent@durham.ac.uk}
 \affiliation{Institute for Particle Physics Phenomenology (IPPP),\\ Department of Physics, Durham University, Durham DH1 3LE, United Kingdom}
\author{Olga Mena}
\email{omena@ific.uv.es}
\affiliation{Instituto de F\'{\i}sica Corpuscular (IFIC)$,$
 CSIC-Universitat de Val\`encia$,$ \\  
 Apartado de Correos 22085$,$ E-46071 Valencia$,$ Spain}

\begin{abstract}
\vspace{1cm}
A full energy and flavor-dependent analysis of the three-year high-energy IceCube neutrino events is presented. By means of multidimensional fits, we derive the current preferred values of the high-energy neutrino flavor ratios, the normalization and spectral index of the astrophysical fluxes, and the expected atmospheric background events, including a prompt component. A crucial assumption resides on the choice of the energy interval used for the analyses, which significantly biases the results. When restricting ourselves to the $\sim$30~TeV$- 3$~PeV energy range, which contains all the observed IceCube events, we find that the inclusion of the spectral information improves the fit to the canonical flavor composition at Earth, ($1:1:1$)$_\oplus$, with respect to a single-energy bin analysis. Increasing both the minimum and the maximum deposited energies has dramatic effects on the reconstructed flavor ratios as well as on the spectral index. Imposing a higher threshold of 60~TeV yields a slightly harder spectrum by allowing a larger muon neutrino component, since above this energy most atmospheric tracklike events are effectively removed. Extending the high-energy cutoff to fully cover the Glashow resonance region leads to a softer spectrum and a preference for tau neutrino dominance, as none of the expected electron antineutrino induced showers have been observed so far. The lack of showers at energies above 2~PeV may point to a broken power-law neutrino spectrum. Future data may confirm or falsify whether or not the recently discovered high-energy neutrino fluxes and the long-standing detected cosmic rays have a common origin. 
 \end{abstract}


\maketitle

\tableofcontents

\section{Introduction}
\label{sec:intro}

The observation of high-energy neutrinos in the IceCube detector at the South Pole~\cite{Aartsen:2013bka, Aartsen:2013jdh, Aartsen:2014gkd} has signaled the beginning of extragalactic high-energy neutrino astronomy. After three years of data taking, 36 neutrino events (plus one event whose energy and direction cannot be reconstructed) with energies between approximately 30~TeV and 2~PeV have provided the evidence for the existence of an extraterrestrial neutrino flux at $5.7\sigma$~\cite{Aartsen:2014gkd}. The discovery of this flux has motivated a large number of studies in the literature to unravel their origin, from different scenarios within standard cosmic-ray sources~\cite{Cholis:2012kq, Kistler:2013my, Kalashev:2013vba, Fox:2013oza, Stecker:2013fxa, Murase:2013ffa, Laha:2013lka, Gao:2013rxa, Murase:2013rfa, Anchordoqui:2013qsi, Neronov:2013lza, Razzaque:2013dsa, Razzaque:2013uoa, Ahlers:2013xia, Liu:2013wia, Fraija:2013cha, Lunardini:2013gva, Anchordoqui:2013dnh, Taylor:2014hya, Tamborra:2014xia, Petropoulou:2014lja, Dado:2014mea, Anchordoqui:2014yva, Padovani:2014bha, Tjus:2014dna, Krauss:2014tna, Chang:2014hua, Ahlers:2014ioa, Dermer:2014vaa, Tavecchio:2014iza, Bai:2014kba, Wang:2014jca, Bhattacharya:2014sta, Winter:2014pya, Fargion:2014jaa, Sahu:2014fua, Anchordoqui:2014rca, Murase:2014tsa, Kalashev:2014vya, Zandanel:2014pva, Tavecchio:2014eia, Kimura:2014jba, Chen:2014gxa, Anchordoqui:2014pca, Razzaque:2014ola, Lunardini:2014wza, Moharana:2015nxa} to more exotic possibilities~\cite{Borriello:2013ala, Feldstein:2013kka, Barger:2013pla, Esmaili:2013gha, Bai:2013nga, Ema:2013nda, Akay:2014tga, Alikhanov:2014uja, Anchordoqui:2014hua, Ioka:2014kca, Ng:2014pca, Zavala:2014dla, Stecker:2014xja, Ibe:2014pja, Bhattacharya:2014yha, Ema:2014ufa, Blum:2014ewa, Araki:2014ona, Akay:2014qka, Aeikens:2014yga, Illana:2014bda, Esmaili:2014rma, Cherry:2014xra, Fong:2014bsa, Stecker:2014oxa, Guo:2014laa}. However, the current angular information and statistics do not allow the identification of the neutrino sources and the flux remains compatible with an isotropic distribution~\cite{Aartsen:2013jdh, Aartsen:2014gkd}. A larger and improved version of the IceCube neutrino detector has among its major goals the resolution of the individual sources of the observed astrophysical neutrinos~\cite{Aartsen:2014njl}.  However, by focusing on observables such as the flavor composition and spectrum, the diagnostic power grows proportionately to the statistics in the available sample of events.

The astrophysical neutrino flavor composition has long been recognized as a powerful tool to disentangle the underlying mechanism of ultrahigh-energy neutrino production~\cite{Rachen:1998fd, Athar:2000yw, Beacom:2003zg, Anchordoqui:2003vc, Kashti:2005qa, Serpico:2005bs, Kachelriess:2006fi, Rodejohann:2006qq, Mena:2006eq, Lipari:2007su, Pakvasa:2007dc, Esmaili:2009dz, Choubey:2009jq, Lai:2009ke, Hummer:2010ai, Meloni:2012nk, Fu:2012zr, Vissani:2013iga, Chatterjee:2013tza, Xu:2014via, Fu:2014isa}, including scenarios of exotic physics~\cite{Enqvist:1998un, Athar:2000yw, Crocker:2001zs, Beacom:2002vi, Barenboim:2003jm, Beacom:2003nh, Beacom:2003eu, Hung:2003jb, Illana:2004qc, Illana:2005pu, Anchordoqui:2005gj, Esmaili:2009fk, Bhattacharya:2009tx, Bhattacharya:2010xj, Baerwald:2012kc, Pakvasa:2012db, Dorame:2013lka, Joshipura:2013yba}. In the standard scenario, ultrahigh-energy neutrinos are produced by the decay of pions and kaons and secondary muons, produced by hadronic interactions in extragalactic cosmic accelerators. This mechanism gives rise to a  flux flavor ratio at the cosmic source of $(\alpha_{e}:\alpha_{\mu}:\alpha_{\tau})_S = (1:2:0)_S$. Current measurements of the neutrino oscillation parameters~\cite{Gonzalez-Garcia:2014bfa, Forero:2014bxa, Capozzi:2013csa} imply that this source composition is transformed into a neutrino flavor ratio at the detector position on Earth of $(\alpha_{e}:\alpha_{\mu}:\alpha_{\tau})_\oplus \simeq (1:1:1)_\oplus$~\cite{Learned:1994wg}. Observationally, in the energy range of interest here, the neutrino flux flavor ratios are tagged via two possible event topologies: muon tracks, and electromagnetic or hadronic showers.  Our previous studies of the neutrino flux flavor ratios~\cite{Mena:2014sja, Palomares-Ruiz:2014zra} showed that the canonical ratio at Earth, $(\alpha_{e}:\alpha_{\mu}:\alpha_{\tau})_\oplus = (1:1:1)_\oplus$, assuming the typical $E^{-2}$ high-energy neutrino spectrum, is disfavored at $81\%$ ($92\%$)~confidence level (C.L.) by a fit to the two-year (three-year) IceCube data, although when accounting for the systematic uncertainties in the backgrounds, the significance is slightly reduced~\cite{Palomares-Ruiz:2014zra}. We pointed out that this mild, albeit compelling, tension could be due to several factors, related to a misunderstanding of the background events, to a misidentification of track events as showers, or to nonstandard mechanisms of neutrino production at and propagation from the sources.

In these previous works we only analyzed the total number of events, discarding the additional information encoded in both the signal and the background energy distributions. Although the energy behavior has already been  exploited in the literature as a diagnosis of the origin of the high-energy neutrino events~\cite{Kistler:2013my, Anchordoqui:2014hua, Barger:2014iua, Winter:2014pya, Watanabe:2014qua, Kalashev:2014vra}, it has never been used in conjunction with the flavor information and by allowing the background normalizations to vary freely. Spectral analyses of the full data have been performed, either fixing the flavor composition to $(1:1:1)_\oplus$~\cite{Aartsen:2013jdh, Aartsen:2014gkd, Chen:2013dza, Kalashev:2014vra} (see also Ref.~\cite{Winter:2014pya} for other cases) or varying the flavor composition but with fixed backgrounds~\cite{Watanabe:2014qua}. In this study, we present the first energy and flavor-dependent analysis of the 36 high-energy IceCube showers and track events where the astrophysical flavor composition, spectral index and normalization (assuming a power-law flux equal for neutrinos and antineutrinos), along with the number of background events are left free to vary. Furthermore, we extend the fit performed in our previous works~\cite{Palomares-Ruiz:2014zra, Mena:2014sja} to higher energies, in order to extract crucial information concerning the astrophysical high-energy neutrino flux and its origin. 

This study includes the spectral index and the normalization of the incoming neutrino flux, the number of conventional atmospheric muon and neutrino backgrounds (considered separately), and a prompt neutrino flux component from charmed particle decays in the atmosphere. We also consider a case in which the extraterrestrial high-energy neutrino flux follows a broken power law with a break of one unit in the spectral index at $E_\nu =1$~PeV. We finally examine the effect of a nonzero probability of misidentifying muon tracks as shower events.

We  present the results from our fits to the 36 events measured by the IceCube experiment considering four possible energy ranges:  the fiducial 28~TeV$- 3$~PeV energy range, which covers the publicly available IceCube high-energy neutrino events; the  60~TeV$- 3$~PeV range, which was used by IceCube for their inference of the properties of the astrophysical flux; and an extension of these ranges to 10~PeV, where a large signal from the well-known Glashow resonance~\cite{Glashow:1960zz} is expected --- but not seen so far. This serves to illustrate the different observational and physical effects that may bias the reconstructed properties of the high-energy neutrino flux. Namely, a larger value of the minimum deposited energy would eliminate most of the expected atmospheric muon background events, while a larger value of the maximum deposited energy would encompass the Glashow resonance at $E_\nu \simeq 6.3$~PeV, which should give rise to yet-unobserved events in the few PeV region (see also Ref.~\cite{Barger:2014iua}). The absence of these events could point, among other possibilities, to a break in the high-energy neutrino spectrum around a few PeV, which could nicely be connected with the behavior of the ultrahigh-energy cosmic ray spectrum~\cite{Kistler:2013my, Anchordoqui:2014hua, Barger:2014iua, Winter:2014pya, Watanabe:2014qua, Kalashev:2014vra}. 

Concerning the fiducial 28~TeV$- 3$~PeV energy range, the canonical ratio of neutrino flavor fluxes at the Earth ($1:1:1$)$_\oplus$ becomes more favored once the information from the event energy distribution is included in our multidimensional fits. For this case, as well as for the other possible energy ranges explored here, the precise values of the best-fit parameters for the neutrino flavor ratios depend on the different quantities involved in the fits. As a general trend, the choice of a larger low-energy threshold (from 28~TeV to 60~TeV) implies a wider allowed region in the flavor parameter spaces that encompass a significant muon neutrino component, even if the best fit shows a slight preference for an electron neutrino component. Increasing the upper limit on deposited energy (from 3~PeV to 10~PeV) results in a favored tau neutrino component and a steeper spectrum. However, in the latter case, the existence of  a break in the high-energy neutrino spectrum could restore the preference for a large electron neutrino component. Finally, we remark that to firmly establish the neutrino flavor ratios from data and therefore unravel the astrophysical neutrino production mechanism, the fraction of tracks that could be misidentified as showers is a critical parameter. Indeed, a large fraction of  tracks being misidentified would skew the reconstructed flavor ratio away from a significant muon neutrino contribution.

The structure of the paper is as follows. We start in Sec.~\ref{sec:rates} by describing our calculations of the shower and track neutrino event rates as a function of the deposited energy, including a description of the main ingredients: the attenuation and regeneration effects in the neutrino propagation through the Earth; the definition of the electromagnetic (EM)- equivalent deposited energy for each type of event in neutrino-nucleon and neutrino-electron interactions and how it depends on the geometry of the detector; and our computation of the IceCube effective mass in terms of the measured deposited energy. Section~\ref{sec:bkg} contains a detailed description of the background treatment followed in our numerical analyses, whose methodology is presented in Sec.~\ref{sec:analysis}.  The results arising from our fits to the three-year IceCube neutrino data are shown and discussed in Sec.~\ref{sec:results}. Finally, we highlight our most important findings and discuss their implications in Sec.~\ref{sec:conclusions}.

\section{Event rates}
\label{sec:rates}

The incoming neutrino fluxes at the IceCube detector are characterized by their energy $E_\nu$ and their flavor composition, $(\alpha_e:\alpha_\mu:\alpha_\tau)_\oplus$, while individual events are characterized via the total EM-equivalent deposited energy $E_{\textrm{dep}}$ and the event topology (tracks or showers). Below PeV energies, interactions with nucleons are the dominant ones. Showers are induced by both $\nu_e$ and $\nu_\tau$ (and $\bar\nu_e$ and $\bar\nu_\tau$) charge current (CC) interactions, as well as by neutral current (NC) interactions of neutrinos of all three flavors. Tracks occur when a muon is produced: either via a $\nu_\mu$ interacting via $W$ exchange (CC) or via tau lepton decay with the tau produced in a $\nu_\tau$ CC event.  For each channel, represented by a subscript $c$, the measured deposited energy rate $dN^c/dE_{\textrm{dep},i}$, which is evaluated for each observed event $i$, is given by
\begin{equation}
\frac{dN^c}{dE_{\textrm{dep}, i}} = \int_0^\infty \frac{dN^c}{dE_{\textrm{true}}} \, R(E_{\textrm{true}},E_{\textrm{dep},i},\sigma(E_{\textrm{true}})) \, d E_{\textrm{true}} ~.
\label{eq:dNdEde}
\end{equation}
The function $R(E_{\textrm{true}},E_{\textrm{dep},i},\sigma(E_{\textrm{true}}))$ is included to account for the energy resolution $\sigma (E_{\textrm{true}})$ and a fit is provided in Appendix~\ref{app:totrates}.

The spectrum of true deposited energies in the detector depends on the attenuation and regeneration factor $Att^f_{\nu_\ell}(E_\nu)$ due to the absorption of neutrinos when traversing the Earth, the detector effective mass $M_{\textrm{eff}}(E_{\textrm{true}})$ as a function of the true deposited energy, the incoming neutrino flux $d \phi^f_{\nu_\ell} (E_\nu)/d E_\nu$ of type $f$ (either of astrophysical or atmospheric origin) and the production cross sections $d\sigma^c/dE_{\textrm{true}}$ of neutrino $\nu_\ell$ for process $c$. For electron and muon neutrinos interacting with nucleons, this can be written in general form as
\begin{equation}
\frac{dN^c}{dE_{\textrm{true}}} = T \, N_A \, \int_0^\infty Att^f_{\nu_\ell}(E_\nu) \, M_{\textrm{eff}}(E_{\textrm{true}}) \, \frac{d\phi^f_{\nu_\ell}(E_\nu)}{d E_\nu} \, \frac{d\sigma^c_{\nu_\ell}(E_\nu,E_{\textrm{true}})}{dE_{\textrm{true}}} \, d E_\nu ~, 
\label{eq:dNdEd}
\end{equation}
where $T=988$~days is the time of data taking and $N_A = 6.022 \times 10^{-23}$~g$^{-1}$.

For tau neutrinos below a few PeV, the produced tau lepton decays inside the detector, so one has to take into account the energy spectra of its decay products. In Appendix~\ref{app:totrates} we provide the detailed expressions for the differential rates for all the processes we consider as a function of the measured EM-equivalent deposited energy, Eq.~(\ref{eq:dNdEde}). Let us note that we also consider interactions of all neutrino flavors with electrons  and we also give the full expressions in Appendix~\ref{app:totrates}. All the relevant cross sections used in this work are provided in Appendix~\ref{app:cs}.

Below, we describe how we compute the attenuation and regeneration factors of the different neutrino fluxes in their passage through the Earth, we explain how to compute the true EM-equivalent deposited energy of hadronic showers and muon tracks and finally, we discuss how to obtain the effective mass $M_{\textrm{eff}}(E_{\textrm{true}})$, which encodes the detector efficiency, as a function of the true EM-equivalent deposited energy. The reader not interested in these technical details can jump to Sec.~\ref{sec:bkg}.

\subsection{Attenuation and regeneration of neutrinos in their passage through the Earth}
\label{sec:attenuation}

The rise of the neutrino-nucleon cross section with energy implies that, for energies above a few TeV, the mean free path inside the Earth becomes comparable to the distance traveled~\cite{Gandhi:1995tf, Gandhi:1998ri}, so the Earth attenuates the flux of neutrinos. The transport equation for $\nu_\ell = \{\nu_e, \nu_\mu, \bar\nu_e, \bar\nu_\mu\}$ is given by~\cite{Nicolaidis:1996qu, Naumov:1998sf, Kwiecinski:1998yf, Iyer:1999wu, Giesel:2003hj, Reya:2005vh, Rakshit:2006yi}
\begin{eqnarray}
\frac{\partial}{\partial X} \left(\frac{d\phi^f_{\nu_\ell} (E_\nu, X)}{dE_\nu}\right) & = & - N_A \, \left(\sigma^{\textrm{NC}}_{\nu_\ell} (E_\nu) + \sigma^{\textrm{CC}}_{\nu_\ell} (E_\nu) \right) \, \frac{d\phi^f_{\nu_\ell}(E_\nu, X)}{dE_\nu} \nonumber \\
& & + N_A \, \int_{0}^{1} \, \frac{dy}{1-y} \, \frac{d\sigma^\textrm{NC}_{\nu_\ell} (E_\nu/(1-y),y)}{dy} \, \frac{d\phi^f_{\nu_\ell} (E_\nu/(1-y), X)}{dE_\nu} ~,
\label{eq:transpN}
\end{eqnarray}
where $\sigma^{\textrm{NC}}_{\nu_\ell} (E_\nu)$ and $\sigma^{\textrm{CC}}_{\nu_\ell} (E_\nu)$ are the NC and CC neutrino-nucleon cross sections of $\nu_\ell$ (see Appendix~\ref{app:cs}),  $X(\theta) = \int_{0}^{L} \rho(x) \, dx$ is the column depth (with $L = 2 R_\oplus \cos\theta$), $R_\oplus$ the Earth radius, and $\theta$ the nadir angle of the direction of the neutrinos with respect to the position of the detector. Throughout this work, the density of the Earth $\rho(x)$ is assumed to be given by the STW105 (also known as reference Earth model, REF) model~\cite{Kutkowski:2008, Trabant:2012}.

For the energies of interest for cosmic neutrinos, in general, interactions with electrons can be neglected. However, for $\bar\nu_e$,  the resonant production of a $W$ boson at energies around $E_\nu = M_W^2/m_e \simeq 6.3$~PeV (the so-called Glashow resonance~\cite{Glashow:1960zz}) has a cross section larger than the neutrino-nucleon cross sections and must be taken into account\footnote{We do not take into account the small effect due to the Doppler broadening of the resonance~\cite{Loewy:2014zva}.}. In all our computations not only do we include this resonant cross section but also all the neutrino-electron interactions.  Therefore, in the transport equation above, Eq.~(\ref{eq:transpN}), we make the substitutions
\begin{eqnarray}
\sigma^{\textrm{NC}}_{\nu_\ell} (E_\nu) + \sigma^{\textrm{CC}}_{\nu_\ell} (E_\nu) & \rightarrow & \sigma^{\textrm{NC}}_{\nu_\ell} (E_\nu) + \sigma^{\textrm{CC}}_{\nu_\ell} (E_\nu) + \sigma^e_{\nu_\ell} (E_\nu) ~, \\[2ex]
\frac{d\sigma^\textrm{NC}_{\nu_\ell} (E_\nu,y)}{dy} & \rightarrow & \frac{d\sigma^\textrm{NC}_{\nu_\ell} (E_\nu,y)}{dy} +  \frac{d\sigma^e_{\nu_\ell, e}(E_\nu, y)}{dy} ~,
\end{eqnarray}
where (see Appendix~\ref{app:cs})
\begin{eqnarray}
\sigma^e_{\nu_e} (E_\nu) & = & \sigma^e_{\nu_e, e} (E_\nu) ~, \\
\sigma^e_{\nu_\mu} (E_\nu) & = & \sigma^e_{\nu_\mu, e} (E_\nu) +  \sigma^e_{\nu_\mu, \mu} (E_\nu) ~, \\
\sigma^e_{\bar\nu_e} (E_\nu) & = & \sigma^e_{\bar\nu_e, e} (E_\nu) +  \sigma^e_{\bar\nu_e, \mu} (E_\nu) +  \sigma^e_{\bar\nu_e, \tau} (E_\nu) + \sigma^e_{\bar\nu_e, h} (E_\nu) ~, \\
\sigma^e_{\bar\nu_\mu} (E_\nu) & = & \sigma^e_{\bar\nu_\mu, e} (E_\nu) ~.
\end{eqnarray}
The notation used throughout this work is as follows: $\sigma^I_{\nu_\ell, \ell'}$ is the cross section for a neutrino of flavor $\ell$ to interact via $I=\{$electron scattering, CC nucleon scattering, NC nucleon scattering$\}$ and produce a lepton $\ell'$. In this work, all the cross sections are defined per nucleon and we assume the Earth to be an isoscalar medium, so the number of electrons is half that of nucleons. This relative factor of 2 is already included in the cross sections above (see Appendix~\ref{app:cs}).

In the case of electron and muon neutrinos, the charged leptons produced in CC interactions in the Earth are quickly brought to rest and either are absorbed or decay at rest, and hence do not contribute to the high-energy flux; in the case of $\nu_\tau$ and $\bar\nu_\tau$, however, the produced tau leptons can decay before being stopped.  Therefore, as long as the $\tau$ decay length is shorter than its mean free path in the Earth, which occurs up to several hundred PeV, $\nu_\tau$'s and $\bar\nu_\tau$'s are not absorbed, but degraded in energy, so the regeneration of the $\nu_\tau$ and $\bar\nu_\tau$ fluxes due to the daughter $\nu_\tau$ and $\bar\nu_\tau$ from $\tau$ decays must be accounted for~\cite{Halzen:1998be, Iyer:1999wu, Dutta:2000jv, Hettlage:2001yf, Tseng:2003pn, Hussain:2003vi, Yoshida:2003js, Reya:2005vh}. For the energies of interest the tau decay length is much shorter than the interaction length and $\tau$ energy losses may be neglected~\cite{Dutta:2000hh, Becattini:2000fj, Tseng:2003pn, Jones:2003zy, Bugaev:2003sw, Dutta:2005yt}. The transport equations for $\nu_\tau$ and $\bar\nu_\tau$ (and tau leptons) are given by the coupled equations~\cite{Iyer:1999wu, Tseng:2003pn, Reya:2005vh, Rakshit:2006yi}
\begin{eqnarray}
\frac{\partial}{\partial X} \left(\frac{d\phi^f_{\nu_\tau} (E, X)}{dE}\right) & = & - N_A \, \left(\sigma^{\textrm{NC}}_{\nu_\tau} (E) + \sigma^{\textrm{CC}}_{\nu_\tau} (E) \right)\, \frac{d\phi^f_{\nu_\tau}(E, X)}{dE} \nonumber \\
& & + N_A \, \int_{0}^{1} \, \frac{dy}{1-y} \, \frac{d\sigma^\textrm{NC}_{\nu_\tau} (E/(1-y),y)}{dy} \, \frac{d\phi^f_{\nu_\tau} (E/(1-y), X)}{dE}  \nonumber \\
 & & + \frac{1}{(E/m_\tau) \, \tau \, \rho(X)} \, \int_{0}^{1}  \, dy \,  \frac{dn(1-y)}{dy} \, \frac{d\phi^f_\tau(E/(1-y), X)}{dE} ~, \\[2ex]
\frac{\partial}{\partial X} \left( \frac{d\phi^f_\tau(E, X)}{dE} \right) & = & - \frac{1}{(E/m_\tau) \, \tau \, \rho(X)} \, \frac{d\phi^f_\tau(E, X)}{dE} + N_A \, \int_{0}^{1} \, \frac{dy}{1-y} \, \frac{d\sigma^\textrm{CC}_{\nu_\tau} (E/(1-y),y)}{dy} \, \frac{d\phi^f_{\nu_\tau} (E/(1-y), X)}{dE} ~, 
\label{eq:transpNtau}
\end{eqnarray}
where $\tau$ and $m_\tau$ are the tau lepton lifetime at rest and the tau lepton mass, respectively, $d\phi^f_\tau(E, X)/dE$ is the flux of tau leptons at $X(\theta)$ produced via CC interactions and $dn(z)/dz$ is the distribution of tau neutrinos after tau lepton decays with $z=E_{\nu_\tau}/E_\tau$, for which we use the parametrization given in Ref.~\cite{Dutta:2000jv}. Analogously to the case of electron and muon neutrinos and antineutrinos, we also include interactions with electrons by making the substitutions, for both $\nu_\tau$ and $\bar\nu_\tau$,
\begin{eqnarray}
\sigma^{\textrm{NC}}_{\nu_\tau} (E_\nu) + \sigma^{\textrm{CC}}_{\nu_\tau} (E_\nu) & \rightarrow & \sigma^{\textrm{NC}}_{\nu_\tau} (E_\nu) + \sigma^{\textrm{CC}}_{\nu_\tau} (E_\nu) + \sigma^e_{\nu_\tau} (E_\nu) ~, \\[2ex]
\frac{d\sigma^\textrm{NC}_{\nu_\tau} (E_\nu,y)}{dy} & \rightarrow & \frac{d\sigma^\textrm{NC}_{\nu_\tau} (E_\nu,y)}{dy} + \frac{d\sigma^e_{\nu_\tau, e}(E_\nu, y)}{dy} ~, 
\end{eqnarray}
and only for $\nu_\tau$, but not for $\bar\nu_\tau$,
\begin{eqnarray}
\frac{d\sigma^\textrm{CC}_{\nu_\tau} (E_\nu,y)}{dy} & \rightarrow & \frac{d\sigma^\textrm{CC}_{\nu_\tau} (E_\nu,y)}{dy} + \frac{d\sigma^e_{\nu_\tau, \tau}(E_\nu, y)}{dy} ~,
\end{eqnarray}
where (see Appendix~\ref{app:cs})
\begin{eqnarray}
\sigma^e_{\nu_\tau} (E_\nu) & = & \sigma^e_{\nu_\tau, e} (E_\nu) + \sigma^e_{\nu_\tau, \tau} (E_\nu) ~, \\
\sigma^e_{\bar\nu_\tau} (E_\nu) & = & \sigma^e_{\bar\nu_\tau, e} (E_\nu) ~.
\end{eqnarray}

\begin{figure}[t]
	\includegraphics[width=\textwidth]{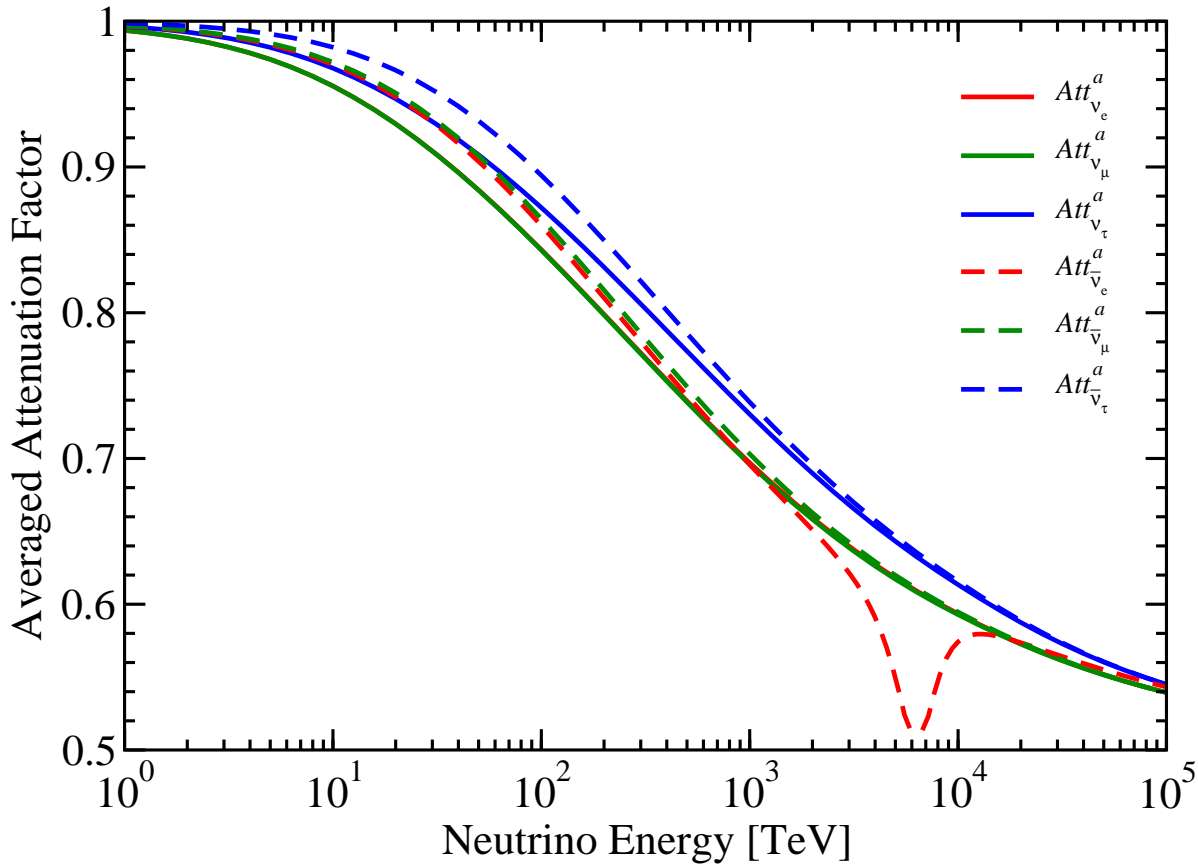}
	\caption{\textbf{\textit{Whole-sky averaged attenuation and regeneration factors for astrophysical neutrinos traversing the Earth}}, for an isotropic power-law spectrum $E^{-\gamma}$, with the IceCube best fit for spectral index $\gamma=2.3$~\cite{Aartsen:2014gkd}.  Neutrinos (solid lines) and antineutrinos (dashed lines) are shown separately, and the attenuation and regeneration factors for $\nu_e$, $\nu_\mu$ and $\nu_\tau$ are represented by red, green and blue lines, respectively. Note that the curves for $\nu_e$ and $\bar\nu_e$ are almost identical to those for $\nu_\mu$  and $\bar\nu_\mu$, except in the case of $\bar\nu_e$ around the Glashow resonance, which is clearly visible.}
	\label{fig:Attastro}
\end{figure}

We additionally consider the secondary $\nu_e$ and $\nu_\mu$ (and $\bar\nu_e$ and $\bar\nu_\mu$) fluxes produced after $\nu_\tau$ (and $\bar\nu_\tau$) CC interactions and the subsequent $\tau$ decay into leptonic channels~\cite{Beacom:2001xn, Dutta:2002zc}, although they are only relevant for very hard spectra. To do so, one has to couple the transport equations for $\nu_\ell = \{\nu_e, \nu_\mu\}$ with those for $\nu_\tau$ by adding the following term to Eq.~(\ref{eq:transpN})~\cite{Dutta:2002zc, Rakshit:2006yi}:
\begin{equation}
G^f_{\nu_\ell} (E, X) = \frac{1}{(E/m_\tau) \, \tau \, \rho(X)} \, \int_{0}^{1}  \, dy \,  \frac{dn_{\tau \rightarrow \nu_\ell}(1-y)}{dy} \, \frac{d\phi^f_\tau(E/(1-y), X)}{dE} ~,
\end{equation}
where for fully polarized tau leptons, the distribution of $\nu_\ell$ for $\tau$ decays, with $z=E_{\nu_\ell}/E_\tau$, is given by~\cite{Lipari:1993hd}
\begin{equation}
 \frac{dn_{\tau \rightarrow \nu_\ell}(z)}{dz} \simeq 0.18 \, \left(4 - 12 z + 12 z^2 - 4 z^3\right) ~.
\end{equation}

To compute the primary neutrino flux after attenuation and regeneration and the flux of secondary electron and muon neutrinos we have followed the approaches of Refs.~\cite{Naumov:1998sf, Iyer:1999wu, Rakshit:2006yi} and alternatively that of Ref.~\cite{Farzan:2014gza} for $\bar\nu_e$ (used in a different context), because around the Glashow resonance the former approach presents some issues of convergence.  We refer the reader to those works for the details of these calculations.

We compute the neutrino and antineutrino fluxes at the detector for all flavors as a function of the nadir angle $\theta$ and the neutrino energy $E_\nu$ and obtain the $4\pi$-averaged attenuation and regeneration factors for isotropic fluxes,
\begin{equation}
Att^a_{\nu_\ell} (E_\nu) = \frac{1}{2} \, \left( 1 + \int_{0}^{1} Att^a_{\nu_\ell} (E_\nu, \theta) \, d\cos\theta \right) ~,
\label{eq:att}
\end{equation}
where $Att^a_{\nu_\ell} (E_\nu, \theta) = (d\phi^a_{\nu_\ell} (E, L)/dE_\nu)/(d\phi^a_{\nu_\ell} (E, 0)/dE_\nu)$ represents the fraction of the initial flux  $d\phi^a_{\nu_\ell} (E_\nu, 0)/dE_\nu$ propagating through the Earth with nadir angle $\theta$ that reaches the detector. The first term in Eq.~(\ref{eq:att}) represents the averaging over downgoing neutrinos that do not cross the Earth.

As an example, these whole-sky averaged suppression factors are shown in Fig.~\ref{fig:Attastro} for an isotropic power-law spectrum $E^{-\gamma}$ with spectral index $\gamma=2.3$. The factors for neutrinos (solid lines) and antineutrinos (dashed lines) are shown separately. One can see the effect of regeneration of $\nu_\tau$ inside the Earth (blue lines) and notice that the factors for $\nu_e$ and $\bar\nu_e$ (red lines) are very similar to those for $\nu_\mu$ and $\bar\nu_\mu$ (green lines). The only large deviation occurs around the Glashow resonance in the case of $\bar\nu_e$ (red dashed line), as can clearly be seen.

\subsection{Deposited  EM-equivalent energies}
\label{sec:deposition}

Following the notation of the IceCube Collaboration, we define the true deposited energy $E_{\textrm{true}}$ as the total true EM-equivalent energy deposited after a neutrino interaction with the vertex within the detector.  Each channel has different efficiencies when it comes to producing a measured energy deposition in the IceCube detector, since what is ultimately measured is the \v{C}erenkov radiation emitted while the charged particles propagate. 

The light yield of the electromagnetic cascades in ice is approximately proportional to the total track length of all the electrons in the cascade, which is used to calculate the number of \v{C}erenkov photons and hence, the energy of the cascade. On the other hand, the total track length of hadronic cascades is not linear with the energy of the cascade. This is due to the presence of more neutral particles like neutrons, to large losses due to the binding energies in hadronic processes and to a higher \v{C}erenkov threshold for hadrons~\cite{Wiebusch, Kowalski}. This always results in a reduction of the EM-equivalent energy (or total track length) of a hadronic shower with respect to that of an electromagnetic shower. Thus, for hadronic showers the EM-equivalent deposited energy is given by 
\begin{equation}
E_h = F_h (E_X) \, E_X ~,
\label{eq:Fhdef}
\end{equation}
where $E_X$ is the shower energy and $F_h (E_X)$ is a suppression factor, which represents the ratio of the track length of a hadronic and an electromagnetic shower of the same energy and is provided in Appendix~\ref{app:totrates}. 

At these energies, the lifetime of a muon is much larger than the time it takes to cross the detector. The total energy deposited by a muon can be described by the mean stopping power. Although at the energies of interest these energy losses can be stochastic and large fluctuations around the mean are expected, it is reasonable to treat them as continuous and approximate the muon energy loss rate by
\begin{equation}
\left< - \frac{dE_\mu}{dx} \right> = a(E_\mu) + b(E_\mu) \, E_\mu
\label{eq:dedx}
\end{equation}
where $E_\mu$ is the muon energy, $a(E_\mu)$ is the electronic stopping power and $b(E_\mu)$ takes into account radiative processes (bremsstrahlung, pair production and photonuclear interactions). Writing the mean stopping power in this way is convenient, since both $a(E_\mu)$ and $b(E_\mu)$ vary slowly in the energy range of interest here. For the sake of simplicity, in this work we use a fit obtained from tabulated data for the muon loss rate in ice~\cite{Agashe:2014kda}, which is provided in Appendix~\ref{app:totrates}.

\begin{figure}
	\includegraphics[width=0.6\textwidth]{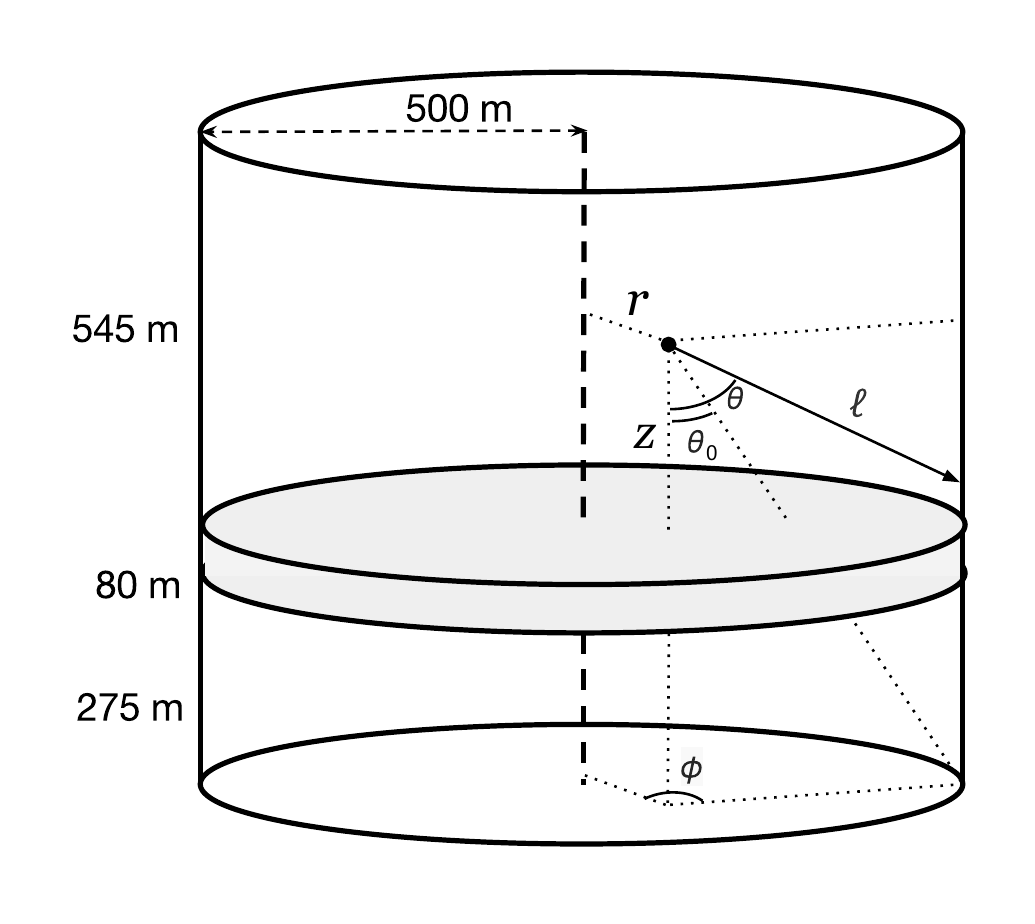}
	\caption{\textbf{\textit{Approximate geometry of the IceCube detector}}, which we use to compute the average distance traveled by a muon generated inside the detector volume in Eqs.~(\ref{eq:distanceintegral})--(\ref{eq:avgL}) and thus the expected energy deposition of a muon track, Eq.~(\ref{eq:avgloss}). From an interaction vertex located at $(r,z)$, the distance $\ell$ represents the distance traveled by a muon propagating in the direction  $(\theta,\phi)$. When $\theta < \theta_0$, $\ell = \ell_z$ and the muon exits via the detector floor; when $\theta > \theta_0$, $\ell = \ell_r$ and the muon exits via the ``wall''. We ignore events generated in the dust zone (gray band), but we do include the energy lost in that zone. }
	\label{fig:ICgeometry}
\end{figure}

To compute the energy deposited along a given muon track, we need to know the position where the muon was produced and its direction, \textit{i.e.}, the interaction vertex in the detector as well as the distance from the vertex to the edge of the detector volume. Bearing in mind that this deposited energy is, in general, much smaller than the energy deposited by the hadronic shower, we compute the vertex position and direction-averaged energy deposition, $\langle \Delta E_\mu \rangle$, along an average muon track in the detector given the initial muon energy, $E_\mu$. To do this, we approximately describe the detector volume as two stacked cylinders of radius 500~m, with respective heights 275~m and 545~m, separated by an 80~m dust zone~\cite{Aartsen:2013jdh}. We consider all the points and directions in the detector to be equally likely and do not count in any muon produced in the dust zone, although we do take into account energy losses in that zone. We illustrate this geometry in Fig.~\ref{fig:ICgeometry}.

For a given interaction vertex location in cylindrical coordinates $(r,z)$ inside the total detector volume, the angle-averaged distance traveled by a muon inside the detector is
\begin{equation}
\langle L(r,z) \rangle = \frac{1}{2 \pi}\int_0^{2\pi} d\phi 
\left( \int_{0}^{\cos\theta_0(r,z,\phi)}  \ell_r(r,z,\theta,\phi) \, d\cos\theta + \int^{1}_{\cos\theta_0(r,z,\phi)}  \ell_z(r,z,\theta,\phi) \, 
d\cos\theta\right) ~,
\label{eq:distanceintegral}
\end{equation}
where $\theta_0$ is the angle to the intersection between the floor and wall of the cylinder and $\ell_r$ is the propagation distance from the vertex to the wall if $\theta > \theta_0$, and $\ell_z$ is the  distance from the vertex to the floor, when $\theta < \theta_0$. Upgoing and downgoing tracks produce an equal contribution to $\langle L(r,z) \rangle$.  Therefore, the integrals in Eq.~(\ref{eq:distanceintegral}) cover only downgoing muons, while an overall factor of two ensures that the full angular range is considered. The full average over all interaction vertex points is then
\begin{equation}
\langle L \rangle = \frac{2 \pi}{V_{\textrm{tot}}}\int dz \int dr \, r \, \langle L(r,z) \rangle ~,
\label{eq:avgL}
\end{equation}
where the volume $V_{\textrm{tot}}$ covers each point in the cylinders, outside the 80~m dust zone. Numerically integrating Eq.~(\ref{eq:avgL}) yields an average distance of $\langle L \rangle = 406$~m.

From Eq.~(\ref{eq:avgL}) it is easy to compute the average muon energy loss per track, $\langle \Delta E_\mu \rangle$, but substituting the lengths $\ell_r$ (or $\ell_z$) by
\begin{equation}
\ell_{\{r,z\}} \rightarrow 1 - e^{- b \, \ell_{\{r,z\}}} ~,
\label{eq:avgexpl}
\end{equation}
where $b$ is defined in Eq.~(\ref{eq:dedx}). Evaluating this integral yields the total averaged deposited energy along an average muon track, in terms of the initial muon energy $E_\mu$,
\begin{equation}
\langle \Delta E_\mu \rangle = \langle 1 - e^{- b \, \ell }\rangle \, \left(E_\mu + a/b\right) \equiv F_\mu \, (E_\mu + a/b) ~,
\label{eq:avgloss}
\end{equation}
where the factor for the average fraction of energy lost along a track of a muon produced at the neutrino interaction vertex is found to be $F_\mu = 0.119$. 

Finally, the detector geometry is also important in the case of $\tau$ lepton production. When the $\tau$ energy exceeds a few hundreds of TeV, its lifetime is long enough that a significant fraction of tau leptons would escape the detector volume before decaying. In a similar way as done for muons, the average fraction of taus with energy $E_\tau$ decaying inside the detector is given by Eq.~(\ref{eq:avgL}), but with the substitution
 \begin{equation}
 \ell_{\{r,z\}} \rightarrow 1 - e^{-(\ell_{\{r,z\}}/\tau) (m_\tau/E_\tau)} ~.
 \end{equation} 
Here we are neglecting the tau lepton energy losses, which scale inversely proportional with the mass and thus are an order of magnitude smaller than for muons~\cite{Dutta:2000hh}. Therefore, we can write the fraction of tau leptons decaying inside the detector as
\begin{equation}
D_\tau(E_\tau) \equiv \langle 1 - e^{-(\ell/\tau) (m_\tau/E_\tau)} \rangle ~.
\label{eq:dtau}
\end{equation}
 
Once the tau decays inside the detector, it has an $\sim 18\%$ probability of producing a muon.  The average energy loss of such a muon is given by the average energy loss, Eq.~(\ref{eq:avgloss}), weighted by the fraction of taus that decay inside the detector,
\begin{equation}
\langle \Delta E_\mu \rangle_\tau = \left< \left(1 - e^{- b \, \ell } \right) \, \left( 1 - e^{-(\ell/\tau) (m_\tau/E_\tau)} \right) \right> \, \left(E_\mu + a/b\right) \equiv F_{\mu, \tau} (E_\tau) \, \left(E_\mu + a/b\right) ~,
\label{eq:avglossmutau}
\end{equation}
where the factor for the average fraction of energy lost along a track of a muon produced in a tau decay after a $\nu_\tau$ or $\bar\nu_\tau$ CC interaction inside the detector is given by $F_{\mu, \tau} (E_\tau)$. Fits to both, $D_\tau(E_\tau)$ and $F_{\mu, \tau} (E_\tau)$, are given in Appendix~\ref{app:totrates}.

\subsection{IceCube effective mass}
\label{sec:ICmass}

We now turn to the computation of the effective mass $M_{\textrm{eff}}(E_{\textrm{true}})$ in terms of the deposited energy in the detector, rather than the neutrino energy, since the former is the actual observable.  This effective mass can be seen as the mass of the target material times the efficiency of converting (EM-equivalent) deposited energy into an observed signal. Obtaining this is not completely straightforward, since the effective masses provided by the IceCube Collaboration are given as a function of the incoming neutrino energy $E_\nu$, for neutrino-nucleon NC and CC interactions involving the three flavors. To perform analyses in terms of the differential deposited energy spectrum, as we present below, we need to express the effective mass in terms of the quantity $E_{\textrm{true}}$ and thus, we need to perform a deconvolution. Using the fact that the effective masses $\tilde M_{a}(E_\nu)$ provided by the IceCube Collaboration are meant to be postinteraction, they are related to the effective mass $M_{\textrm{eff}}(E_{\textrm{true}})$ via 
\begin{eqnarray}
\left(\sigma^{\textrm{NC}}_{\nu_\ell}(E_\nu) + \sigma^{\textrm{NC}}_{\bar\nu_\ell}(E_\nu) \right) \, \tilde M^{\textrm{NC}}_{\nu_\ell}(E_\nu) & = & 
\int_{0}^{1} \left(\frac{d \sigma^{\textrm{NC}}_{\nu_\ell}(E_\nu, y)}{dy} + \frac{d \sigma^{\textrm{NC}}_{\bar\nu_\ell}(E_\nu, y)}{dy}\right) \, M_{\textrm{eff}}(E^{\textrm{NC}}) \, dy ~,  
\label{eq:MeffNC} \\
\left(\sigma^{\textrm{CC}}_{\nu_e}(E_\nu) + \sigma^{\textrm{CC}}_{\bar\nu_e}(E_\nu)\right) \, \tilde M^{\textrm{CC}}_{\nu_e}(E_\nu) & = & 
\int_{0}^{1} \left(\frac{d \sigma^{\textrm{CC}}_{\nu_e}(E_\nu, y)}{dy} + \frac{d \sigma^{\textrm{CC}}_{\bar\nu_e}(E_\nu, y)}{dy}\right) \, M_{\textrm{eff}}(E^{\textrm{CC}}_{e}) \, dy ~, 
\label{eq:MeffCCe} \\
\left(\sigma^{\textrm{CC}}_{\nu_\mu}(E_\nu) + \sigma^{\textrm{CC}}_{\bar\nu_\mu}(E_\nu)\right)\, \tilde M^{\textrm{CC}}_{\nu_\mu}(E_\nu) 
& = & 
\int_{0}^{1} \left(\frac{d \sigma^{\textrm{CC}}_{\nu_\mu}(E_\nu, y)}{dy} + \frac{d \sigma^{\textrm{CC}}_{\bar\nu_\mu}(E_\nu, y)}{dy}\right)\, M_{\textrm{eff}}(E^{\textrm{CC}}_{\mu}) \, dy ~,
\label{eq:MeffCCm} \\
\left(\sigma^{\textrm{CC}}_{\nu_\tau}(E_\nu) + \sigma^{\textrm{CC}}_{\bar\nu_\tau}(E_\nu) \right)\, \tilde M^{\textrm{CC}}_{\nu_\tau}(E_\nu) & = & 
\int_{0}^{1} \left(\frac{d \sigma^{\textrm{CC}}_{\nu_\tau}(E_\nu, y)}{dy} + \frac{d \sigma^{\textrm{CC}}_{\bar\nu_\tau}(E_\nu, y)}{dy}\right) \nonumber \\
& & \times \int_{0}^{1} \sum_{k=h,e,\mu} \, \left(D_\tau(E_\tau) \,   M_{\textrm{eff}}(E^{\textrm{CC}}_{\tau, k}) +  (1-D_\tau (E_\tau)) \, M_{\textrm{eff}}(E_h) \right) \frac{dn_k (z)}{dz} \, dz  \, dy ~, 
\label{eq:MeffCCt}
\end{eqnarray}
where $dn_k(z)/dz$~\cite{Dutta:2000jv} is the energy distribution of the daughter $\nu_\tau$, $e$, or $\mu$ with energy $E_{\nu_\tau, \tau}$ ($z = E_{\nu_\tau, \tau}/E_\tau$), $E_{e, \tau}$ ($z = E_{e, \tau}/E_\tau$), or $E_{\mu, \tau}$ ($z = E_{\mu, \tau}/E_\tau$) from $\tau$ decay via the hadronic, electronic, or muonic channel ($k = \{h, \, e, \, \mu\}$), respectively, and the true EM-equivalent deposited energies for each case are given in Appendix~\ref{app:totrates}. Note that the function $M_{\textrm{eff}}(E_{\textrm{true}})$ is common to all flavors and channels, since it only depends on the true EM-equivalent deposited energy $E_{\textrm{true}}$.  However, notice that $E_{\textrm{true}}$ is not given by the same expression for the different types of interactions. Let us also point out that we do not include the energy deposition along tau tracks, which is negligible, as the losses in this case are much smaller than in the case of muon tracks.

\begin{figure}[t]
	\includegraphics[width=\textwidth]{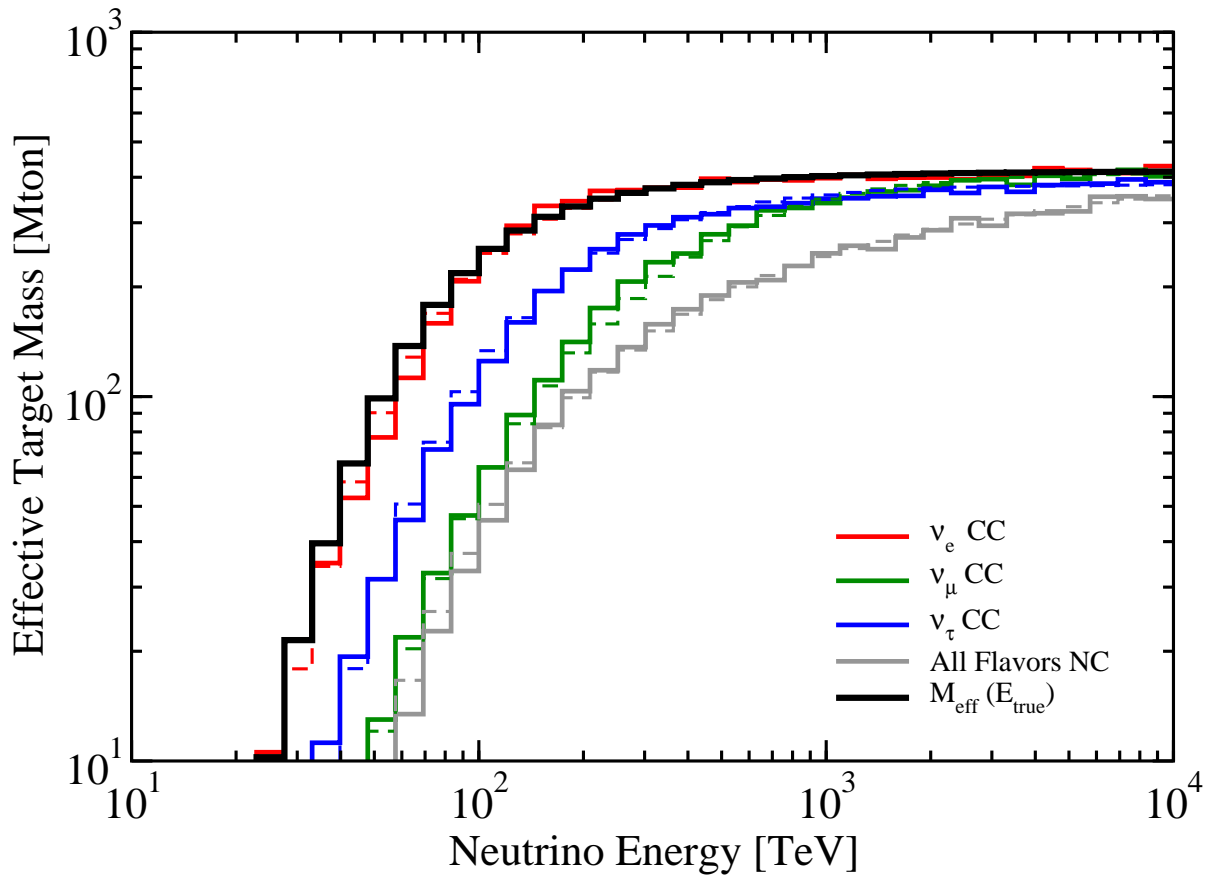}
	\caption{\textbf{\textit{Effective masses as a function of the neutrino energy.}} Solid lines are the effective masses provided by the IceCube Collaboration, whereas dashed lines represent the effective masses computed with Eqs.~(\ref{eq:MeffNC})-(\ref{eq:MeffCCt}) using the best-fit values for $\{p,q,c\}$ for the effective mass as a function of the deposited energy, Eq.~(\ref{eq:Mparam}). Red, green, blue and gray histograms represent CC $\nu_e$, $\nu_\mu$, $\nu_\tau$ and NC (all flavors) interactions. The black histogram depicts the effective mass function $M_{\textrm{eff}}(E_{\textrm{true}})$, Eq.~(\ref{eq:Mparam}), as a function of the true  deposited energy, $E_{\textrm{true}}$.}
	\label{fig:Meff}
\end{figure}

To obtain $M_{\textrm{eff}}(E_{\textrm{true}})$, we need to find a suitable parametrization for it. We consider a simple functional form with three independent parameters, given by
\begin{equation}
M_{\textrm{eff}}(x) = \begin{cases}
\rho_{\textrm{ice}} \, \dfrac{c \, x^q}{1 + d \, x^q} & \textrm{if  } x \geq 0 \\[3ex]
0 &  \textrm{if  } x < 0 ~,
\end{cases}
\label{eq:Mparam}
\end{equation}
where $\rho_{\textrm{ice}} = 0.9167 \, \textrm{g/cm}^3$ is the ice density and
\begin{equation}
x \equiv \log_{10}\left(\frac{E_{\textrm{true}}}{E_{\textrm{th}}}\right) ~,
\end{equation}
with $E_{\textrm{th}} = 10$~TeV.

Then we perform a least squares analysis for the three parameters $\{c, d, q\}$,
\begin{equation}
\chi^2 = \sum_{i, j} \left(1 - \frac{\tilde M^{\textrm{fit}}_j(E_{\nu, i}; c, d, q)}{\tilde M_j(E_{\nu, i})}\right)^2 ~,
\label{eq:chi2Veff}
\end{equation}
where $i$ indicates the neutrino energy bin (we use all bins where the effective mass is different from zero) and $j = \{\textrm{NC}, \nu_e-\textrm{CC}, \nu_\mu-\textrm{CC}, \nu_\tau-\textrm{CC}\}$ indicates the type of interaction. The actual effective masses $\tilde M_j(E_{\nu, i})$ are taken from Ref.~\cite{Aartsen:2013jdh} and $\tilde M^{\textrm{fit}}_j(E_{\nu, i}; a, b, c)$ are obtained from Eqs.~(\ref{eq:MeffNC}) - (\ref{eq:MeffCCt}). Up to two significant digits the best-fit parameters are found to be $c=0.50$, $d=1.1$, and $q=4.6$.

The accuracy of the fit can be seen in Fig.~\ref{fig:Meff}, where we show the IceCube effective masses with solid lines and our reconstructed results, computed with Eqs.~(\ref{eq:MeffNC})-(\ref{eq:MeffCCt}) using the best-fit values for $\{c,d,q\}$ for $M_{\textrm{eff}}(E_{\textrm{true}})$, Eq.~(\ref{eq:Mparam}), are shown with dashed lines. The result for $\nu_e$ CC ($\nu_\mu$ CC, $\nu_\tau$, NC) interactions is depicted in red (green, blue, gray). We also illustrate the effective mass as a function of the deposited energy. As can be seen, the accuracy of the fit is very good over the whole energy range, and, in particular, below the 10\% level for the relevant energy range. We note that the resulting mass as a function of the deposited energy is very similar to the effective mass for $\nu_e$ (plus $\bar\nu_e$) CC interactions as a function of the neutrino energy. This is expected, as the energy deposited after $\nu_e$ CC interactions is very close to the actual neutrino energy (one electromagnetic shower that fully deposits its energy and one hadronic shower that deposits most of its energy). We also point out that the effective mass for $\nu_\tau$ CC interactions is larger than for $\nu_\mu$ CC below a few hundred TeV, due to the fact that the produced tau lepton in the former case, after decaying, deposits more energy than the produced muon in the latter case. There is also an interesting feature when comparing these two histograms. At energies close to a PeV, they cross each other. This has to do with the fact that at these high energies some tau leptons exit the detector, which decreases the efficiency for this type of interaction. Finally, the effective mass for NC interactions is the smallest one because only one hadronic shower could be detected and  an important fraction of the neutrino energy escapes the detector in the form of daughter neutrinos.

\section{Backgrounds}
\label{sec:bkg}

The sources of background for cosmic neutrinos in IceCube are due to the interactions of cosmic rays with the nuclei of the atmosphere. These interactions produce fluxes of secondary muons and neutrinos from all directions, the so-called atmospheric muon and neutrino fluxes.  

Although the rate of atmospheric muons is huge, about 3~kHz in IceCube, it can be significantly reduced by using part of the detector as a veto for entering muon tracks~\cite{Aartsen:2013jdh}.  The final event selection allows a reduction at the level of one part in $10^5$ for muons with at least 6000 photoelectrons detected (approximately, deposited energies higher than 30~TeV), so the final expected event rate is of a few events per year.  This is estimated based on data by tagging muons in one region of the detector and by measuring the detection rate in a separate region equivalent to the veto region. The expected number of events above the threshold after 988 days of data taking is $8.4 \pm 4.2$. Nevertheless, to perform a statistical analysis based on the energy distribution of the detected events, we need to know the spectrum of the veto-passing atmospheric muons.  To good approximation, the convolution of the atmospheric muon spectrum and the veto probability is well described by a power law in the range of study~\cite{Whitehorn}. To determine the spectral index of this distribution we use the IceCube expected rates in two energy bins~\cite{Aartsen:2014gkd}. For deposited energies below 60~TeV the expected number of muons after 988~days is $8.0$, whereas for energies above 60~TeV, only $0.4$ atmospheric muons are expected to contribute to the background. Therefore, the spectral index $\gamma_\mu$ of the $E_\mu^{-\gamma_\mu}$ distribution is given by\footnote{Although it has no impact on our statistical analyses, to plot event spectra in Sec.~\ref{sec:results}, we assume a broken power law with breaking energy at 28~TeV, such that the number of atmospheric muons in the range $10^{1.2}$~TeV$- 10^{1.4}$~TeV after 988~days in IceCube is 0.2~\cite{Aartsen:2014gkd}.} 
\begin{equation}
\gamma_\mu = \frac{\log(21)}{\log(60 \, \textrm{TeV}/\Emin)} + 1 ~,
\label{eq:gammamu}
\end{equation} 
where $\Emin$ is the minimum measured EM-equivalent deposited energy considered in the sample. However,  the actual threshold is given in terms of the minimum number of detected photoelectrons (6000), not in terms of the minimum deposited energy. The conversion from the number of detected photoelectrons to deposited energy is not straightforward, since the scaling factor depends on the region of the detector where each event occurs. However, a detailed and careful analysis in terms of detected photoelectrons requires the full Monte Carlo that describes the detector and, therefore, it has to be performed by the IceCube Collaboration. Here, we use $\Emin = 28$~TeV, which is the round value below the energy of the less energetic of the events in the full sample. Let us note that we have checked that small variations in this minimum energy do not result in significant changes in our final results. For this value of $\Emin$, $\gamma_\mu \simeq 5$. On the other hand, the misidentification of tracks as showers for the muon atmospheric background is quoted to be $\lesssim 10\%~$\cite{Aartsen:2014gkd}; we take it to be exactly 10\%. This choice does not have a significant impact on the results.

In addition to the veto-passing atmospheric muons, atmospheric neutrinos constitute the other source of background in this search. In this work, we consider the conventional (mainly from $\pi$ and $K$ decays) and prompt (from charmed meson decays) atmospheric neutrino flux calculations of Refs.~\cite{Sinegovsky:2011ab, Petrova:2012qf, Sinegovskaya:2013wgm, Sinegovskaya:2014pia} based on the Hillas and Gaisser (HGm) cosmic-ray approximation~\cite{Hillas:2006ms, Gaisser:2012zz} and the hadronic model of Kimel and Mokhov (KM)~\cite{Kalinovsky:1989kk}, with updated parameters~\cite{Naumov:2001uc, Fiorentini:2001wa, Naumov:2002dm}, for the conventional flux, and on the Zatsepin and Sokolskaya cosmic-ray model (ZS)~\cite{Zatsepin:2006ci} and the quark-gluon string model (QGSM)~\cite{Kaidalov:1984ne, Kaidalov:1986zs, Kaidalov:1985jg, Bugaev:1989we} for the prompt flux. We use the  (conventional and prompt) atmospheric $\nu_e$, $\nu_\mu$, $\bar\nu_e$ and $\bar\nu_\mu$ fluxes as a function of the nadir angle and computed up to 100~PeV.  This calculation results in similar spectra to those computed by Ref.~\cite{Honda:2006qj} in the case of the conventional flux, which the IceCube Collaboration uses. However, the latter calculation only extends up to 10~TeV, so it has to be extrapolated up to PeV energies~\cite{Aartsen:2013eka}. Let us note that the ZS+QGSM prompt flux results in a larger number of expected events than the flux obtained with the dipole model~\cite{Enberg:2008te, Bhattacharya:2015jpa}, although important differences in shape are only present above PeV energies due to the ZS cosmic-ray model. Nevertheless, at those energies  prompt neutrinos are not expected to have any impact on the discussion.  

\pagebreak

In this analysis we also include the suppression of the flux of downgoing atmospheric $\nu_\mu$'s and $\bar\nu_\mu$'s ($f=\nu$ for conventional and $f=p$ for prompt\footnote{Rigorously, one should compute the veto-passing probability for prompt atmospheric neutrinos using their corresponding fluxes. We simply use the same result as that for conventional neutrinos.}) contributing to the background by tagging the muon produced by the same parent meson decaying in the atmosphere which can trigger the muon veto of the detector. To compute the passing probability we have followed Refs.~\cite{Schonert:2008is, Gaisser:2014bja}. This probability is defined as~\cite{Gaisser:2014bja}
\begin{equation}
P^f_{\nu_\mu}(E_\nu, \theta_z) = 1 - \frac{d\phi_{\nu_\mu}^{f,*}(E_\nu, \theta_z, E_{\mu, \textrm{min}})/dE_\nu d\cos\theta_z}{d\phi^f_{\nu_\mu}(E_\nu, \theta_z)/dE_\nu d\cos\theta_z} ~,
\label{eq:Patm}
\end{equation}
where $d\phi_{\nu_\mu}^{f,*}(E_\nu, \theta_z, E_{\mu, \textrm{min}})/dE_\nu d\cos\theta_z$ is the flux of neutrinos that reach the detector with zenith angle $\theta_z = \pi - \theta$ and are accompanied by the muon produced by the decay of the same meson, and it is given by
\begin{equation}
\frac{d\phi_{\nu_\mu}^{f,*}(E_\nu, \theta_z, E_{\mu, \textrm{min}})}{dE_\nu \, d\cos\theta_z} \simeq \frac{0.14 \, E_\nu^{-(\gamma_\nu + 1)}}{1 - Z_{NN}} \, \sum_{i=\pi, K} \, Br_i \, \frac{A_i}{1 + B_i \, E_\nu \, \cos\theta_z^*/\epsilon_i} ~, 
\label{eq:Patm1}
\end{equation}
where 
\begin{eqnarray}
A_i = \frac{Z_{N, i}}{1-r_i} \, \frac{1}{1 + \gamma_\nu} \, \frac{1}{z_{i, \textrm{min}}^{\gamma_\nu+1}} ~, & \hspace{5mm} &
B_i  =  z_{i, \textrm{min}} \, \frac{\gamma_\nu +2}{\gamma_\nu +1} \, \frac{1 - \Lambda_N/\Lambda_i}{\ln{\Lambda_i/\Lambda_N}} ~, 
\label{eq:Patm2} \\
z_{i, \textrm{min}} = \max \{ \frac{1}{1-r_i}, 1 + \frac{E_{\mu, \textrm{min}}}{E_\nu} \} ~,  & \hspace{5mm} & r_i = \frac{m_\mu^2}{m_i^2} ~, 
\label{eq:Patm3} 
\end{eqnarray}
with $\gamma_\nu = 1.7$, $Z_{NN} = 0.298$, $\Lambda_N = 120$~g/cm$^2$, $\Lambda_\pi= 160$~g/cm$^2$, $\Lambda_K = 180$~g/cm$^2$, $\epsilon_\pi = 115$~GeV, $\epsilon_K = 850$~GeV, $Br_\pi=1$, $Br_K=0.6355$, $Z_{N, \pi} =0.079$ and $Z_{N, K} = 0.0118$~\cite{Gaisser:2012zz, Gaisser:2002jj}. The minimum muon energy at production that is required for the muon to reach depth $X_{\textrm{IC}} = d_{\textrm{IC}}/\cos\theta_{\textrm{IC}}$ with at least energy $E_{\textrm{th}} = 10$~TeV is 
\begin{equation}
E_{\mu, \textrm{min}} = E_{\textrm{th}} \, e^{b \, X_{\textrm{IC}}} + \left( e^{b \, X_{\textrm{IC}}}-1 \right) \, \frac{a}{b} ~,
\label{eq:Emumin}
\end{equation}
where $a$ and $b$ are given in Eq.~(\ref{eq:dedx}), we take $d_{\textrm{IC}} = 1.45$~km as the vertical depth of IceCube and 
\begin{equation}
\cos \theta_{\textrm{IC}} = \sqrt{1-\frac{\sin^2\theta_z}{(1+d_{\textrm{IC}}/(R_\oplus-d_{\textrm{IC}}))^2}} 
\label{eq:cosIC}
\end{equation}
is the local angle at the surface of the Earth with respect to the position of the detector taking into account the curvature of the Earth.

\begin{figure}[t]
	\includegraphics[width=\textwidth]{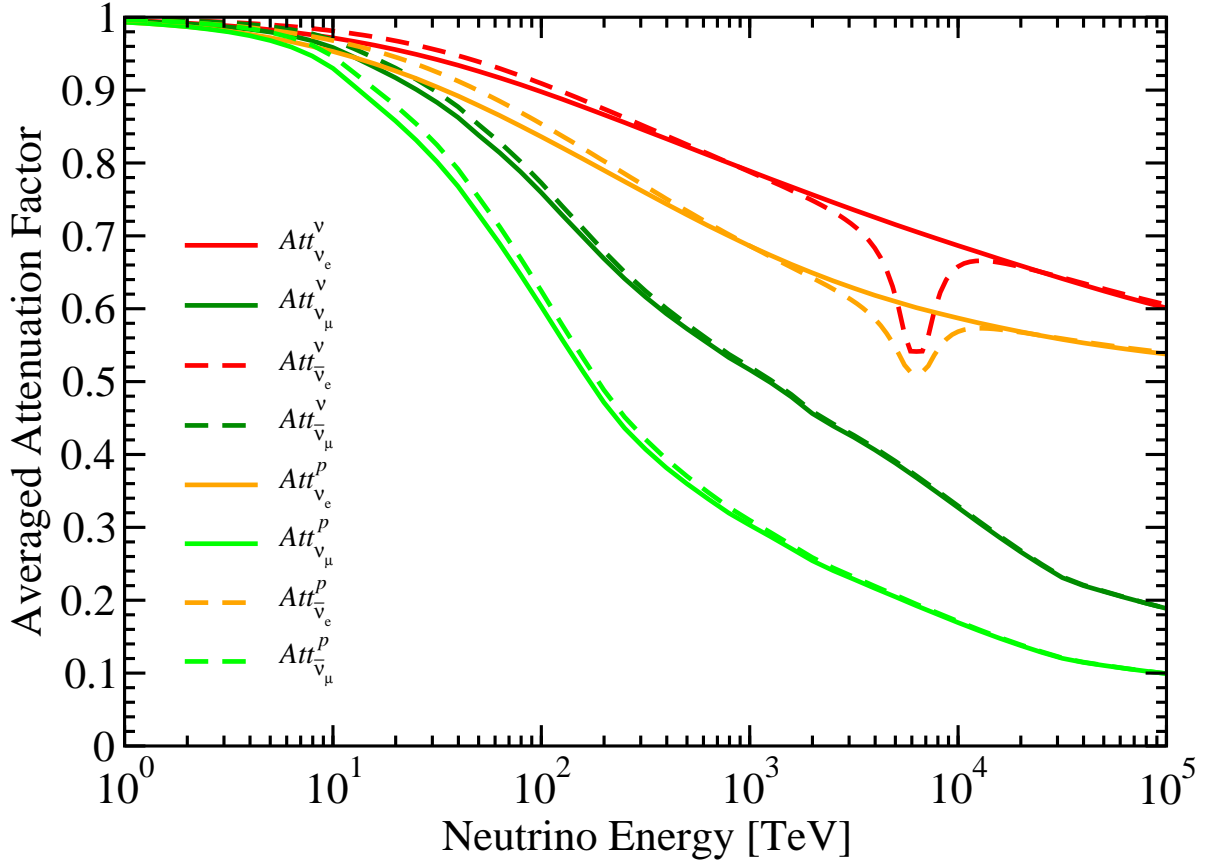}
	\caption{\textbf{\textit{Whole-sky averaged attenuation and regeneration factors for atmospheric neutrinos traversing the Earth}}, $Att^f_{\nu_\ell} (E_\nu)$, using the HGm+KM model (red lines for $\nu_e$ and $\bar\nu_e$ and dark green lines for $\nu_\mu$ and $\bar\nu_\mu$) for the conventional neutrinos ($f=\nu$) and the ZS+QGSM model (orange lines for $\nu_e$ and $\bar\nu_e$ and light green lines for $\nu_\mu$ and $\bar\nu_\mu$) for prompt neutrinos from charmed meson decays ($f=p$), from the calculations of Refs.~\cite{Sinegovsky:2011ab, Petrova:2012qf, Sinegovskaya:2013wgm, Sinegovskaya:2014pia}. Neutrinos (solid lines) and antineutrinos (dashed lines) are shown separately.}
	\label{fig:Attnuatm}
\end{figure}

The quantity $\cos\theta_z^*$ is the cosine of the local zenith angle at the altitude of the first interaction\footnote{Although the average altitude of the first interaction depends on $\theta_z$, we take its value at $\theta_z \simeq \pi/2$~\cite{Chirkin:2004ic}.}, $h_{\textrm{atm}} = 32$~km, and takes into account the curvature of the Earth, representing a non-negligible correction for angles larger than $70^\circ$~\cite{Lipari:1993hd}. Analogously to Eq.~(\ref{eq:cosIC}), 
\begin{equation}
\cos\theta_z^* = \sqrt{1-\frac{\sin^2\theta_z}{(1 + h_{\textrm{atm}}/R_\oplus)^2}} ~.
\label{eq:cos*}
\end{equation}

Finally, to compute Eq.~(\ref{eq:Patm}), the parametrization for $d\phi^f_{\nu_\mu}(E_\nu, \theta_z)/dE_\nu d\cos\theta_z$ is given by Eqs.~(\ref{eq:Patm1})-(\ref{eq:cos*}), but with $z_{i, \textrm{min}} = 1/(1-r_i)$. Moreover, as done by the IceCube Collaboration~\cite{Aartsen:2013jdh, Aartsen:2014gkd}, the suppression factor, \textit{i.e.}, the second term in the right-hand side in Eq.~(\ref{eq:Patm}), is bounded from above at 0.9 to cover uncertainties in hadronic interaction models.

Once we have the passing probability for atmospheric $\nu_\mu$ and $\bar\nu_\mu$ (conventional and prompt), we can compute the whole-sky averaged attenuation and regeneration factor, which is given by
\begin{equation}
Att^f_{\nu_\mu} (E_\nu) =  \dfrac{ \int_{-1}^{0} P^f_{\nu_\mu}(E_\nu, \theta) \, \frac{d\phi^f_{\nu_\mu} (E_\nu, \theta)}{dE_\nu d\cos\theta} \, d\cos\theta + \int_{0}^{1} Att^f_{\nu_\mu} (E_\nu, \theta) \, \frac{d\phi^f_{\nu_\mu}(E_\nu, \theta)}{dE_\nu d\cos\theta} \, d\cos\theta}{\int_{-1}^{1} \frac{d\phi^f_{\nu_\mu} (E_\nu, \theta)}{dE_\nu d\cos\theta} \, d\cos\theta} ~.
\end{equation}
On the other hand, for the nadir (or zenith) angle-dependent (conventional and prompt) atmospheric $\nu_e$ and $\bar\nu_e$ fluxes, the corresponding $4\pi$-averaged attenuation and regeneration factor is
\begin{equation}
Att^f_{\nu_e} (E_\nu) =  \dfrac{ \int_{-1}^{0} \frac{d\phi^f_{\nu_e} (E_\nu, \theta)}{dE_\nu d\cos\theta} \, d\cos\theta + \int_{0}^{1} Att^f_{\nu_e} (E_\nu, \theta) \, \frac{d\phi^f_{\nu_e}(E_\nu, \theta)}{dE_\nu d\cos\theta} \, d\cos\theta}{\int_{-1}^{1} \frac{d\phi^f_{\nu_e} (E_\nu, \theta)}{dE_\nu d\cos\theta} \, d\cos\theta} ~.
\end{equation}

In both cases, we have explicitly written the nadir (or zenith) angle dependence, which we use in our calculations. The results, using the calculations of  Refs.~\cite{Sinegovsky:2011ab, Petrova:2012qf, Sinegovskaya:2013wgm, Sinegovskaya:2014pia}, for the conventional atmospheric neutrinos with the HGm+KM model (red lines for $\nu_e$ and $\bar\nu_e$ and dark green lines for $\nu_\mu$ and $\bar\nu_\mu$) and for the prompt atmospheric neutrinos with the ZS+QGSM model (orange lines for $\nu_e$ and $\bar\nu_e$ and light green lines for $\nu_\mu$ and $\bar\nu_\mu$), are shown in Fig.~\ref{fig:Attnuatm}. Again, we depict the suppression factors for neutrinos (solid lines) and antineutrinos (dashed lines) separately and the Glashow resonance is clearly visible for both types of fluxes. Note that the attenuation and regeneration factors for prompt $\nu_e$ and $\bar\nu_e$ atmospheric neutrinos are similar to the factors for astrophysical $\nu_e$ and $\bar\nu_e$ neutrinos shown in Fig.~\ref{fig:Attastro}, unlike what happens for conventional neutrinos and for $\nu_\mu$ and $\bar\nu_\mu$ prompt and conventional fluxes. The former is due to the fact that the conventional flux is nonisotropic, with the maximum around the horizon, where very little absorption takes place. The latter is due to the extra veto applied to $\nu_\mu$ and $\bar\nu_\mu$ atmospheric neutrino fluxes, as discussed above.

\section{Statistical analysis}
\label{sec:analysis}

We perform an unbinned extended maximum likelihood analysis, using the energy and the event topology information of all of the 36 high-energy IceCube events. The full likelihood is given by
\begin{equation}
\like = e^{-N_a - N_\nu - N_\mu - N_p} \prod_{i = 1}^{N^{\textrm{sh}}_{\textrm{obs}}} \like^{\textrm{sh}}_{i} \prod_{i = 1}^{N^{\textrm{tr}}_{\textrm{obs}}} \like^{\textrm{tr}}_{i} ~,
\end{equation}
where $N_a$, $N_\nu$, $N_\mu$ and $N_p$ refer to the expected astrophysical (signal) neutrino, conventional atmospheric neutrino, atmospheric muon and (when used) prompt atmospheric neutrino total number of events, respectively, and $N^{\textrm{sh}}_{\textrm{obs}}$ and $N^{\textrm{tr}}_{\textrm{obs}}$ are the total number of observed showers and tracks, all in the energy interval under analysis. The partial likelihood for each type of event topology, $k=\{\textrm{tr, sh}\}$, is
\begin{equation}
\like^k_i = N_a \, \pdf_{i}^{k, a} (\{\alpha\}, \gamma) + N_\nu \, \pdf_{i}^{k, \nu} + N_\mu \, \pdf_{i}^{k, \mu} + N_p \, \pdf_{i}^{k, p} ~,
\label{eq:fullLike}
\end{equation}
where for each event observed at IceCube, we obtain the probability density function (PDF) for each type of neutrino flux and define the normalized PDF, $\pdf_{i}^{k, f}$, as the probability distribution for an observed event with energy $E_{\textrm{dep}, i}$ and topology $k$ caused by a flux of type $f$ of incoming neutrinos.  For an isotropic flux of astrophysical neutrinos with flavor combination at Earth $\{\alpha_e \, : \, \alpha_\mu \, : \, \alpha_\tau\}_\oplus$ and spectrum $d\phi^a/dE_\nu \propto E_\nu^{-\gamma}$ (assuming the same spectrum and normalization for antineutrinos),
\begin{equation}
\pdf_{i}^{k, a} (\{\alpha\}, \gamma) = \frac{1}{\sum_{\ell, j} \alpha_\ell \int_\Emin^\Emax dE_{\textrm{dep}} \,  \frac{dN_{\ell}^{j, a}}{dE_{\textrm{dep}} }} \, \sum_\ell \alpha_\ell \, \frac{dN_{\ell}^{k, a}}{dE_{\textrm{dep},i}}   ~.
\end{equation}
The integral in the denominator covers the EM-equivalent deposited energy region considered for each analysis and the sum goes over the three neutrino flavors, $\ell = \{e, \mu, \tau\}$, and the type of event topology, $j=\{\textrm{tr, sh}\}$. The spectra $dN_{\ell}^{\textrm{sh}, a}/dE_{\textrm{dep}}$ and $dN_{\ell}^{\textrm{tr}, a}/dE_{\textrm{dep}}$ resulting from the sum of all the partial contributions from different processes to showers and tracks from neutrinos and antineutrinos of flavor $\ell$ are detailed in Appendix~\ref{app:totrates}. 

Likewise, for the atmospheric backgrounds we have
\begin{eqnarray}
\pdf_{i}^{k, \nu} & = & \frac{1}{\sum_{j} \int_\Emin^\Emax dE_{\textrm{dep}} \,  \frac{dN^{j, \nu}}{dE_{\textrm{dep}} }} \, \frac{dN^{k, \nu}}{dE_{\textrm{dep},i}}   ~, \\
\pdf_{i}^{k, \mu} & = & \frac{1}{\sum_{j} \int_\Emin^\Emax dE_{\textrm{dep}} \,  \frac{dN^{j, \mu}}{dE_{\textrm{dep}} }} \, \frac{dN^{k, \mu}}{dE_{\textrm{dep},i}}   ~, \\
\pdf_{i}^{k, p} & = & \frac{1}{\sum_{j} \int_\Emin^\Emax dE_{\textrm{dep}} \,  \frac{dN^{j, p}}{dE_{\textrm{dep}} }} \, \frac{dN^{k, p}}{dE_{\textrm{dep},i}}   ~,
\end{eqnarray}
where the conventional and prompt atmospheric neutrino fluxes are used to compute the event distributions just as in the astrophysical case. However, note that for the conventional and prompt atmospheric neutrino fluxes the relative contributions from each flavor, and for neutrinos and antineutrinos, are obtained by using the corresponding flux for each case. The event distribution of atmospheric muons was discussed in Sec.~\ref{sec:bkg}. For the three background sources, we write the sum over flavors of the (neutrino-plus-antineutrino-induced) differential event spectra.

We perform our analyses of the IceCube events using the \textsc{MultiNest} nested sampling algorithm \cite{Feroz:2007kg, Feroz:2008xx, 2013arXiv1306.2144F}, with the likelihoods given above. We let the following parameters vary freely:
\begin{equation}
\{\alpha_e, \alpha_\mu, \gamma, N_a, N_\nu, N_\mu \} ~,
\label{eq:6Pparams}
\end{equation}
where $\alpha_\tau \equiv 1 - \alpha_e - \alpha_\mu$ and $N_p = 0$. We refer to this parameter set as ``6P''.  We compare these results with \textsc{MultiNest} runs where we fix $N_\nu$ and $N_\mu$ to the expected values provided by IceCube: $N_\nu = 6.6$ (2.4 when the energy range is restricted to $E_{\textrm{dep}} > 60$~TeV) and $N_\mu = 8.4$ (0.4 above 60~TeV), $N_p=0$, and the energy power-law index of the astrophysical flux to their IceCube best-fit value, $\gamma = 2.3$~\cite{Aartsen:2014gkd}. This parameter set is called ``3P''.

To probe the effect of a possible prompt atmospheric contribution from the decay of charmed mesons, we define
\begin{equation}
7P \equiv  \{\alpha_e, \alpha_\mu, \gamma, N_a, N_\nu, N_\mu, N_p\} ~,
\label{eq:7Pparams}
\end{equation}
with the further imposition of a half-Gaussian prior on the number of prompt atmospheric neutrinos $N_p$ widths $\sigma_{N_p} = 5.5$ (3.2) for the $> 28$~TeV ($> 60$~TeV) energy ranges. This reflects the $1\sigma$ C.L. exclusions provided by the lower energy $\nu_\mu$ fluxes quoted in Ref.~\cite{Aartsen:2014gkd}. For consistency, this case also includes a continuous Poisson prior on $N_\mu$ centered at the expected fluxes given above.

Finally, for the cases of a broken power-law spectrum and when including track misidentification, we also show results varying $\{\alpha_e, \alpha_\mu, \gamma, N_a\}$ and setting the other parameters to their IceCube best-fit values. These analyses are referred to as ``4P''.

For our analyses we allow the full range of flavor compositions as seen at Earth: $\alpha_\ell$ may go from 0 to 1 for each flavor $\ell = \{e,\mu,\tau\}$, as long as their sum is unity. In reality, neutrino oscillations are averaged out after propagation over large distances, leading to a much smaller allowed region in flavor space as the flavor composition averages to a small part of the parameter space at Earth, around $(1:1:1)_\oplus$. If the origin of the high-energy neutrinos is purely charged pion decay, the expected composition $(1:2:0)_S$ would oscillate to an observed $\simeq (1.04:0.99:0.97)_\oplus$ ratio, while neutron decay sources yielding $(1:0:0)_S$ give rise to a $\simeq (0.55:0.24:0.21)_\oplus$ flavor ratio at Earth. For reference, the full space of allowed \textit{source} neutrinos is shown in our ternary plots as a thin blue triangle. At present, the paucity of observed events gives a very flat likelihood within this area, resulting in very little statistical power with regards to the IceCube neutrino composition at the sources. Furthermore, we find that the Bayes factor comparing the full flavor range to the averaged-oscillation triangle is close to one, meaning that there is no evidence at present for such a restriction to be particularly good or bad. 

We compare the results with different choices for the minimum and maximum EM-equivalent deposited energies for the analysis, $\Emin$ and $\Emax$, and show how these choices affect what one can conclude about the data sample. We consider four distinct energy ranges: 28~TeV$- 3$~PeV, the range which covers all published high-energy IceCube events, 60~TeV$- 3$~PeV which eliminates most of the background atmospheric muon events and is the energy interval considered by the IceCube Collaboration to present their results, as well as 28~TeV$- 10$~PeV and 60~TeV$- 10$~PeV. By extending the analysis to 10~PeV, we cover the energy region of the Glashow resonance at $E_\nu \simeq 6.3$~PeV where a few events are expected if the astrophysical spectrum extends beyond a few PeV. The absence of any observed events above 2~PeV could indicate a small electron neutrino component, a very soft spectrum, or a break in the spectrum around a few PeV. This could nicely be connected with the ultrahigh-energy cosmic-ray paradigm~\cite{Anchordoqui:2013qsi, Anchordoqui:2013dnh, Winter:2014pya, Murase:2014tsa}. Conversely, the main effect of removing events below 60~TeV is a vast reduction in the atmospheric event contamination. While this yields a much cleaner astrophysical signal, it also removes 16 of the 36 observed events, reducing the statistical power of the analysis. 

To avoid subtleties related to the problem's particular geometry in the large parameter space, we present most of our results as profile likelihoods of the plotted parameter space, rather than as Bayesian posteriors. To do this, we bin the samples produced by \textsc{MultiNest} with respect to the relevant parameters, and within each bin $j$ in the parameter space, we find the point of maximum likelihood. Then, we define the $\lambda_j \equiv -2 \log (\like_j/\like_{\textrm{max}})$, where $\like_{\textrm{max}}$ is the likelihood of the overall best-fit point. The test statistic $\lambda_j$ is assumed to be distributed like a $\chi^2$ with $n$ degrees of freedom, with $n$ equal to the number of parameters in the corresponding analysis. When instructive we also present quantities obtained with the Bayesian posteriors.

\begin{sidewaystable}[htbp]
	\caption{\textbf{\textit{Summary of the best-fit points (Bayesian posterior means in parentheses) in each model we considered to analyze the IceCube data, with 1$\boldsymbol{\sigma}$ errors.}} Fixed quantities are indicated by italics. 6P refers to the six-parameter set defined in Eq.~(\ref{eq:6Pparams}), while 3P fixes $\gamma$ to 2.3, and the background counts $N_\nu$ and $N_\mu$ to the rates estimated by IceCube. The 4P case allows the spectral index to vary; ``4P+br'' indicates a break of one unit in the spectral index of the astrophysical neutrino spectrum at $E_\nu=1$~PeV, as discussed in Sec.~\ref{sec:glashow}. Finally the rows labeled ``20\% mis-ID'' (``30\% mis-ID'') include a 20\% (30\%) fraction of tracks misidentified as showers, as discussed in Sec.~\ref{sec:misid}.  The final column indicates the $p$ value of the flavor composition  $(1:1:1)_\oplus$, assuming the test statistic $-2 \log(\like/\like_{\textrm{max}})$ to follow a $\chi^2$ distribution. }\label{tab:resulttable}
	\begin{tabular*}{\textwidth}{|c | @{\extracolsep{\fill}} c | c c c c c c |} 
		\hline \hline
		Energy range & Params. &$(\alpha_e:\alpha_\mu:\alpha_\tau)_\oplus$ & $\gamma$ & $N_a$ & $N_\nu$ & $N_\mu$ & $p(1:1:1)_\oplus$ \\ \hline
		\multirow{ 3}{*}{28~TeV$- 3$~PeV} & {6P}& $(0.75:0.25:0.00)$ &2.96 $^{+ 0.34}_{- 0.37}$ (2.86 $\pm$ 0.28) & 26.2 $^{+ 8.8}_{- 8.9}$ (25.3 $\pm$ 5.7) & 4.8 $^{+ 9.1}_{- 4.4}$ (7.9 $\pm$ 4.7) & 4.7 $^{+ 4.4}_{- 3.7}$ (6.0 $\pm$ 3.1) & 0.84 \\
		& 4P & $(0.86:0.14:0.00)$ &2.82 $^{+ 0.31}_{- 0.31}$ (2.85 $\pm$ 0.26) & 23.6 $^{+ 6.3}_{- 5.7}$ (24.8 $\pm$ 5.2) &  \textit{6.6} & \textit{8.4} &0.42 \\
		& 3P & $(0.92:0.08:0.00)$ & \textit{2.3} & 20.6$^{+ 6.6}_{- 4.8}$ (22.2 $\pm$ 5.0) &  \textit{6.6} & \textit{8.4} & 0.29 \\ 
		20\% mis-ID					  & 4P &$(0.77:0.23:0.00)$ &2.76 $^{+ 0.31}_{- 0.33}$ (2.78 $\pm$ 0.27) & 22.4 $^{+ 6.7}_{- 5.3}$ (23.8 $\pm$ 5.2) &  \textit{6.6} & \textit{8.4} &0.71	\\					  \hline
		\multirow{ 3}{*}{28~TeV$- 10$~PeV} & 6P   &$(0.63:0.27:0.10)$ &3.02 $^{+ 0.38}_{- 0.35}$ (2.95 $\pm$ 0.25) & 26.9 $^{+ 9.5}_{- 9.8}$ (25.9 $\pm$ 5.6) & 4.1 $^{+ 9.5}_{- 9.8}$ (7.5 $\pm$ 4.5) & 4.9 $^{+ 9.5}_{- 9.8}$ (5.9 $\pm$ 3.0) & 0.89 \\
		& 4P & $(0.85:0.14:0.01)$ &2.90 $^{+ 0.32}_{- 0.31}$ (2.92 $\pm$ 0.24) & 23.7 $^{+ 6.7}_{- 5.5}$ (25.1 $\pm$ 5.2) &  \textit{6.6} & \textit{8.4} &0.48 \\
		& 3P &$(0.00:0.00:1.00)$ & \textit{2.3} & 21.1 $^{+ 5.9}_{- 5.4}$ (21.9 $\pm$ 4.8) &  \textit{6.6} & \textit{8.4} & 0.16  \\			 
		20\% mis-ID					  & 4P & $(0.75:0.25:0.00)$ &2.87 $^{+ 0.27}_{- 0.41}$ (2.86 $\pm$ 0.25) & 23.2 $^{+ 6.0}_{- 6.3}$ (24.1 $\pm$ 5.1) &  \textit{6.6} & \textit{8.4} &0.79\\ \hline
		\multirow{ 4}{*}{60~TeV$- 3$~PeV} & 6P & $(0.98:0.00:0.02)$ & 2.34 $^{+ 0.39}_{- 0.31}$ (2.40 $\pm$ 0.29) & 13.7 $^{+ 7.2}_{- 4.2}$ (16.0 $\pm$ 4.0) & 6.5 $^{+ 4.1}_{- 5.5}$ (4.6 $\pm$ 3.1) & 0.1 $^{+ 4.8}_{- 0.0}$ (3.0 $\pm$ 2.0) & 0.50 \\
		& 4P & $(0.77:0.23:0.00)$ &2.48 $^{+ 0.31}_{- 0.33}$ (2.52 $\pm$ 0.27) & 16.6 $^{+ 4.8}_{- 4.9}$ (17.6 $\pm$ 4.1) &  \textit{2.4} & \textit{0.4} &0.69 \\
		& 4P+br &$(0.76:0.24:0.00)$ &2.35 $^{+ 0.36}_{- 0.34}$ (2.37 $\pm$ 0.31) & 16.5 $^{+ 4.7}_{- 4.9}$ (17.6 $\pm$ 4.1) &  \textit{2.4} & \textit{0.4} &0.58\\
		& 3P &$(0.82:0.18:0.00)$ & \textit{2.3} & 16.2 $^{+ 5.5}_{- 4.2}$ (17.4 $\pm$ 4.2) &  \textit{2.4} & \textit{0.4} & 0.60 \\ 
		20\% mis-ID					  & 4P &$(0.68:0.32:0.00)$ &2.48 $^{+ 0.30}_{- 0.34}$ (2.49 $\pm$ 0.28) & 16.4 $^{+ 4.7}_{- 5.0}$ (17.4 $\pm$ 4.1) &  \textit{2.4} & \textit{0.4} &0.88 \\ \hline			  
		\multirow{ 4}{*}{60~TeV$- 10$~PeV} & 6P& $(0.01:0.01:0.98)$ &2.48 $^{+ 0.33}_{- 0.34}$ (2.58 $\pm$ 0.25) & 16.6 $^{+ 4.9}_{- 6.1}$ (16.4 $\pm$ 4.0) & 1.5 $^{+ 7.0}_{- 1.1}$ (4.3 $\pm$ 3.0) & 2.2 $^{+ 2.8}_{- 2.2}$ (2.9 $\pm$ 2.0) & 0.61 \\
		& 4P &$(0.00:0.02:0.98)$ &2.50 $^{+ 0.36}_{- 0.28}$ (2.65 $\pm$ 0.25) & 16.4 $^{+ 4.8}_{- 4.8}$ (17.8 $\pm$ 4.1) &  \textit{2.4} & \textit{0.4} &0.69\\
		& 4P+br &$(0.75:0.25:0.00)$ &2.43 $^{+ 0.31}_{- 0.34}$ (2.44 $\pm$ 0.29) & 16.5 $^{+ 4.8}_{- 4.8}$ (17.6 $\pm$ 4.1) &  \textit{2.4} & \textit{0.4} &0.65\\
		& 3P &$(0.00:0.00:1.00)$ & \textit{2.3} & 16.2 $^{+ 5.5}_{- 4.0}$ (17.3 $\pm$ 4.1) &  \textit{2.4} & \textit{0.4} & 0.33  \\
		20\% mis-ID					&4P   &$(0.00:0.11:0.89)$ &2.50 $^{+ 0.35}_{- 0.29}$ (2.62 $\pm$ 0.25) & 16.7 $^{+ 4.8}_{- 4.9}$ (17.5 $\pm$ 4.1) &  \textit{2.4} & \textit{0.4} &0.82 \\
		30\% mis-ID					&4P   & $(0.00:0.18:0.82)$ &2.49 $^{+ 0.35}_{- 0.30}$ (2.61 $\pm$ 0.25) & 16.3 $^{+ 5.8}_{- 3.9}$ (17.4 $\pm$ 4.1) &  \textit{2.4} & \textit{0.4} &0.84   \\
		\hline \hline
	\end{tabular*}
	%
	\caption{\textbf{\textit{Same as Table~\ref{tab:resulttable} but for the 7P analyses}}, \textit{i.e.}, including the number of prompt atmospheric neutrinos $N_p$ associated with charmed meson decays, as well as a prior on the $N_p$ and $N_\mu$, as explained after Eq.~(\ref{eq:7Pparams}). } \label{tab:charmtable}
	\begin{tabular*}{\textwidth}{|c | @{\extracolsep{\fill}} c | c  c c c c  c |} 
		\hline \hline
		Energy range & $(\alpha_e:\alpha_\mu:\alpha_\tau)_\oplus$ & $\gamma$ & $N_a$ & $N_\nu$ & $N_\mu$ & $N_p$ & $p(1:1:1)_\oplus$  \\ \hline
		\multirow{ 1}{*}{28~TeV$- 3$~PeV} &$(0.75:0.25:0.00)$ &2.93 $^{+ 0.32}_{- 0.39}$ (2.80 $\pm$ 0.40) & 24.6 $^{+ 10.0}_{- 7.2}$ (20.7 $\pm$ 6.4) & 4.3 $^{+ 6.9}_{- 4.0}$ (6.8 $\pm$ 3.9) & 6.6 $^{+ 2.6}_{- 2.2}$ (7.1 $\pm$ 2.0) & 0.2 $^{+ 3.9}_{- 0.2}$ (4.7 $\pm$ 3.1) & 0.80 \\
		
		\multirow{ 1}{*}{28~TeV$- 10$~PeV} &$(0.61:0.30:0.09)$ &2.97 $^{+ 0.31}_{- 0.35}$ (2.91 $\pm$ 0.33) & 26.5 $^{+ 8.3}_{- 8.3}$ (21.6 $\pm$ 6.2) & 2.9 $^{+ 7.4}_{- 2.9}$ (6.3 $\pm$ 3.8) & 6.8 $^{+ 2.6}_{- 2.2}$ (7.0 $\pm$ 2.0) & 0.2 $^{+ 3.8}_{- 0.2}$ (4.5 $\pm$ 3.0) & 0.89\\
		
		\multirow{ 1}{*}{60~TeV$- 3$~PeV} & $(0.99:0.00:0.01)$ &2.23 $^{+ 0.44}_{- 0.31}$ (2.24 $\pm$ 0.36) & 11.9 $^{+ 7.3}_{- 3.5}$ (12.4 $\pm$ 4.2) & 6.8 $^{+ 3.4}_{- 4.2}$ (5.3 $\pm$ 2.9) & 0.1 $^{+ 0.7}_{- 0.1}$ (0.8 $\pm$ 0.6) & 0.7 $^{+ 3.2}_{- 0.4}$ (3.4 $\pm$ 1.8) & 0.43 \\	
		
		\multirow{ 1}{*}{60~TeV$- 10$~PeV} & $(0.01:0.01:0.98)$ &2.39 $^{+ 0.40}_{- 0.28}$ (2.47 $\pm$ 0.31) & 14.3 $^{+ 4.9}_{- 5.7}$ (12.9 $\pm$ 4.1) & 4.5 $^{+ 4.2}_{- 2.8}$ (4.9 $\pm$ 2.8) & 0.1 $^{+ 0.7}_{- 0.1}$ (0.8 $\pm$ 0.6) & 1.0 $^{+ 2.6}_{- 0.7}$ (3.2 $\pm$ 1.8) & 0.55 \\
		
		\hline \hline
	\end{tabular*}
\end{sidewaystable}

\begin{figure}
	\begin{tabular}{c c}
		\includegraphics[width=0.5\textwidth]{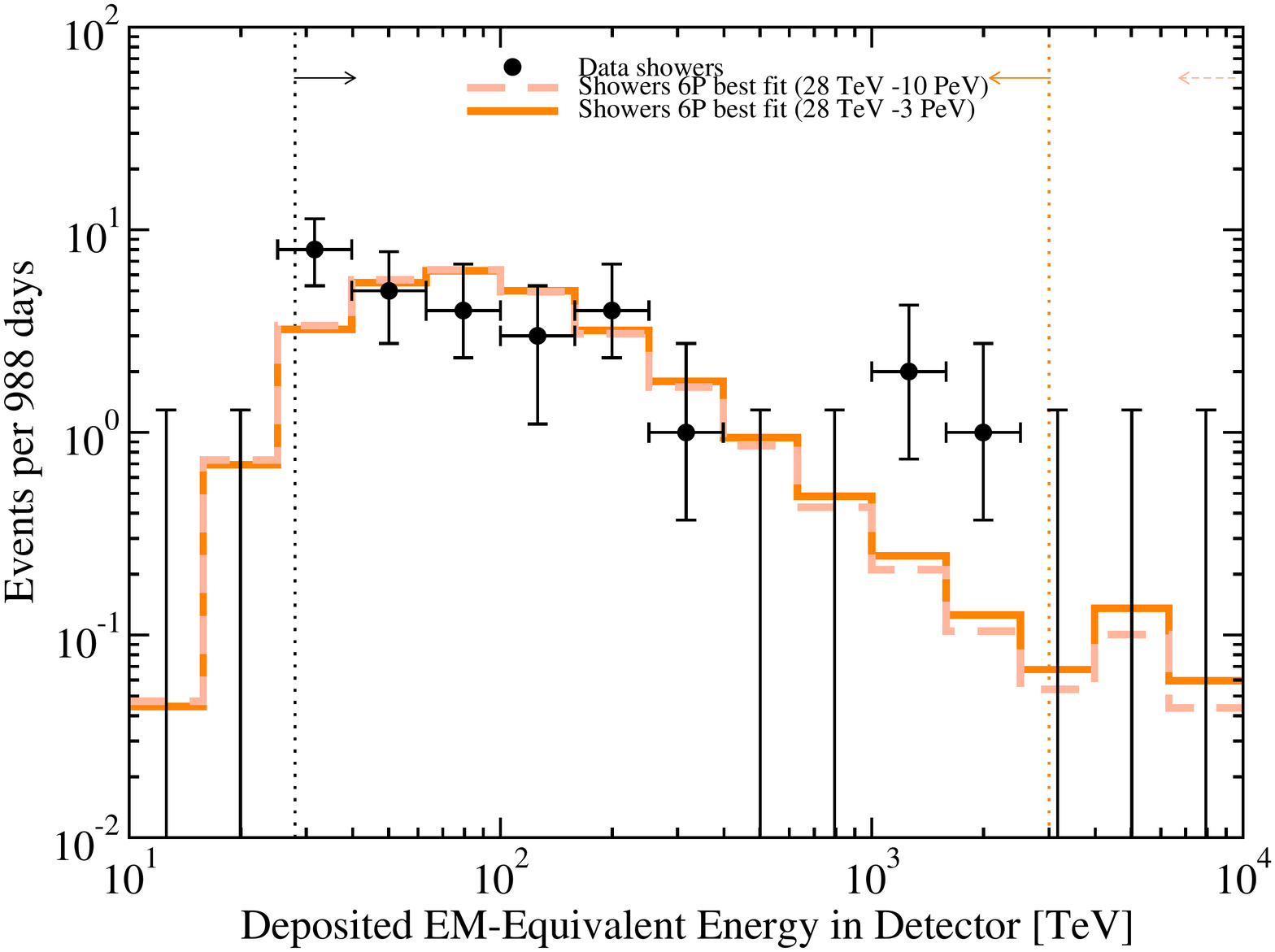} & \includegraphics[width=0.5\textwidth]{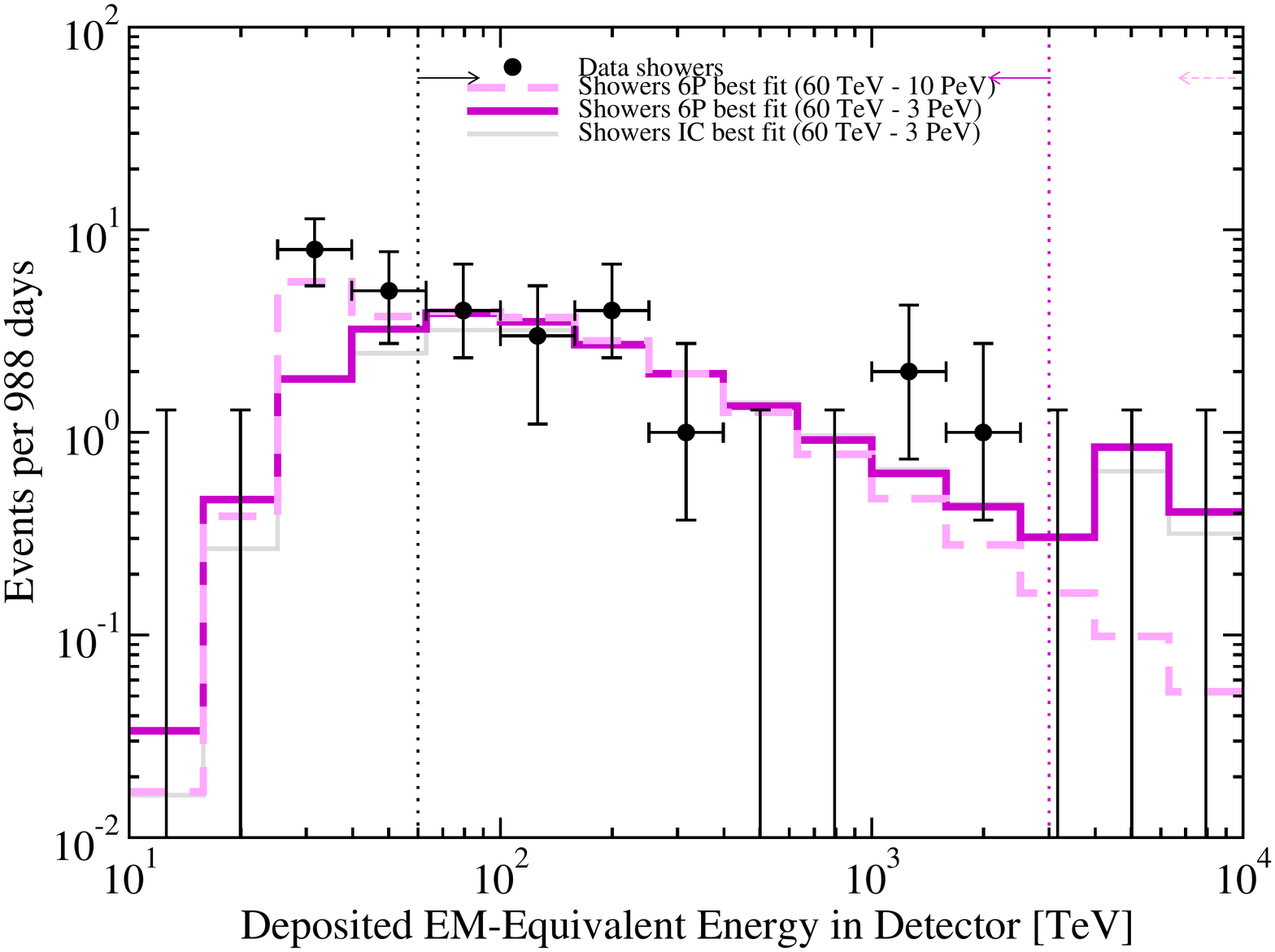} \\
		\includegraphics[width=0.5\textwidth]{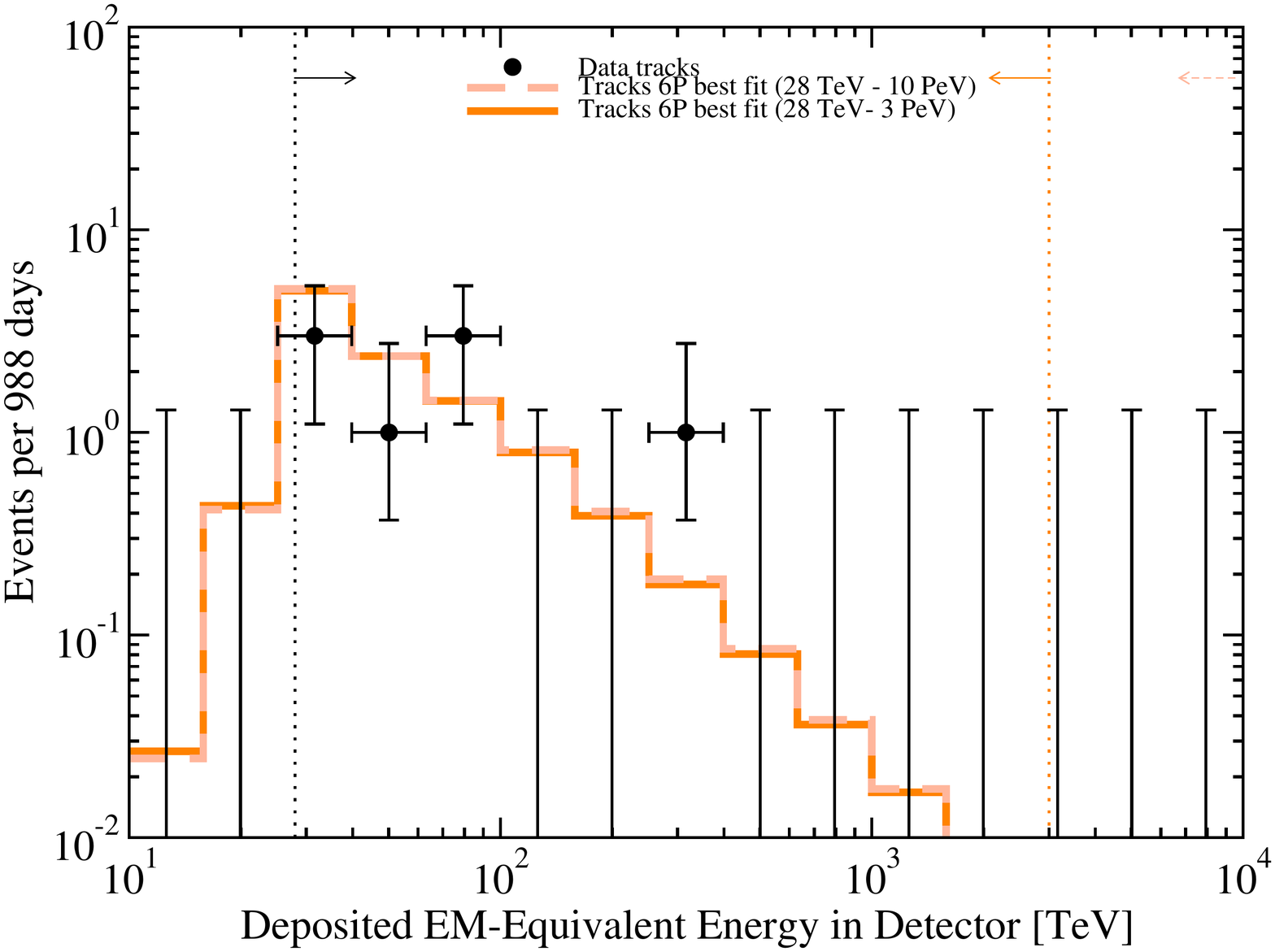} & \includegraphics[width=0.5\textwidth]{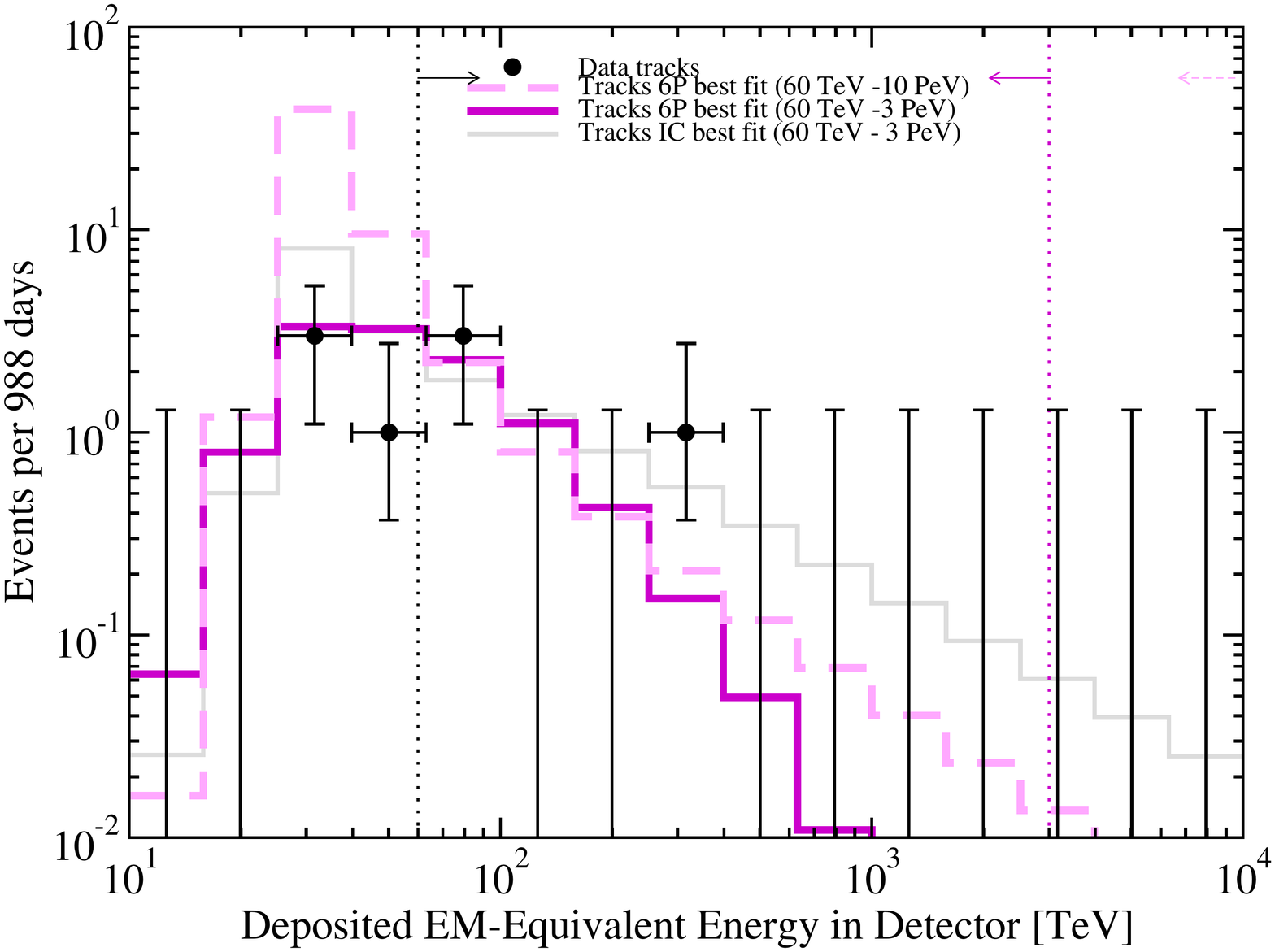} \\
		\includegraphics[width=0.5\textwidth]{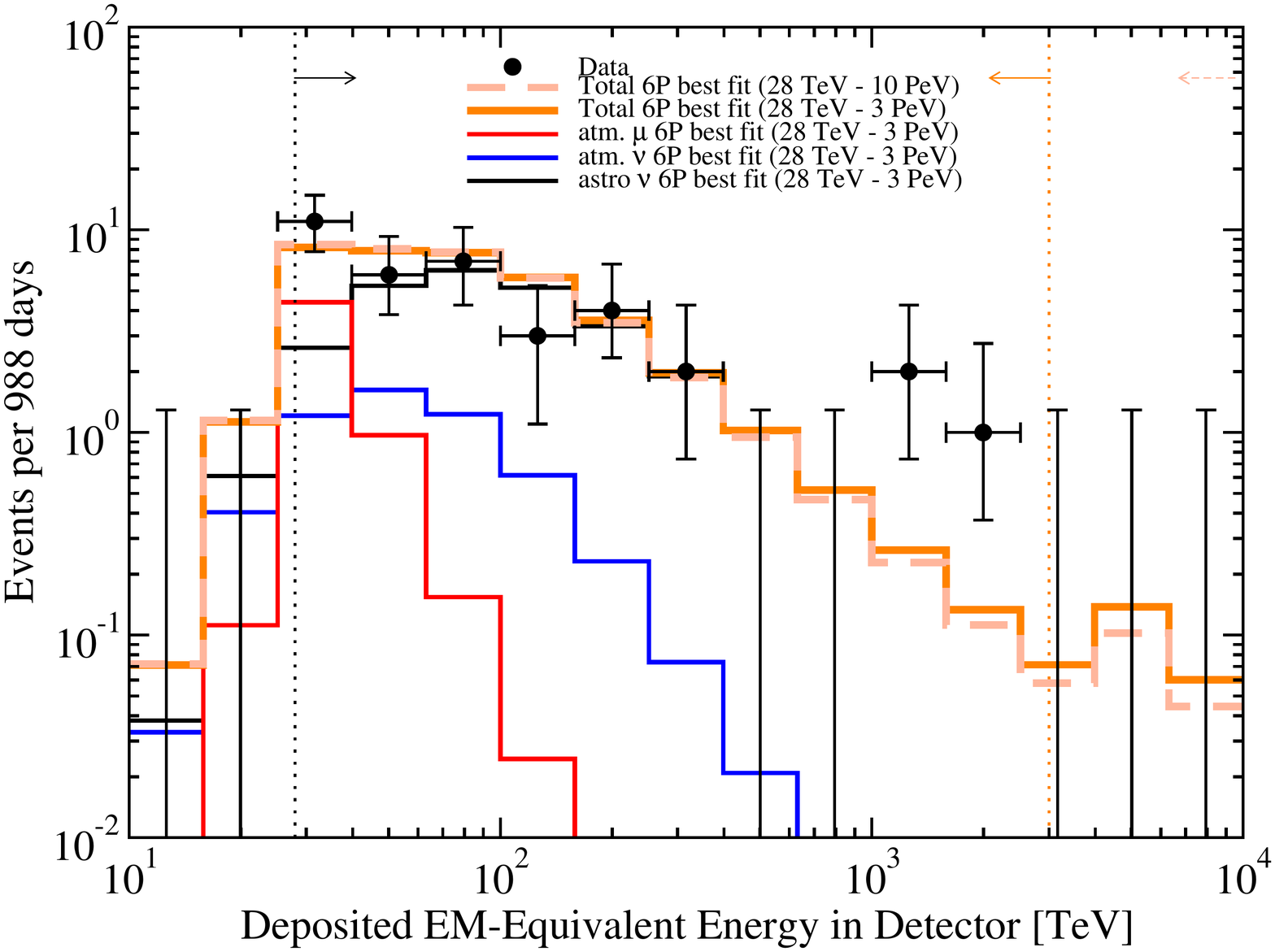} & \includegraphics[width=0.5\textwidth]{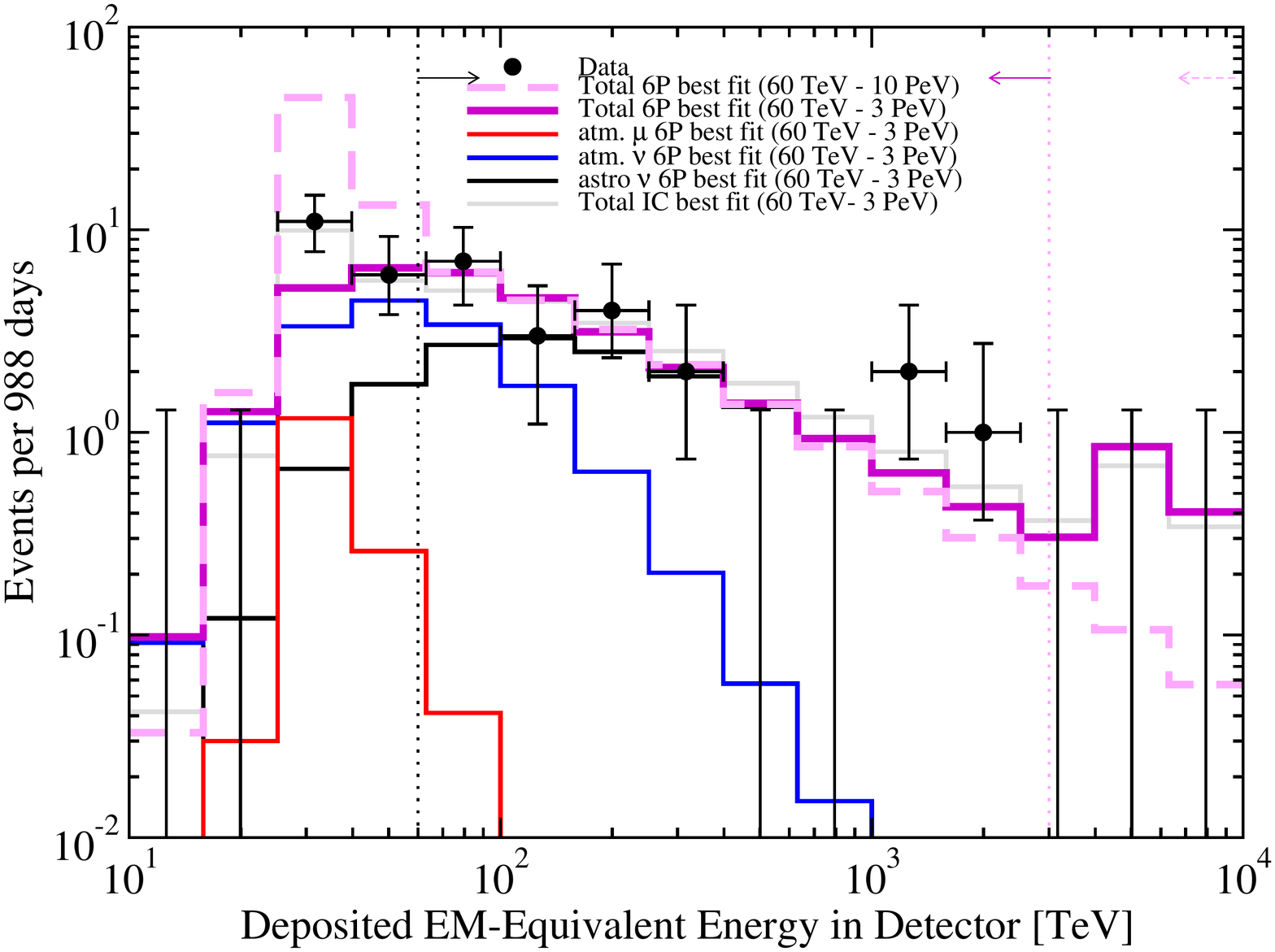} 
	\end{tabular}
	\caption{\textbf{\textit{Spectra of the best fits of our six-parameter (6P) analyses}}, overlaid with the best fits provided by the IceCube Collaboration and the binned high-energy neutrino event data with Feldman-Cousins errors~\cite{Feldman:1997qc}. Left panels: the spectrum of the 36 events above 28~TeV. Right panels: the spectrum of the 20 events detected above 60~TeV. Top panels show the shower component only; middle panels show the track component, and bottom panels show the total number of events (showers + tracks), in addition to explicitly showing the atmospheric event contributions to the 28~TeV$- 3$~PeV (60~TeV$- 3$~PeV) case in the left (right) panels. }
	\label{fig:BFspectra}
\end{figure}

\section{Results}
\label{sec:results}

In Table~\ref{tab:resulttable} we show the best-fit points from our analyses, as well as the Bayesian posterior means. In the following subsections, we present our results and summarize the conclusions that may be drawn from our analyses of the high-energy IceCube events. In Fig.~\ref{fig:BFspectra} we show the spectra of the best-fit points for the 6P analyses, divided into shower-only (top), track-only (middle), and total (bottom) contributions, along with the binned IceCube data with Feldman-Cousins errors~\cite{Feldman:1997qc}. The bottom panels also show the atmospheric neutrino and muon contributions to our best fit for the cases with an upper cut at 3~PeV. On the left we show the spectra of the different contributions for events observed above 28~TeV, while the panels on the right show fits to events above 60~TeV, where the background rates are much lower. On the right panels we also show the result for the IceCube best fit in the energy interval 60~TeV$- 3$~PeV, which assumed $(1 : 1: 1)_\oplus$ and obtained $\gamma = 2.3$ and $N_\phi = 4.5 \times 10^{-18} \, \textrm{GeV}^{-1} \, \textrm{cm}^{-2} \, \textrm{s}^{-1} \, \textrm{sr}^{-1}$, where $N_\phi$ is the total astrophysical flux at 100~TeV. From these plots, one can see how the events below 60~TeV push the astrophysical flux toward a softer spectrum. Let us also stress the importance of the flavor composition, by noting that cases with similar total spectra in the range where they are fitted, give rise to different shower and track spectra that allow the breaking of the degeneracy.

\begin{figure}
	\begin{tabular}{c c}
		\includegraphics[width=0.5\textwidth]{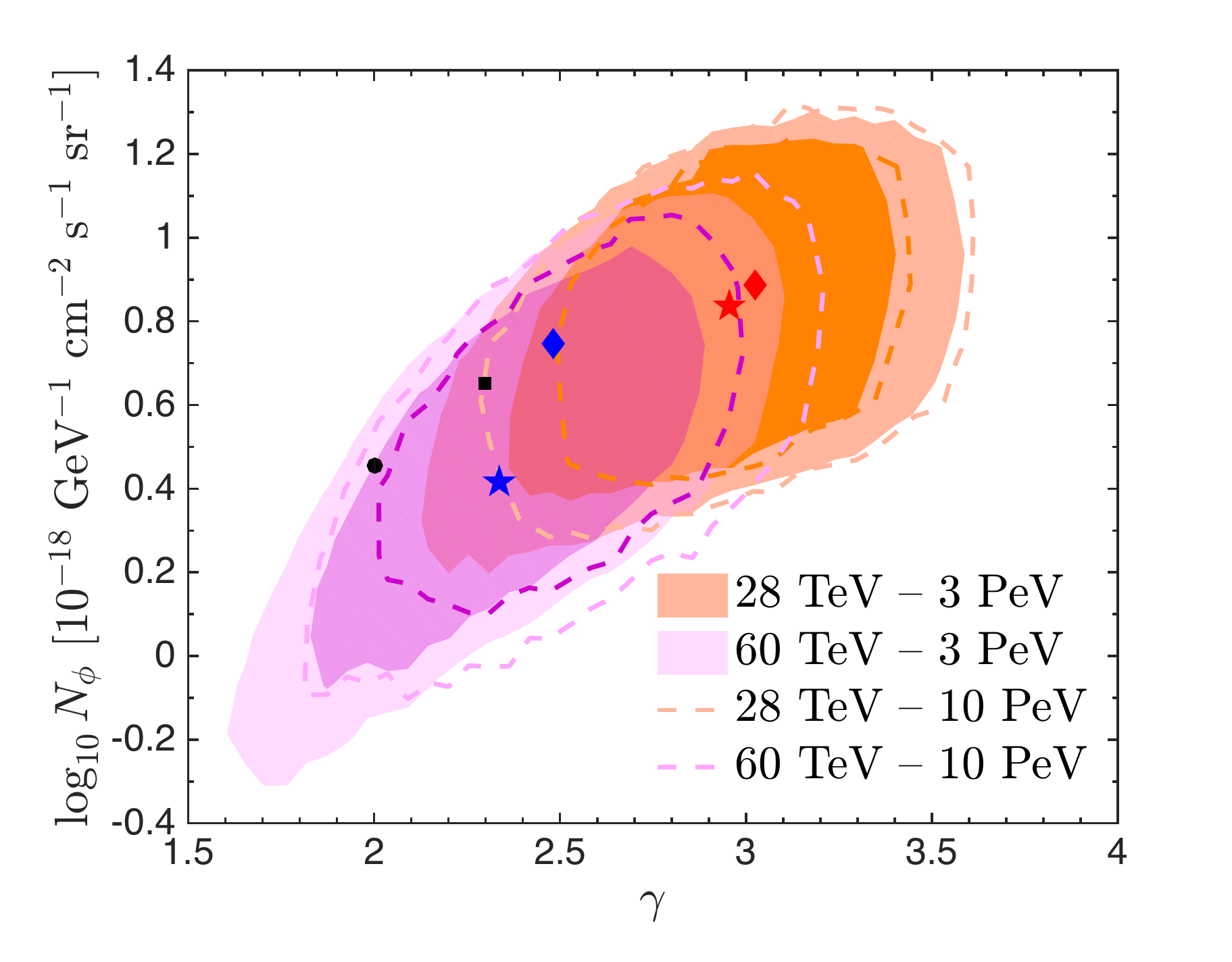} & \includegraphics[width=0.5\textwidth]{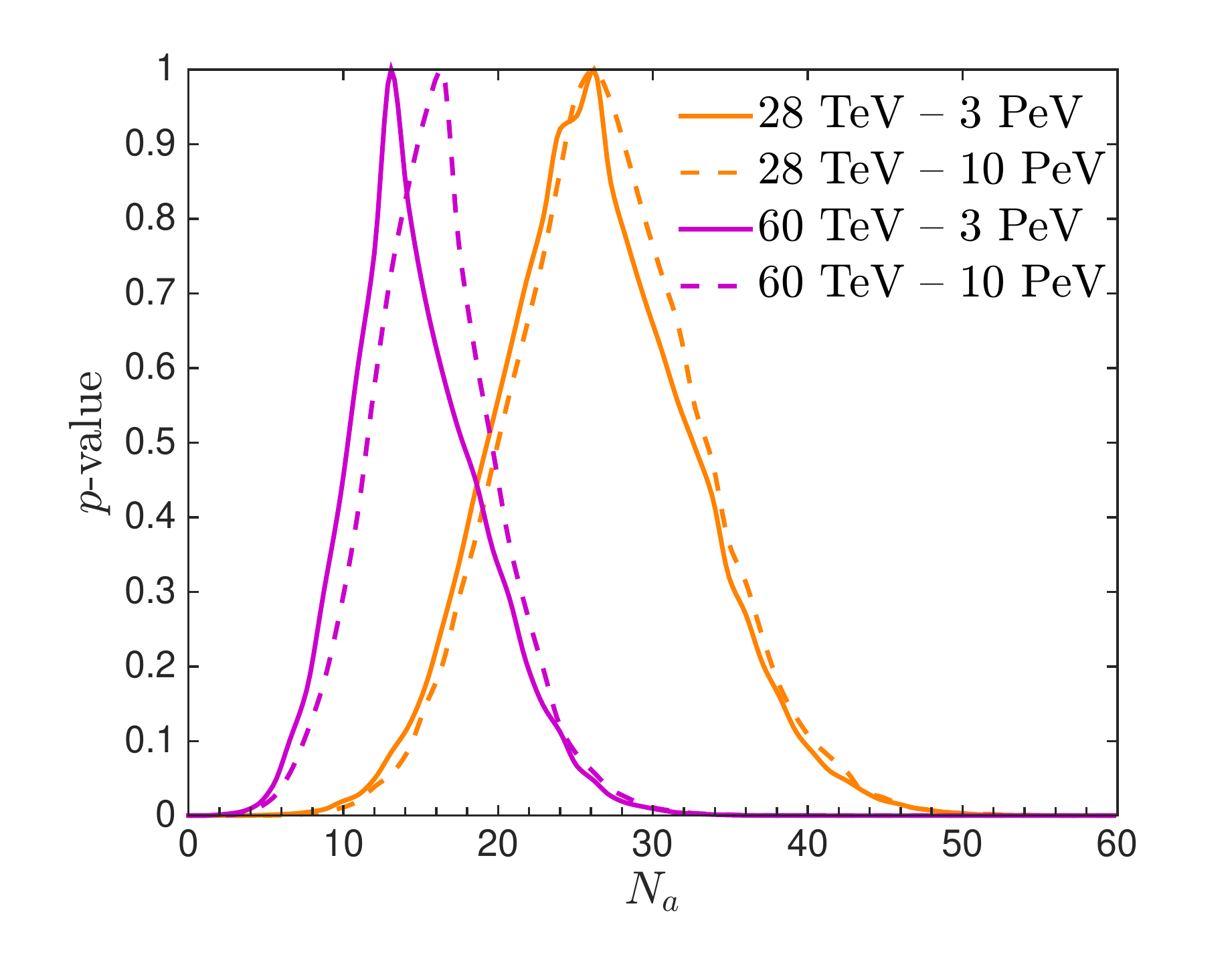} 
	\end{tabular}
	\caption{\textbf{\textit{Left panel: contours in the $\boldsymbol{\gamma}$ - $\boldsymbol{N_\phi}$ plane}}, at $1\sigma$ C.L. (dark colors) and $2\sigma$ C.L. (light colors). Filled contours (closed curves) represent the cases with a high-energy cut in the deposited energy of 3~PeV (10~PeV). Including the full energy range yields a preference for a larger spectral index due to the lack of events from the Glashow resonance. The blue star (diamond) is the best-fit point, $N_\phi = 2.6 \times 10^{-18} \, \textrm{GeV}^{-1} \, \textrm{cm}^{-2} \, \textrm{s}^{-1} \, \textrm{sr}^{-1}$ ($N_\phi = 5.6 \times 10^{-18} \, \textrm{GeV}^{-1} \, \textrm{cm}^{-2} \, \textrm{s}^{-1} \, \textrm{sr}^{-1}$), for 60~TeV$- 3$~PeV (60~TeV$- 10$~PeV), while the red star (diamond) is the best-fit point,  $N_\phi = 6.9 \times 10^{-18} \, \textrm{GeV}^{-1} \, \textrm{cm}^{-2} \, \textrm{s}^{-1} \, \textrm{sr}^{-1}$ ($N_\phi = 7.7 \times 10^{-18} \, \textrm{GeV}^{-1} \, \textrm{cm}^{-2} \, \textrm{s}^{-1} \, \textrm{sr}^{-1}$), for 28~TeV$- 3$~PeV (28~TeV$- 10$~PeV). The black square is the IceCube Collaboration's quoted best fit, $N_\phi = 4.5 \times 10^{-18} \, \textrm{GeV}^{-1} \, \textrm{cm}^{-2} \, \textrm{s}^{-1} \, \textrm{sr}^{-1}$ and $\gamma=2.3$, whereas the black circle is their best fit when the spectral index is fixed to $\gamma = 2$, $N_\phi = 2.85 \times 10^{-18} \, \textrm{GeV}^{-1} \, \textrm{cm}^{-2} \, \textrm{s}^{-1} \, \textrm{sr}^{-1}$~\cite{Aartsen:2014gkd}. \textbf{\textit{Right panel: $\boldsymbol{p}$ values along the range of astrophysical events $\boldsymbol{N_a}$.}} The results in both panels refer to 6P analyses.}
	\label{fig:eindex}
\end{figure}

\pagebreak

\subsection{Spectral index}

By letting the power-law index $\gamma$ of the astrophysical neutrino flux vary freely, for the 6P analysis we find a best fit $\gamma = 2.96^{+0.34}_{-0.37}$ for the energy range 28~TeV$- 3$~PeV. When events below 60~TeV are omitted, this is lowered to $\gamma = 2.34^{+0.39}_{-0.31}$, in perfect agreement with the best-fit value that is obtained by the IceCube Collaboration~\cite{Aartsen:2014gkd}. This is an indication of contamination by the steeper background spectra (subdominant above 60~TeV), which can explain a large fraction of the events below 60~TeV and force the astrophysical spectrum to be softer. However, this results in a lower probability for events above a few hundred TeV. Conversely, preference for a steeper spectrum when energies above 3~PeV are included can be interpreted as an indication of the missing events due to the Glashow resonance. We discuss this more thoroughly in Sec.~\ref{sec:glashow}. Let us also note that a different analysis, with a low-energy threshold at 1~TeV, also obtained a steep spectrum with $\gamma = 2.46$ as its best fit~\cite{Aartsen:2014muf}. A potential problem of such a soft spectrum, for hadronuclear scenarios where neutrinos are produced from pion decays along with gamma rays, is the violation of the isotropic gamma-ray background bounds when extrapolated to lower energies~\cite{Murase:2013rfa}. Such a limit would imply a power-law index $\gamma \lesssim 2.2$ and could in turn help distinguishing different mechanisms of neutrino production.

In the left panel of Fig.~\ref{fig:eindex} we show the $1\sigma$ and $2\sigma$ contours (for 2 degrees of freedom) in the $\gamma$ -- $N_\phi$ plane, where the total flux normalization  $N_\phi$, in units of $10^{-18} \, \textrm{GeV}^{-1} \, \textrm{cm}^{-2} \, \textrm{s}^{-1} \, \textrm{sr}^{-1}$, is defined as
\begin{equation}
\frac{d\phi^a}{dE_\nu} = N_\phi \, \left(\frac{E_\nu}{100~\textrm{TeV}}\right)^{-\gamma} ~.
\end{equation}
We show with filled contours (closed dashed curves) the two cases with $\Emax = 3$~PeV (10~PeV). The contours for the low-energy cut at $\Emin = 28$~TeV (60~TeV) are depicted in orange (purple). We also show the best-fit combinations of  ($\gamma$, $N_\phi$) inside our contours, indicated by stars (diamonds) when the highest deposited energy in the analysis is 3~PeV (10~PeV), as well as the best-fit values quoted by the IceCube Collaboration for fixed spectral index $\gamma = 2$ (black circle) and for their best fit, $\gamma = 2.3$ (black square)~\cite{Aartsen:2014gkd}. The IceCube analysis assumed the flavor combination expected from pion sources, \textit{i.e.}, $(1 : 1 : 1)_\oplus$, and considered the energy range 60~TeV$- 3$~PeV, so it has to be compared with our purple contours. The major difference comes from the fact that we do not fix the flavor composition, but let it freely float. It turns out that both IceCube points are within the $1\sigma$ C.L. contour, although close to the edges. From this figure, one can see the behavior, just pointed out, that extending the energy range beyond the  window considered in the IceCube analysis has a significant impact on the inferred spectral index of the astrophysical neutrino flux: adding events below 60~TeV (orange regions) or above 3~PeV (dashed lines) steepens the spectrum. This was already noted in an independent analysis of the IceCube data~\cite{Winter:2014pya}.

When one adds the potential contribution of atmospheric neutrinos from charmed meson decays [7P analyses; see Eq.~(\ref{eq:7Pparams})and the description of priors below it], the best-fit astrophysical flux tends to be slightly harder in all cases. This mainly has to do with the fact that the background from atmospheric muons is forced to be smaller than in the 6P case (where no priors are applied), and hence, a slightly harder astrophysical spectrum is needed to account for the number of observed events.

In the right panel of Fig.~\ref{fig:eindex} we show the $p$ value for the total number of astrophysical neutrinos in the four energy intervals. Above 28~TeV we expect $\sim 26-27$ of the 36 observed events to be of astrophysical origin, but with a large $1\sigma$ spread of $\sim 9-10$ events. Above 60~TeV, the number is $\sim$16 out of 20, with an uncertainty of $\sim$5 events. The larger number of background events obtained for the best fit in the 60~TeV$- 3$~PeV case reflects in a slightly smaller number of total events from astrophysical neutrinos. Nevertheless, this is not statistically significant at present.

When we perform the 3P analyses, \textit{i.e.}, fixing the spectral index and the atmospheric background to the IceCube analysis best-fit values, our best-fit number of astrophysical events for the interval 28~TeV$- 3$~PeV is reduced to $\simeq 20-21$. This is due to the fact that by fixing the spectral index to 2.3 a larger number of background events are needed to explain the data. Otherwise, this can be compensated by a steeper spectrum, as can be seen from Table~\ref{tab:resulttable}. For the cases with a low-energy cut at 60~TeV, there is basically no difference because the value of the fixed spectral index is very close to the best fit.

\begin{figure}
	\begin{tabular}{l l}
		\includegraphics[width=0.5\textwidth]{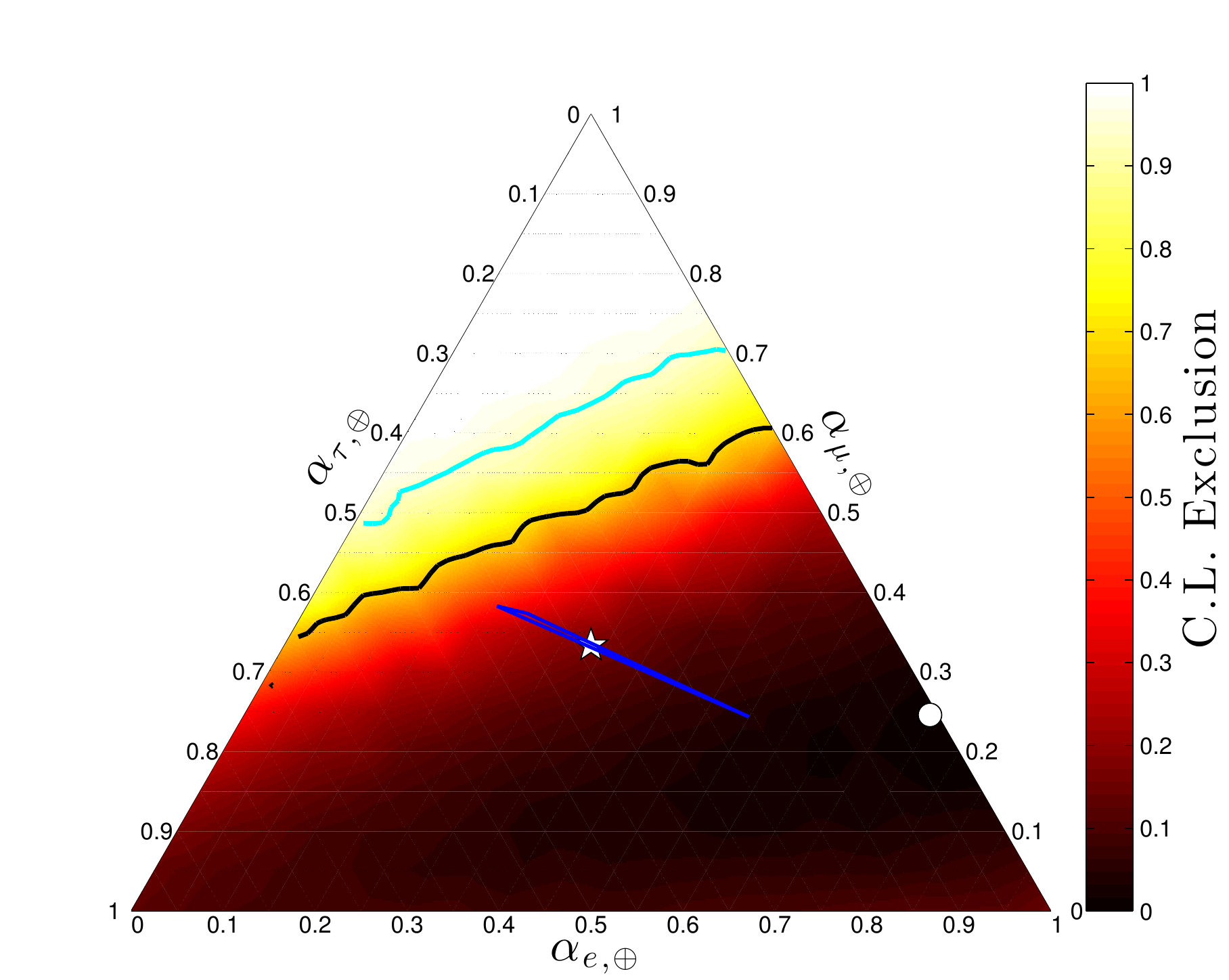} & \includegraphics[width=0.5\textwidth]{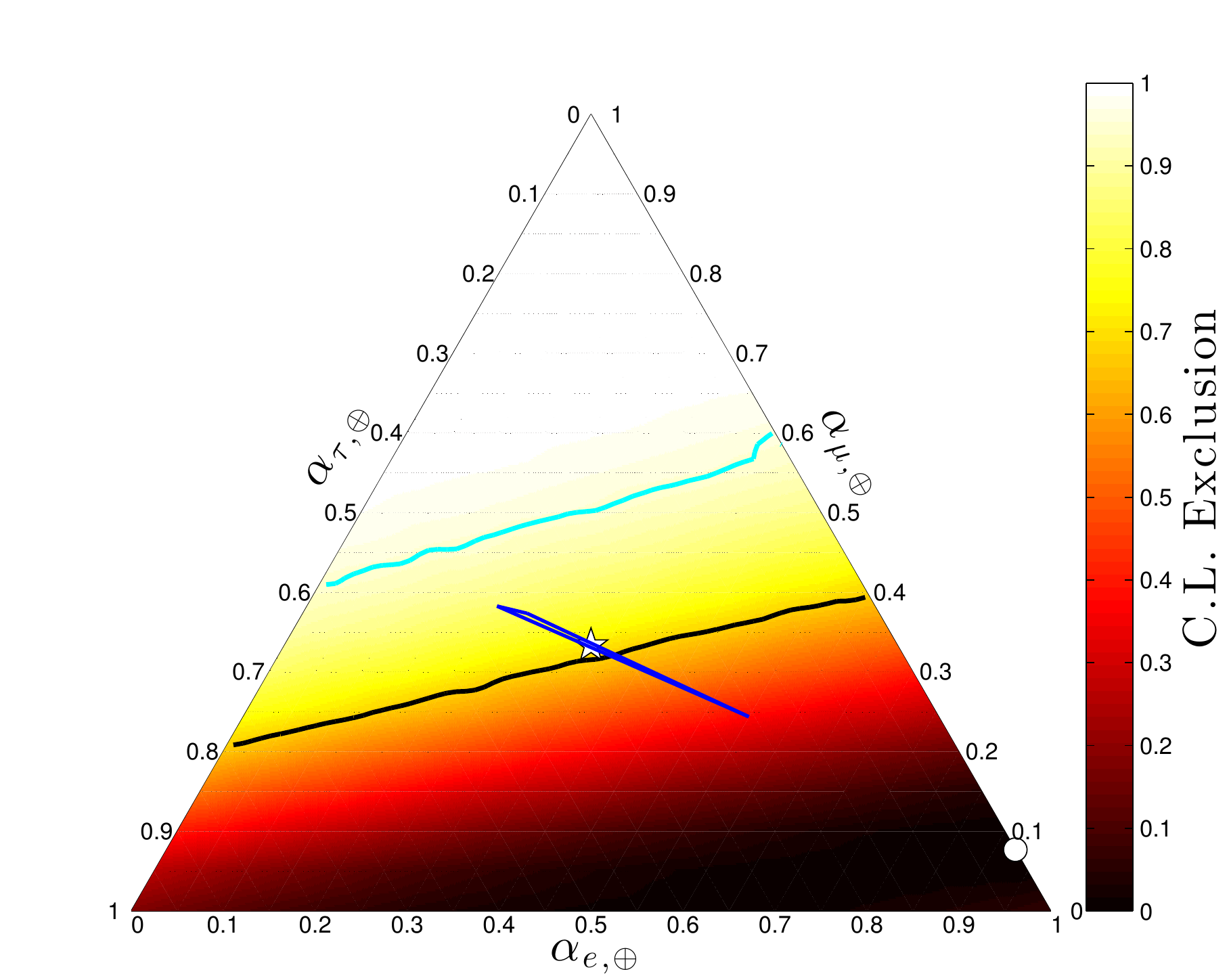} 
	\end{tabular}
	\caption{\textbf{\textit{Effect of accounting for systematic errors.}} Ternary plots of the profile likelihood exclusions of the neutrino flavor composition of the 36 high-energy events seen at IceCube after 988~days, in the energy range 28~TeV$- 3$~PeV. Left panel: varying the full parameter space (6P analysis). Right panel: fixing the spectral index to $\gamma = 2.3$ and the backgrounds to the expected total number of events given by the IceCube Collaboration \cite{Aartsen:2014gkd}: $N_\mu = 8.4$, $N_\nu = 6.6$ and $N_p=0$ (3P analysis). The black (cyan) lines represent the $1\sigma$ ($2\sigma$) C.L. allowed regions. The best fits, indicated by the white circles, are $(0.75 : 0. 25 : 0)_\oplus$ (left panel) and $(0.92 : 0. 08 : 0)_\oplus$ (right panel). The thin blue triangle represents the space of allowed astrophysical neutrinos assuming averaged oscillations during propagation from the sources and the star represents the canonical $(1:1:1)_\oplus$.}
	\label{fig:28-3}
\end{figure}

\subsection{Flavor composition}

As was concluded from the results of our previous analysis in the 28~TeV$- 3$~PeV range~\cite{Mena:2014sja, Palomares-Ruiz:2014zra}, which omitted spectral information and included some simplifying assumptions, a significant $\nu_\mu$ component in the astrophysical flux is disfavored (see, however, Sec.~\ref{sec:misid}). This is due to the paucity of tracks in the observed event sample. We show here that conclusions about the flavor composition at Earth $(\alpha_e:\alpha_\mu:\alpha_\tau)_\oplus$ of the high-energy events are strongly dependent on the energy interval considered for the analysis and whether the spectral index of the astrophysical flux and the number of background events are allowed to vary freely.

We first consider the range 28~TeV$- 3$~PeV, which is shown in Fig.~\ref{fig:28-3} for the case of the 6P (left panel) and the 3P (right panel) analyses. From the right panel, we note that our previous results~\cite{Mena:2014sja, Palomares-Ruiz:2014zra}, using a single energy bin, fixing the backgrounds to the IceCube expectations and only using the event topology information, is qualitatively confirmed when the energy distribution is included (see also Refs.~\cite{Watanabe:2014qua, Kalashev:2014vra}), although adding the spectral information as we do here ameliorates the fit for $(1:1:1)_\oplus$, bringing the $p$ value of that point, for $\gamma=2.3$, from 0.14 to 0.29. When the backgrounds and the spectral index (left panel) are left free to vary, the canonical $(1:1:1)_\oplus$ scenario is well within the $1\sigma$ C.L. contour and strong conclusions cannot be drawn. The expected number of background events is allowed to be sufficiently small so that a significant contribution to the track sample from astrophysical neutrinos is required.  Consequently, this also implies a larger number of the total number of astrophysical events. This is not surprising, since the disfavored astrophysical $\nu_\mu$ component arose from the large number of tracks expected from the atmospheric muon and neutrino backgrounds, $\sim$12 in the 28~TeV$- 3$~PeV deposited energy range, compared with the actual 8 tracks observed after three years. At the same time, as discussed above, a steeper spectrum is preferred. 

\begin{figure}
	\begin{tabular}{c c}
		\includegraphics[width=0.5\textwidth]{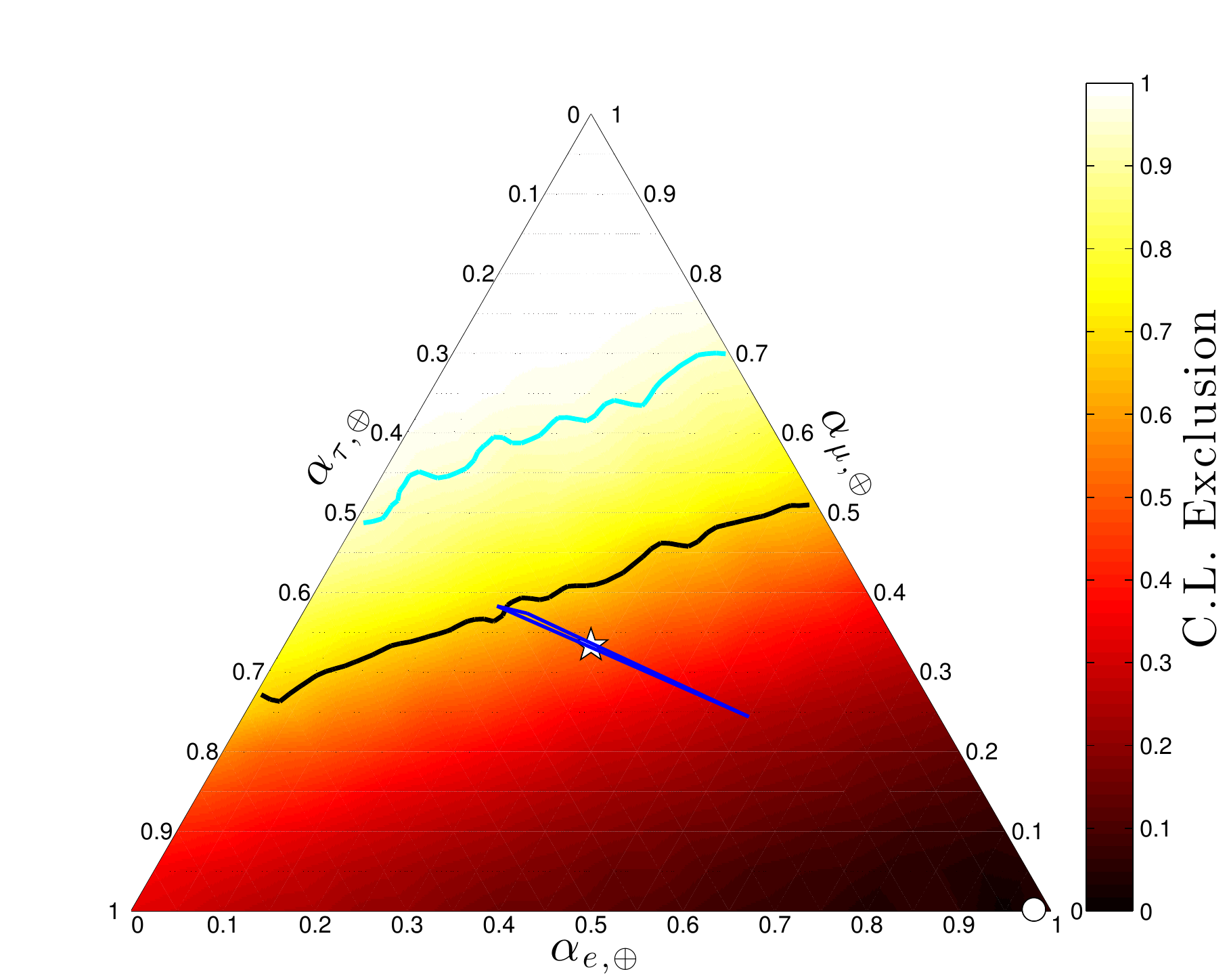} & \includegraphics[width=0.5\textwidth]{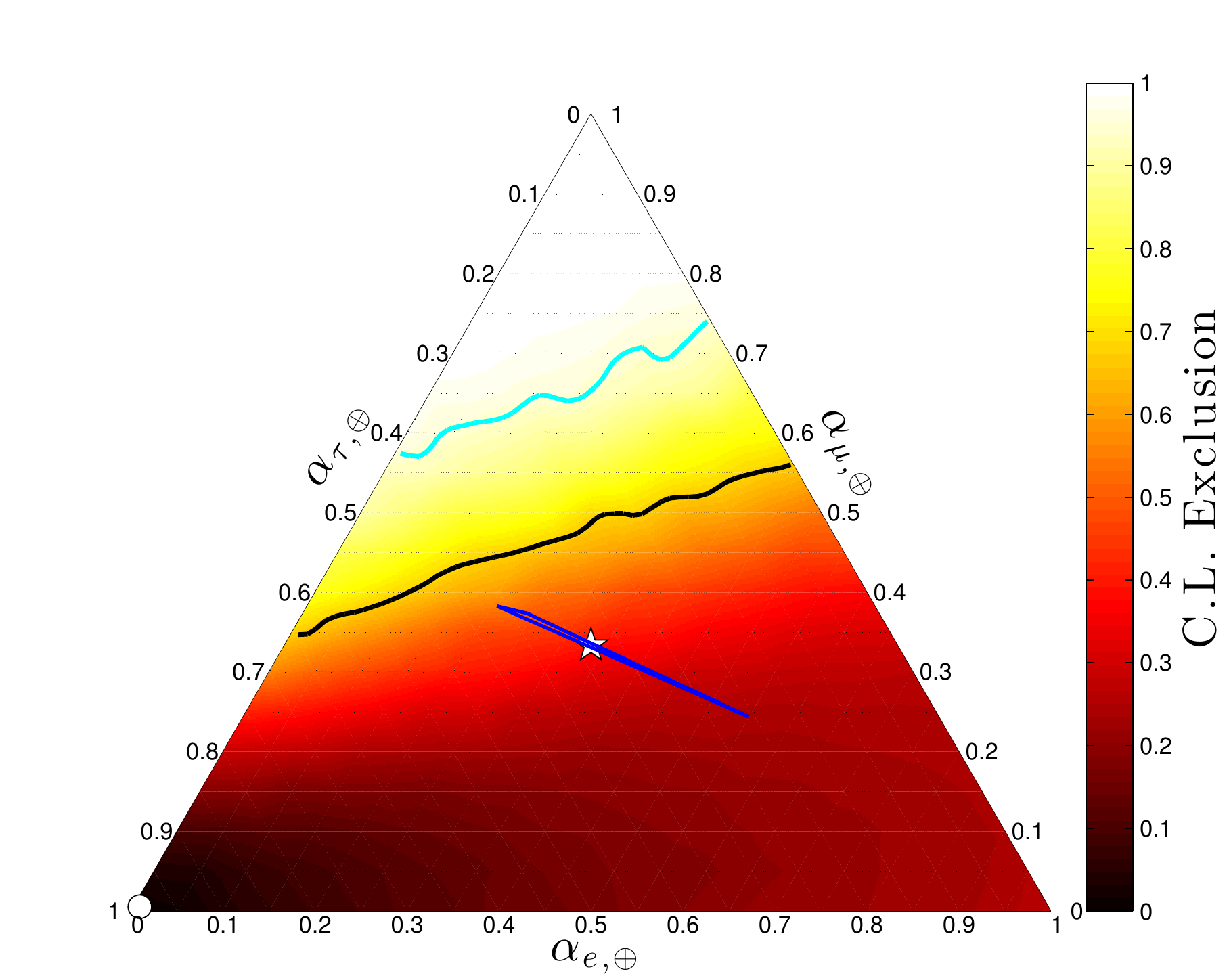} 
	\end{tabular}
	\caption{\textbf{\textit{Effect of extending the energy interval of the analysis up to 10~PeV.}}
		Ternary plots of the profile likelihood exclusions of the neutrino flavor composition of the 20 high-energy events seen at IceCube above $E_{\textrm{dep}} = 60$~TeV with the 6P analyses, in the energy ranges 60~TeV$- 3$~PeV (left panel) and 60~TeV$- 10$~PeV (right panel). The best fits are $(0.98 : 0. 02 : 0)_\oplus$ (left panel) and $(0.01 : 0. 01 : 0.98)_\oplus$ (right panel). The consequence of not having observed any event above 2~PeV is the shift of the best-fit point from a flux with a dominant $\nu_e$ component (for $\Emax = 3$~PeV) to an almost pure $\nu_\tau$ flux (for $\Emax = 10$~PeV). Same format as Fig.~\ref{fig:28-3}.}
	\label{fig:60tev}
\end{figure}

Moving the minimum deposited energy to 60~TeV alleviates much of this tension, since the number of expected background tracks only accounts for only $\sim$2 out of the 4 observed events. This can be seen in Fig.~\ref{fig:60tev} where we depict the flavor contours for the 60~TeV$- 3$~PeV (left panel) and the 60~TeV$- 10$~PeV (right panel) cases obtained with the 6P analyses. As can also be seen in Table~\ref{tab:resulttable}, extending the analysis above 3~PeV yields a dramatic shift in the location of the best fit: rather than preferring a dominant $\nu_e$ component, a strong $\nu_\tau$ component becomes favored. As we will return to in Sec.~\ref{sec:glashow}, this can be understood by the increasing effect of the Glashow resonance as we consider energies around $\sim$6.3~PeV. All in all, the low statistics in this range, only 20 observed events (16 showers and 4 tracks), does not allow us to reach any strong conclusion regarding the flavor composition and the entire parameter space for the standard scenario with averaged oscillations during propagation from the sources (the blue sliver) is allowed within $1\sigma$ C.L.

Finally, let us note that the flavor composition remains almost identical to that obtained with the 6P analyses when performing a 7P fit, \textit{i.e.}, when the prompt atmospheric neutrino flux is included with a prior using IceCube limits and an additional prior on the atmospheric muon background is implemented.

\begin{figure}
	\begin{tabular}{c c}
		\includegraphics[width=0.5\textwidth]{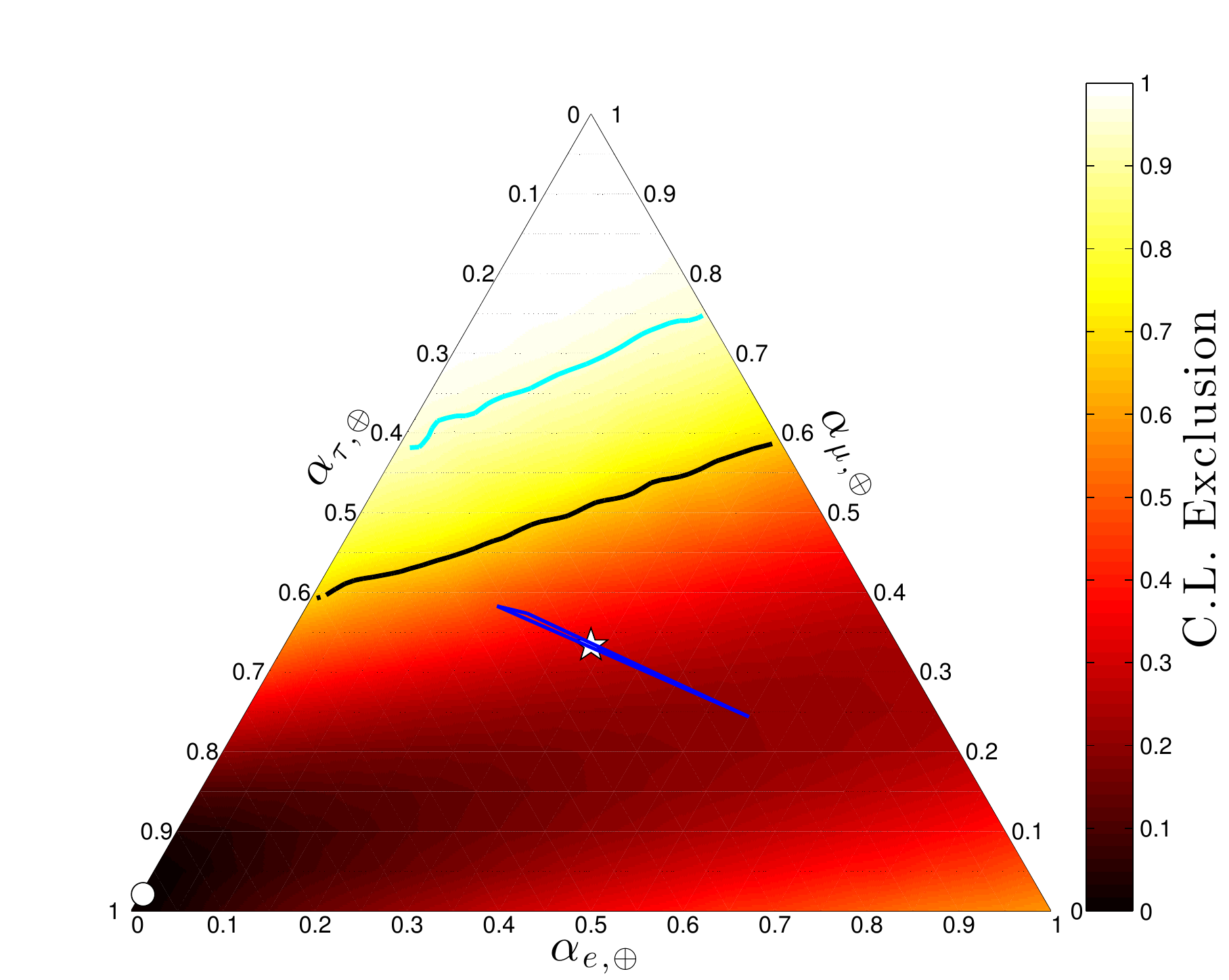} & \includegraphics[width=0.5\textwidth]{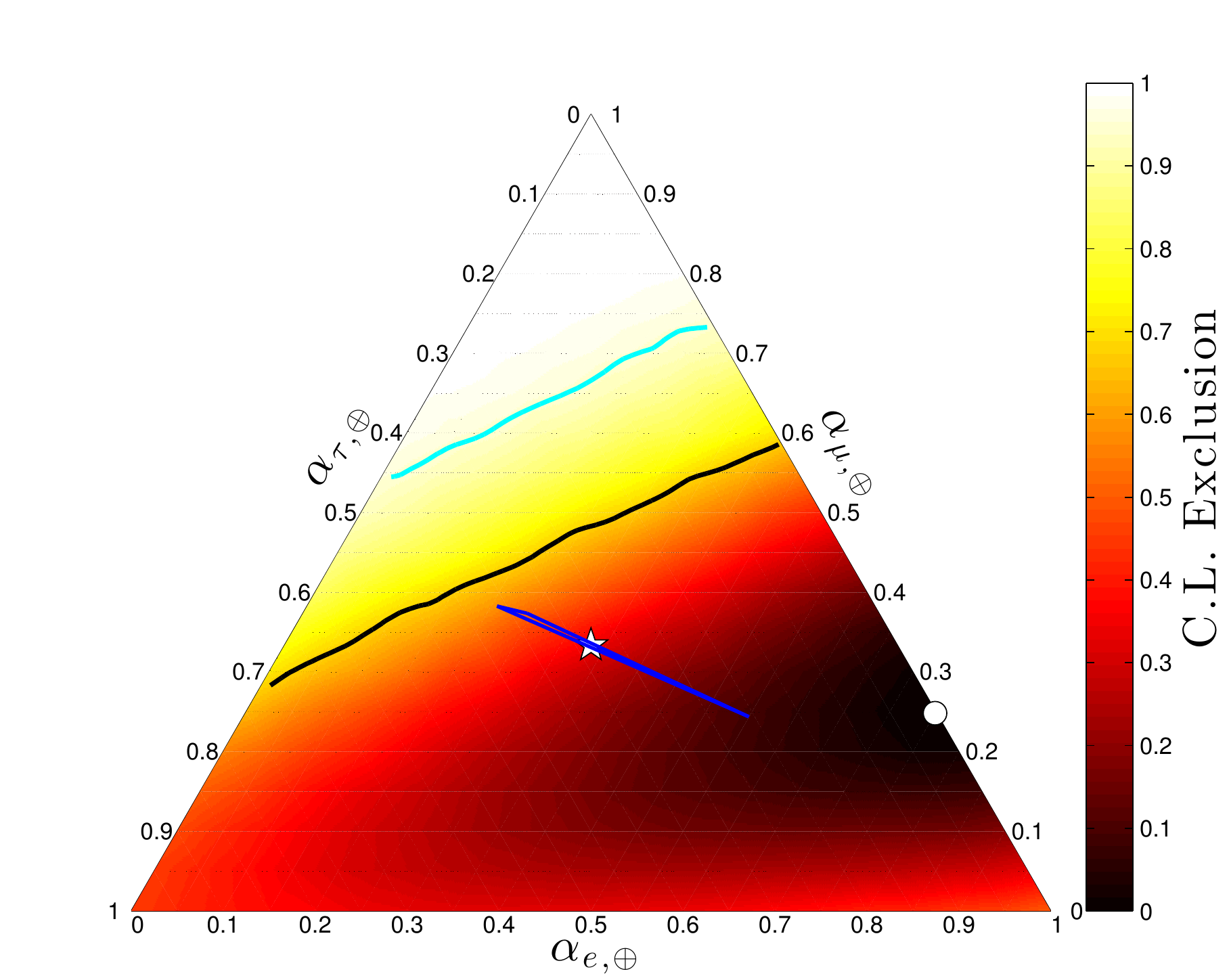} 
	\end{tabular}
	\caption{\textbf{\textit{Effect of adding a break of one unit in the spectral index of the astrophysical neutrino spectrum at $\boldsymbol{E_\nu =}$ 1~PeV.}} Ternary plots of the profile likelihood exclusions of the neutrino flavor composition of the 20 high-energy events seen at IceCube in the energy range 60~TeV$- 10$~PeV with no break in the spectrum (left panel) and with a break of $\delta \gamma = 1$ at $E_\nu = 1$~PeV (right panel). We perform 4P analyses: atmospheric backgrounds are fixed as in the 3P case, but we allow the spectral index $\gamma$ to vary in each case. Same format as Fig.~\ref{fig:28-3}. Table~\ref{tab:resulttable} shows the best-fit values for these cases, as well as for the 60~TeV$- 3$~PeV energy range.}
	\label{fig:break}
\end{figure}

\subsection{The Glashow resonance and broken power laws}
\label{sec:glashow}

In Fig.~\ref{fig:60tev}, by considering the cases with $\Emin = 60$~TeV, we show the effect of the upper deposited energy limit of our analysis region. By extending this limit from 3~TeV (left panel) to 10~PeV (right panel), the best fit shifts dramatically, from $(\alpha_e:\alpha_\mu:\alpha_\tau)_\oplus \simeq (1:0:0)_\oplus$ to $\simeq (0:0:1)_\oplus$. This shift also occurs in the $> 28$~TeV case --- albeit less dramatically, as the lower-energy events remove some statistical weight from high energies. This difference in the best fit (and in the spectral index) is mainly due to the absence of events observed near the Glashow resonance peak. For the IceCube best fit assuming the canonical $(1:1:1)_\oplus$ one expects about 1.2 events above 3~PeV. On the other hand, for the best fit in the 6P case for the energy range 60~TeV$- 10$~PeV (almost a purely $\nu_\tau$ flux), only $\sim$0.3 events are expected above 3~PeV (see bottom right panel of Fig.~\ref{fig:BFspectra}). We further tested this by performing \textsc{MultiNest} runs with interactions with electrons artificially removed. This leads to a return of the best-fit region toward $(1:0:0)_\oplus$.

The importance of the lack of events around 6.3~PeV has already been discussed in the literature~\cite{Barger:2014iua}. It could be an indication of a broken power law~\cite{Anchordoqui:2014hua} or even a cutoff~\cite{Learned:2014vya, Winter:2014pya} in the astrophysical neutrino spectrum. We show the effect of adding a break in the power-law spectrum, by setting it at 1~PeV and by modifying the spectral index by one unit for higher energies, \textit{i.e.}, such that $\delta \gamma \equiv \gamma(E_\nu > 1 \, \mathrm{PeV}) - \gamma(E_\nu < 1 \, \mathrm{PeV}) = 1$. We perform 4P analyses, wherein we fix the atmospheric fluxes, but allow $\gamma$ to vary.

Adding a break of one unit in the spectral index dramatically reduces the expected event rate around the Glashow resonance, bringing the best fit closer in line with expectation. It furthermore allows for a harder spectrum in the region where the events were observed: when a break is added, the best-fit $\gamma (E_\nu < 1 \, \mathrm{PeV})$ goes from 2.48 to 2.34 in the 60~TeV$- 3$~PeV energy range (2.49 to 2.43 for 60~TeV$- 10$~PeV). The most striking effect is on the flavor composition, though. In Fig.~\ref{fig:break} we show this for the two cases with a minimum deposited energy of 60~TeV and by performing 4P fits. When the energy range is extended up to 10~PeV, adding a break moves the best-fit location from $(0:0.02:0.98)_\oplus$ to $(0.75:0.25:0)_\oplus$, as a higher $\bar \nu_e$ flux becomes allowed due to the flux suppression above $E_\nu = 1$~PeV. This shows that if such a break does indeed exist, omitting it when fitting the high-energy events can lead to a major mischaracterization of the flavor composition. However, with the current statistics, performing a full spectral analysis including the two extra parameters to describe the broken power law (the position of the break and the spectral index at energies above it) would be somewhat fruitless.

\begin{figure}
	\begin{tabular}{c c}
		\includegraphics[width=0.5\textwidth]{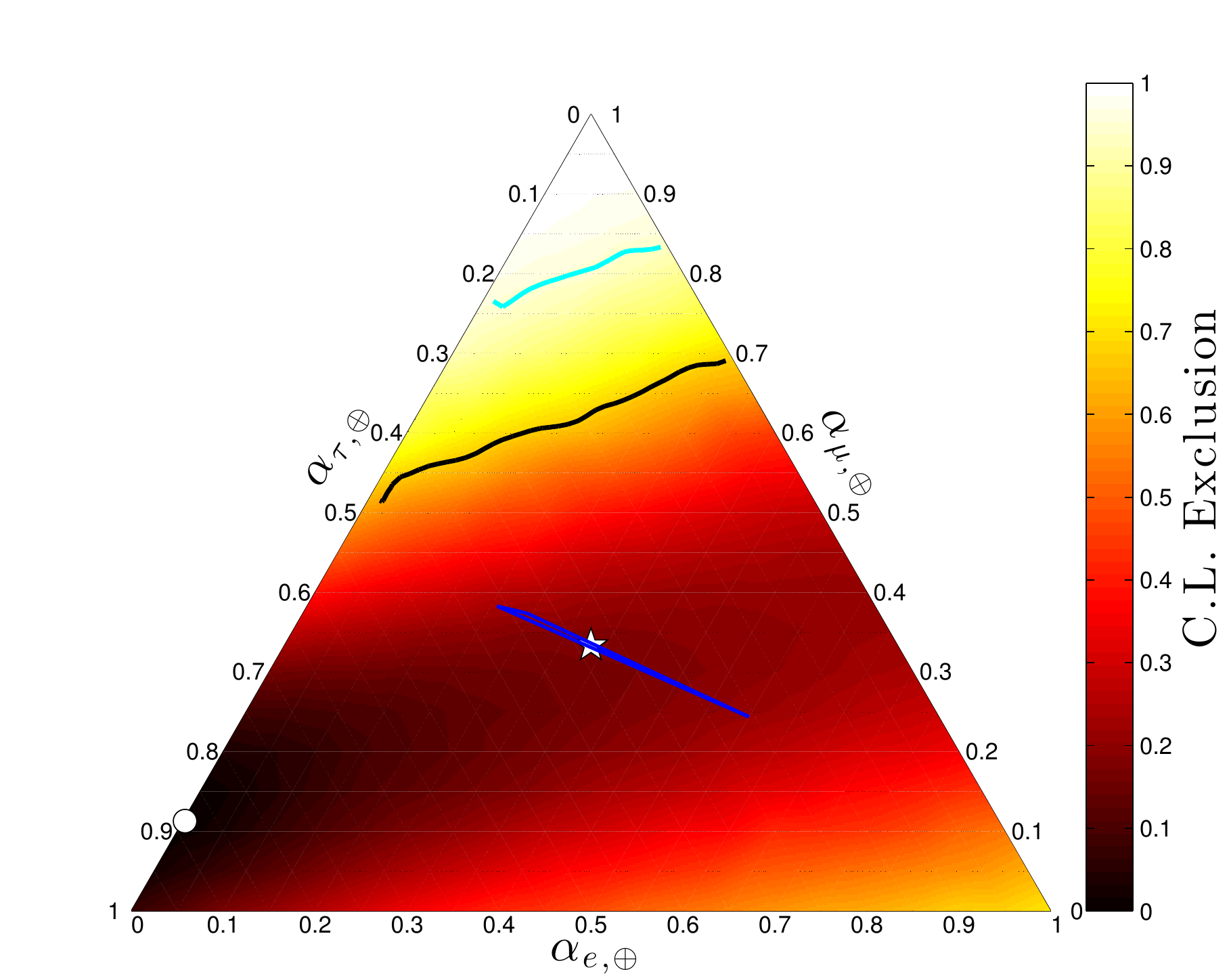} & \includegraphics[width=0.5\textwidth]{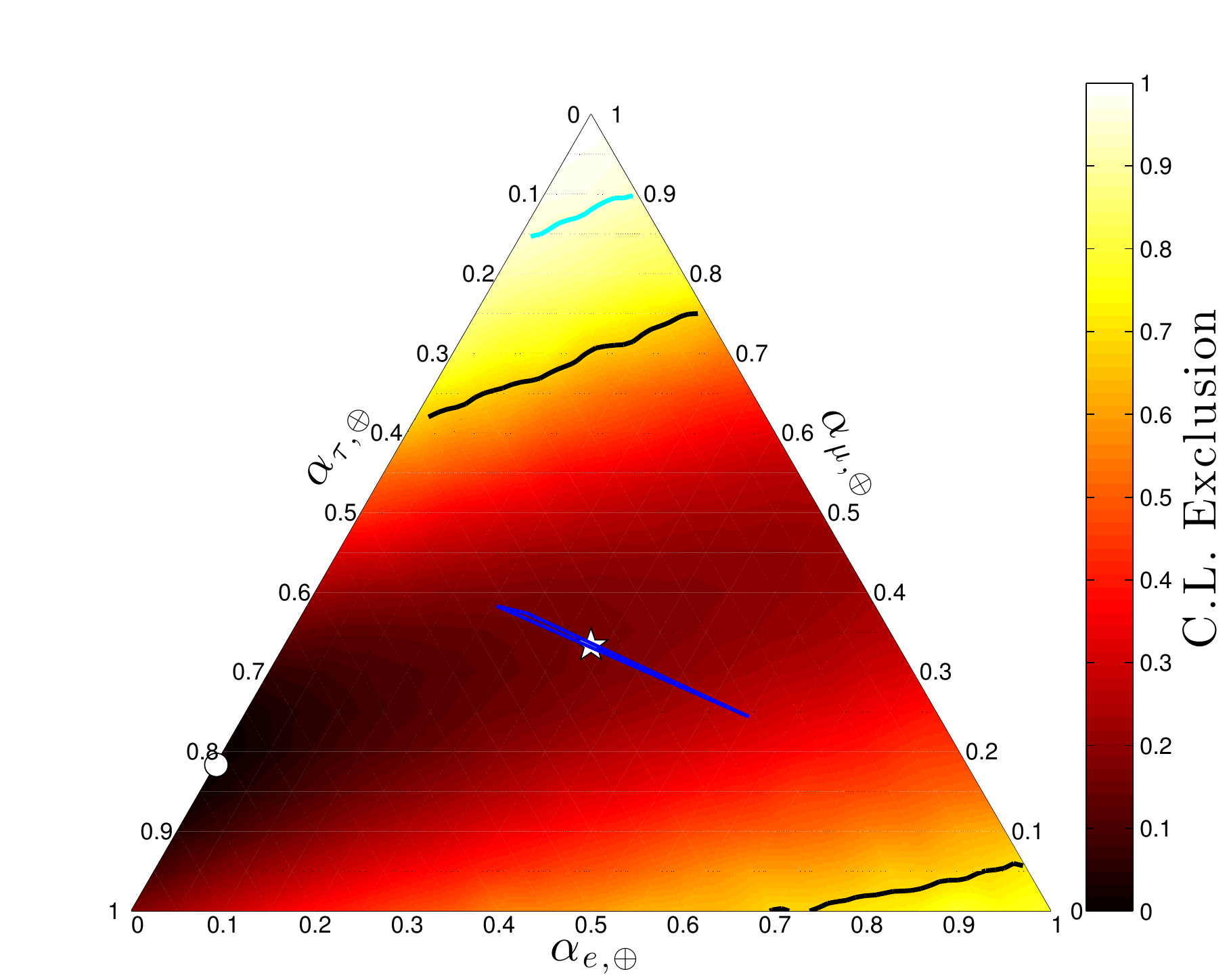} 
	\end{tabular}
	\caption{\textbf{\textit{Effect of including muon track misidentification as shower events.}} Ternary plots of the profile likelihood exclusions of the neutrino flavor composition of the 20 high-energy events seen at IceCube in the energy range 60~TeV$- 10$~PeV with a 20\% fraction of muon track misidentification (left panel) and with a 30\% misclassification fraction (right panel). We perform 4P analyses. This shows that a less extreme flavor ratio is easily obtained if some of the muon tracks are missed and the associated events are identified as showers. These should be compared with the left panel of Fig.~\ref{fig:break}, which shows the same case but with no misidentification of tracks. Same format as Fig.~\ref{fig:28-3}. In Table~\ref{tab:resulttable}, we also indicate the best fits for the other energy intervals for the case of 20\% track misidentification.}
	\label{fig:misID}
\end{figure}

\subsection{Track/shower misidentification}
\label{sec:misid}

We also consider the possibility that some fraction of the tracks produced in the detector could be misidentified as showers. This could occur for instance when a muon neutrino produces a muon  close to the borders of the detector, but the muon escapes undetected and only the hadronic shower is seen. Although this has been mentioned in a recent IceCube analysis with a low-energy threshold at 1~TeV~\cite{Aartsen:2014muf}, we could reproduce the numbers of the two- and three-year analyses above $\sim$30~TeV without including this information~\cite{Aartsen:2013jdh, Aartsen:2014gkd}. Let us note that this was one of the potential explanations to alleviate the tension for the $(1:1:1)_\oplus$ case that was put forward in our previous works~\cite{Mena:2014sja, Palomares-Ruiz:2014zra}.

Here, we show the effect of the misidentification of muon tracks as showers, for both incoming astrophysical and atmospheric neutrino events\footnote{The reverse, \textit{i.e.}, a shower being misclassified as a track, is very rare~\cite{Aartsen:2014muf}. We also note that a 10\% misidentification fraction for atmospheric muons is already included in all our fits.}. We illustrate the effect in the energy range 60~TeV$- 10$~PeV in Fig.~\ref{fig:misID} by performing 4P fits (leaving the spectral index free), which should be compared with the left panel of Fig.~\ref{fig:break}, where no misidentification of tracks was included. We depict the ternary plots of the profile likelihoods in flavor space for the case with 20\% track/shower misclassification (left panel), and in the case where this fraction reaches 30\%\footnote{After the first version of this work, the IceCube Collaboration released a flavor study that quoted the fraction of track misidentification as 30\%~\cite{Aartsen:2015ivb}. Note that differences with respect to the case of 20\% track misidentification, which was presented in our work before the IceCube preprint appeared on the arXiv, are very small, and IceCube results agree with our findings. Another recent analysis also appeared after our first version~\cite{Palladino:2015zua}, and it also agrees with our main conclusions.} (right panel). The shift toward a larger muon neutrino component in the astrophysical flux is clear. In this case, a pure $\nu_e$ flux, which is very close to the best fit when the considered energy range is 60~TeV$- 3$~PeV and if no track misidentification takes place, is disfavored at more than $1\sigma$ C.L. As can also be seen in Table~\ref{tab:resulttable}, where we provide the best fit for other analyses, track/shower misidentification allows for a larger astrophysical muon neutrino component, as the expected tracks from atmospheric events no longer necessarily swamp the signal events which, on the other hand, consist of a larger relative number of showers. And for the 4P analysis, for 20\% (30\%) track misclassification as showers, this results in a 0.82 (0.84) $p$ value for $(1:1:1)_\oplus$ when the energy interval is 60~TeV$- 10$~PeV (as compared to 0.69 when no misidentification occurs). Let us also note that track misidentification does not change the best fit for the spectral index, however.

\begin{figure}
	\begin{tabular}{c c}
		\includegraphics[width=0.5\textwidth]{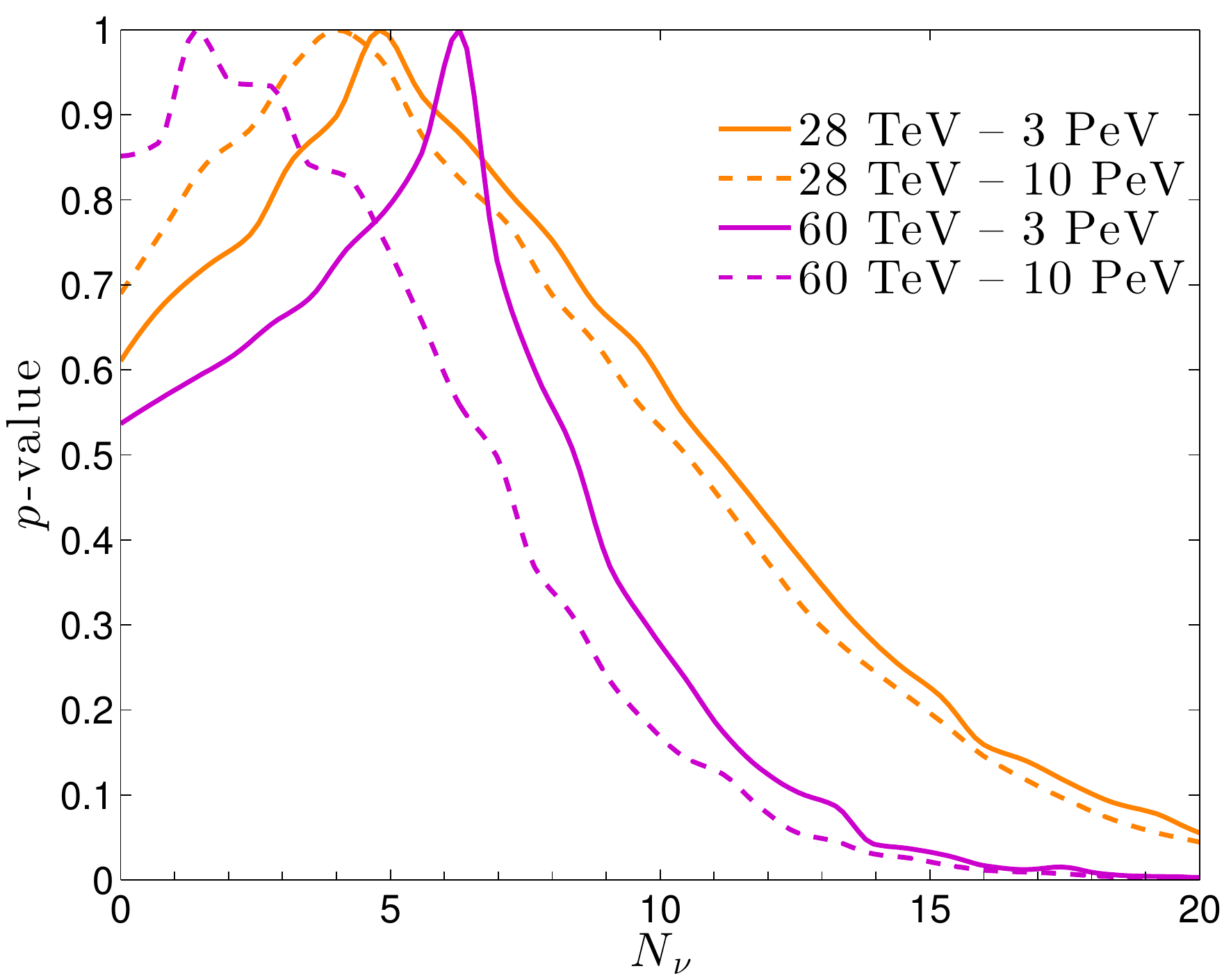} & \includegraphics[width=0.5\textwidth]{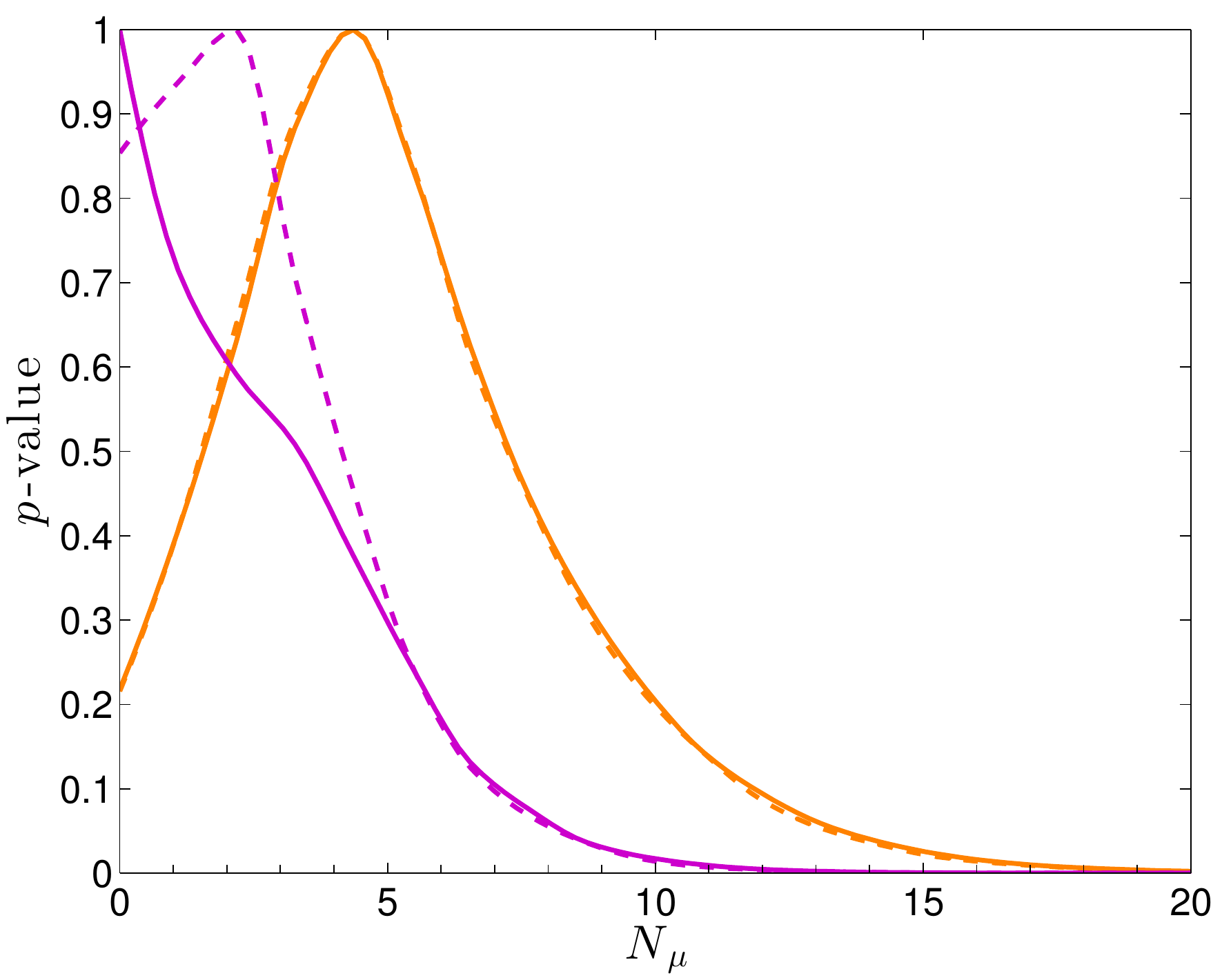}   
	\end{tabular}
	\caption{\textbf{\textit{The $\boldsymbol{p}$ values for the total number of events induced by conventional atmospheric neutrinos (left panel) and that of veto-passing atmospheric muons (right panel)}} in each of the energy ranges considered in this work. They are obtained after performing 6P fits.}
	\label{fig:atmosphericrates}
\end{figure}
\begin{figure}
	\begin{tabular}{c c c}
		\includegraphics[width=0.33\textwidth]{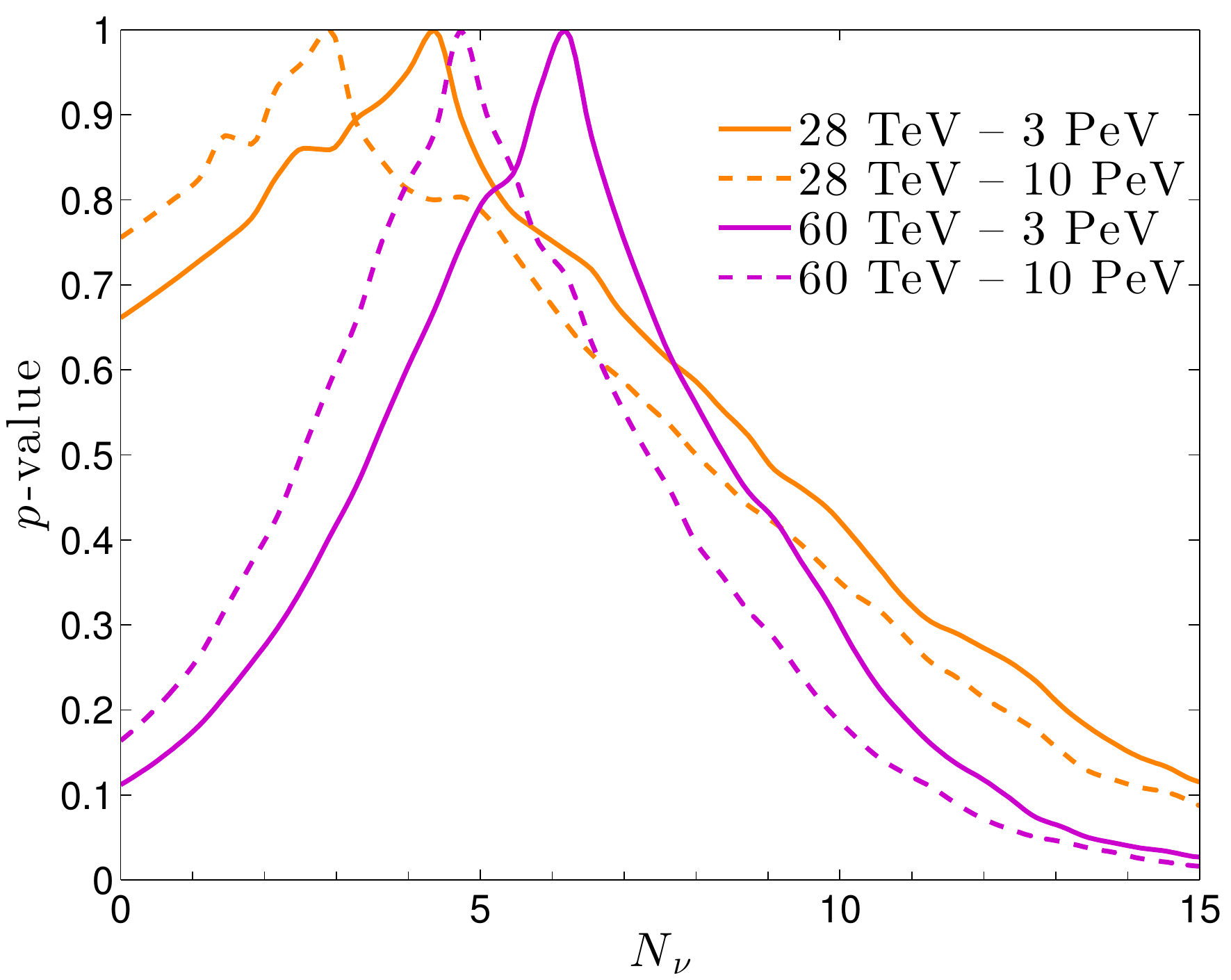} & \includegraphics[width=0.33\textwidth]{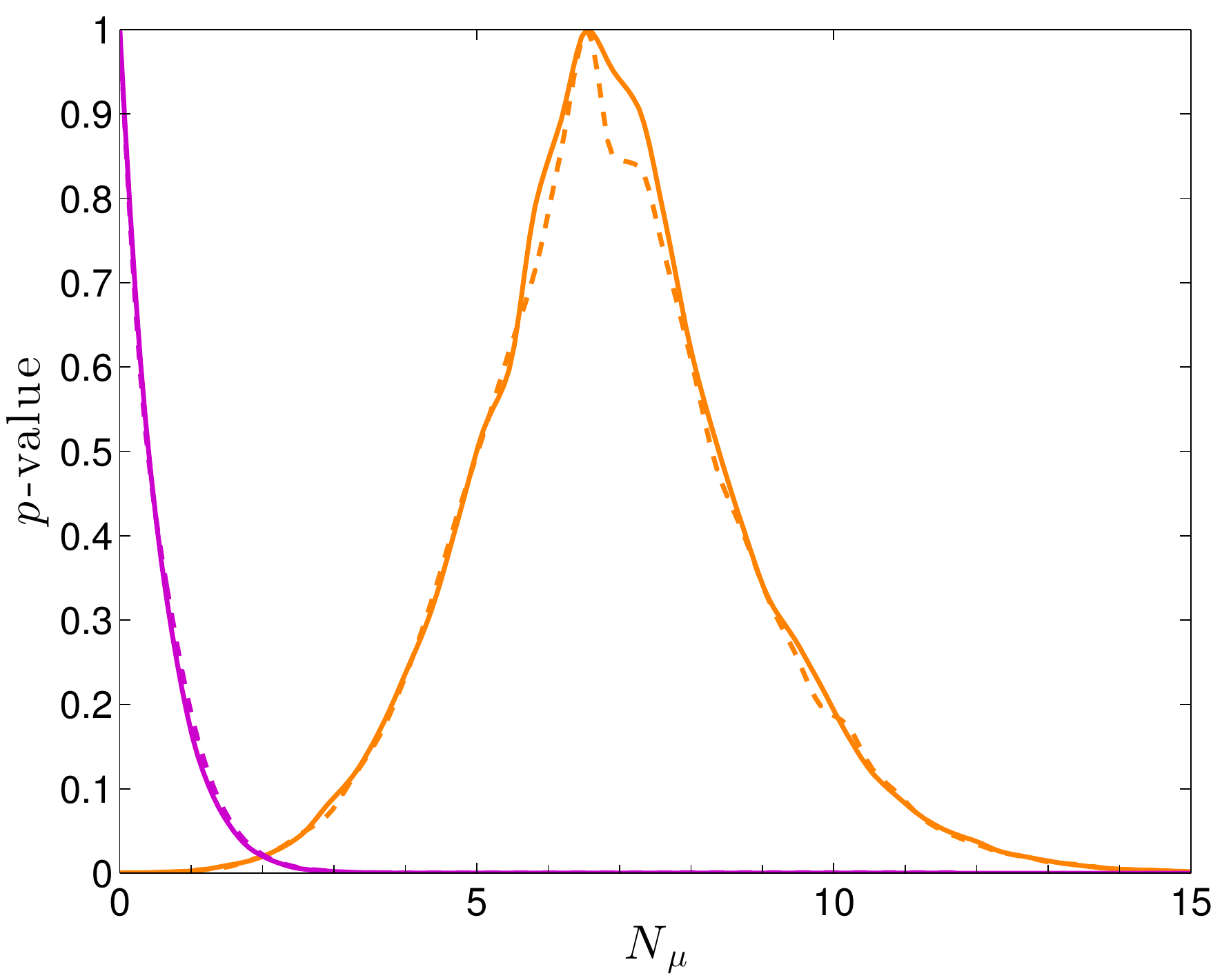}   & \includegraphics[width=0.33\textwidth]{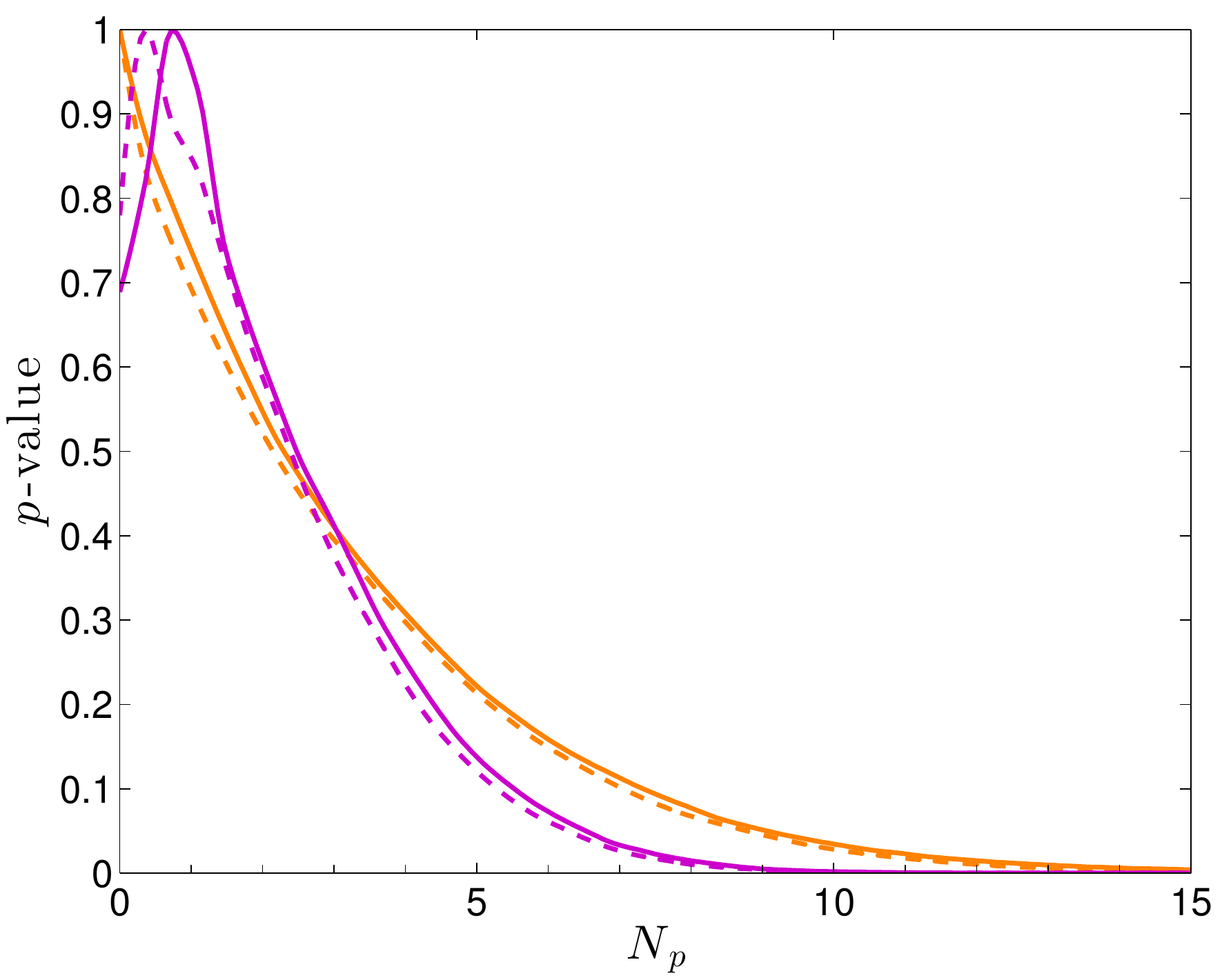}   
	\end{tabular}
	\caption{\textbf{\textit{The $\boldsymbol{p}$ values for the number of conventional atmospheric neutrinos (left panel), atmospheric muons (middle) and prompt atmospheric neutrinos from charmed meson decays (right panel), for the 7P analyses}}, \textit{i.e.}, when a prompt component is added to our fit and priors on $N_\mu$ and $N_p$ are added as discussed in Sec.~\ref{sec:analysis}.}
	\label{fig:prompts}
\end{figure}

\subsection{Atmospheric backgrounds}
\label{sec:atmos}

In our 6P fits, we show what happens when the atmospheric background event rates are allowed to freely float.  In Fig.~\ref{fig:atmosphericrates} we show the one-dimensional profile likelihoods of the background atmospheric neutrinos and muons when all six parameters are allowed to vary freely for the four energy intervals under discussion. The best-fit values, posterior means and errors are given in Table~\ref{tab:resulttable}. In most cases, a better fit can be obtained with a lower atmospheric component than that expected by the IceCube Collaboration, although the $1\sigma$ errors include the expected values in every case. The fit for the 60~TeV$- 3$~PeV interval results in a particularly large number of atmospheric neutrino events. On the other hand, a large number of atmospheric muons is predicted in the 60~TeV$- 10$~PeV interval. In both cases, the required fluxes would swamp the signal below 60~TeV. This is an indication of the low statistical power of the data sample to determine the number of background events and of the fact that we do not use the angular information.

We then consider the case when a prompt atmospheric neutrino component is added to the analysis, corresponding to the 7P case, wherein $N_\mu$ and $N_p$ are subject to priors based on independent measurements. Without this prior, the data show a preference for a large prompt component over the astrophysical signal. However, in this work we are not including the directional information of the events. Using the veto probability discussed in Sec.~\ref{sec:bkg}, the rate of prompt neutrinos from the Southern Hemisphere must be approximately half of that from the Northern Hemisphere, and no evidence of such a suppression is supported by data~\cite{Aartsen:2014gkd}. Therefore, the prior on $N_p$ is not only well justified but required.

In Fig.~\ref{fig:prompts}, profile likelihoods for the total number of events of conventional atmospheric neutrino (left panel), atmospheric muon (middle panel) and prompt atmospheric neutrino (right panel) backgrounds in the 7P case are shown. The best-fit points, Bayesian posterior means, and uncertainties are given in Table~\ref{tab:charmtable}. In most cases, a very small component from charmed meson decays is favored: fewer than one event in the three-year data set, although the 1$\sigma$ C.L. errors can accommodate up to four extra prompt events. Whereas the number of atmospheric muons slightly changes as a consequence of the prior imposed on it, the number of events induced by conventional neutrinos remains statistically the same as that obtained with the 6P fits. The main effect of an added prompt component on other model parameters is to allow for a slightly harder astrophysical component, with a slightly smaller number of astrophysical  events. All in all, 6P and 7P analyses provide very similar results for the rest of the parameters.

\section{Conclusions}
\label{sec:conclusions}

The first evidence for a high-energy neutrino flux of extraterrestrial origin has been obtained after three years of observation by the IceCube experiment~\cite{Aartsen:2013bka, Aartsen:2013jdh, Aartsen:2014gkd}. In this period of time, a total of 36 events (plus one event whose energy and direction cannot be reconstructed) have been detected with deposited energies between $\sim$30~TeV and $\sim$2~PeV. In Refs.~\cite{Mena:2014sja, Palomares-Ruiz:2014zra}, we showed that by looking only at the flavor composition within a single energy bin (28~TeV$- 3$~PeV), there is an apparent tension between the canonical astrophysical signal  $(1:1:1)_{\oplus}$ and the observed ratio of tracks to showers, leading to a best fit for the signal at $(1:0:0)_{\oplus}$. In spite of the relatively small number of observed events, interesting conclusions can be drawn from a spectral analysis of the observations. Spectral analyses of the full data have been performed, either fixing the flavor composition to $(1:1:1)_\oplus$~\cite{Aartsen:2013jdh, Aartsen:2014gkd, Chen:2013dza, Kalashev:2014vra} or varying the flavor composition but with fixed backgrounds~\cite{Watanabe:2014qua}. In this work we performed, for the first time, a detailed spectral analysis of the 36 high-energy neutrino events detected by IceCube, where the astrophysical flavor composition, spectral index, and normalization, along with the number of background events were left free to vary. Our results are summarized in Tables~\ref{tab:resulttable} and~\ref{tab:charmtable}, along with a series of figures, which drive the main discussion of the spectral shape of the spectrum as well as of the flavor composition of the astrophysical flux. 

We have shown that the picture includes many more subtleties than what a single-energy bin analysis could reveal. By adding spectral information, we have shown that an astrophysical $\nu_\mu$ or $\nu_\tau$ component may be required for a good fit to the higher-energy tracks that were observed. Indeed, these tracks are not well explained by the expected atmospheric muon and neutrino fluxes due to their steep spectra. The current data prefer a slightly larger astrophysical --- and thus smaller atmospheric --- component, although there are still too few events for this to be statistically significant.

Concurrently, we have confirmed that the lack of observed events above 2~PeV carries important information about the $\bar \nu_e$ astrophysical spectrum in the form of missing events around the Glashow resonance (see also Ref.~\cite{Barger:2014iua}). We have done so by means of comparing the results for the energy intervals with a maximum deposited energy of 3~PeV and those with maximum deposited energy of 10~PeV. Moreover, we have also considered the possibility of a break in the astrophysical spectrum and how this would impact our results. In general, considering the lack of events above 2~PeV for unbroken power-law spectra implies both a larger $\nu_\tau$ flux and a steeper spectrum. This is more pronounced in the case of a higher minimum energy, which leaves background events with higher statistical power out of the analysis. Nevertheless, if these trends become significant with future data, we are left with a puzzle: a preference for a large electron neutrino component between 28~TeV and a few PeV due to a paucity of tracks, combined with a lack of electron antineutrinos at high energies, reflected by a deficit of events around the Glashow resonance. The latter could have a simple explanation, such as a break in the power law around 1~PeV, which could have a connection with the ultrahigh-energy cosmic-ray paradigm~\cite{Anchordoqui:2013qsi, Anchordoqui:2013dnh, Winter:2014pya, Murase:2014tsa}. This would not be surprising, since the dominant production mechanism of high-energy neutrinos is thought to be spallation of high-energy cosmic rays by photons or protons in the intergalactic medium, yielding neutrinos that are $\sim$10 times less energetic. A different dominant production mechanism could also be partly responsible. For example, $p\gamma$-only scattering vastly reduces the expected $\bar \nu_e$ flux, and $\pi^\pm$ decays in dense environments could furthermore lead to a suppressed electron neutrino component, as the resulting muon can have time to lose a significant fraction of its energy to the source environment before decaying~\cite{Barger:2014iua}. Conversely, a principal origin in the form of neutron decays leads to a higher $\bar \nu_e$ component, compounding the problem. Such scenarios have been invoked to explain the lack of tracks in the observed data and could plausibly dominate~\cite{Anchordoqui:2014pca}. This tension could be reduced by a two-component astrophysical flux~\cite{Chen:2014gxa}, though. At present all these possibilities are still allowed by observations.

Although at present it is not statistically significant, if the best-fit point in flavor space continues to lie outside of the \textit{source} triangle (blue sliver in our ternary plots) where flavor compositions allowed by averaged neutrino oscillations lie, we could be confronted with an even bigger puzzle. In principle, the lack of expected tracks could suggest either a misunderstanding of the atmospheric backgrounds or a misidentification of tracks as showers.  Indeed, in Fig.~\ref{fig:misID} we have shown that a misclassification of tracks as showers would imply a higher likelihood for the \textit{source} triangle and would point to a standard physics origin.  More compellingly, this signal could be a hint of exotic physics such as in some scenarios of neutrino decay~\cite{Crocker:2001zs, Beacom:2002vi, Baerwald:2012kc, Pakvasa:2012db, Dorame:2013lka}, pseudo-Dirac neutrinos~\cite{Beacom:2003eu, Esmaili:2009fk, Pakvasa:2012db, Joshipura:2013yba}, sterile neutrino altered dispersion relations due to shortcuts in an extra dimension~\cite{Aeikens:2014yga}, TeV gravity~\cite{Illana:2014bda}, $T$ violation~\cite{Alikhanov:2014uja}, scenarios with a color octet neutrino~\cite{Akay:2014tga, Akay:2014qka}, Lorentz violation~\cite{Borriello:2013ala, Anchordoqui:2014hua, Stecker:2014xja, Stecker:2014oxa}, leptoquarks~\cite{Barger:2013pla} or neutrino-secret interactions~\cite{Ioka:2014kca, Ng:2014pca, Ibe:2014pja, Blum:2014ewa, Araki:2014ona, Cherry:2014xra}. 

The era of neutrino astronomy has just started, triggered by the first ever detected high-energy neutrinos in IceCube. After three years of data, a rate of $10-20$ events per year is not high. However, even with such a small sample, compelling questions are raised on the production, propagation and detection of neutrinos at high energies. More events are required: even though the next data release might shed some light on some of the questions elicited by our analyses, we see an important need for experiments with higher sensitivity and exposure. In this sense a future high-energy extension of the IceCube detector~\cite{Aartsen:2014njl} and the planned KM3NeT telescope~\cite{KM3NeT} will be of crucial importance.

\section*{Acknowledgments}
We thank Sergei Sinegovsky for providing us with tabulated high-energy atmospheric neutrino fluxes and David Cerde\~no for useful discussions. S.~P.~R. is supported by a Ram\'on y Cajal contract, by the Spanish MINECO under Grant No. FPA2011-23596 and by the Generalitat Valenciana under Grant No. PROMETEOII/2014/049. O.~M. is supported by the Consolider Ingenio Project No. CSD2007--00060 and by the Grant No. FPA2011--29678 of the Spanish MINECO, and by the Generalitat Valenciana under Grant No. PROMETEO/2009/116.  The authors are also partially supported by PITN-GA-2011-289442-INVISIBLES.  S.~P.~R. is also partially supported by the Portuguese FCT through Projects No. CERN/FP/123580/2011 and No. PTDC/FIS-NUC/0548/2012, which are partially funded through POCTI (FEDER).

\appendix

\section{Neutrino cross sections}
\label{app:cs}

For the neutrino-nucleon and antineutrino-nucleon differential cross sections we use the \textsc{nusigma} Monte Carlo code~\cite{Blennow:2007tw}, which uses the CTEQ6-DIS parton distribution functions~\cite{Pumplin:2002vw, Pumplin:2005rh}. The NC and CC differential cross sections for neutrinos (and similarly for antineutrinos) of a given flavor $\nu_\ell$ off a target molecule with mass, atomic and neutron numbers $A$, $Z$, and $N$, respectively, is
\begin{equation}
\frac{d\sigma^{\textrm{NC/CC}}_{\nu_\ell}}{dy} = \frac{1}{A} \, \left(Z \, \frac{d\sigma^{\textrm{NC/CC}}_{p, \nu_\ell}}{dy} + N \, \frac{d\sigma^{\textrm{NC/CC}}_{n, \nu_\ell}}{dy}\right)
\end{equation}
where $A=18$, $Z=10$ and $N=8$ correspond to water (for an isoscalar target, $A/2=Z=N$) and $d\sigma^{\textrm{NC/CC}}_{p(n), \nu_\ell}/dy$ are the NC/CC neutrino-proton(neutron) cross sections. 

For the case of neutrino interactions with electrons per molecule, one needs to include a factor $Z/A$ (as for the neutrino-proton cross section).  The relevant cross sections are given by~\cite{Gandhi:1995tf, Mikaelian:1980vd}

\begin{equation}
\frac{A}{Z} \, \frac{d\sigma^{e}_{\nu_e, e}(E_\nu,y)}{dy} \equiv \frac{d\sigma(\nu_e e \rightarrow \nu_e e)}{dy} = \frac{G_F^2 \, s}{\pi} \,  \left(\frac{g_R^2 \, (1-y)^2}{(1 + y \, s/M_Z^2)^2}+ \left(\frac{g_L ^2}{1 + y \, s/M_Z^2} + \frac{1}{1 + (1-y) \, s/M_W^2}\right)^2\right) ~,
\label{eq:nueelastic}
\end{equation}

\begin{equation}
\frac{A}{Z} \, \frac{d\sigma^{e}_{\bar\nu_e, e}(E_\nu,y)}{dy} \equiv \frac{d\sigma(\bar\nu_e e \rightarrow \bar\nu_e e)}{dy} = \frac{G_F^2 \, s}{\pi} \,  \left(\frac{g_R^2}{(1 + y \, s/M_Z^2)^2}+ \left| \frac{g_L ^2}{1 + y \, s/M_Z^2} + \frac{1}{1-s/M_W^2 + i \, \Gamma_W/M_W}\right|^2 \, (1-y)^2\right) ~, 
\label{eq:anueelastic}
\end{equation}

\begin{equation}
\frac{A}{Z} \, \frac{d\sigma^{e}_{\bar\nu_e, \ell}(E_\nu,y)}{dy} \equiv\frac{d\sigma(\bar\nu_e e \rightarrow \bar\nu_\ell \ell)}{dy} = \frac{G_F^2 \, s}{\pi} \, \frac{ (1-y)^2 \, \left(1-(m_\ell^2 - m_e^2)/s\right)^2}{(1 - s/M_W^2)^2 + \Gamma_W^2/M_W^2} ~,
\label{eq:ael}
\end{equation}

\begin{equation}
\frac{A}{Z} \, \frac{d\sigma^{e}_{\bar\nu_e, h}(E_\nu,y)}{dy} \equiv
\frac{d\sigma(\bar\nu_e e \rightarrow \textrm{hadrons})}{dy} = \frac{d\sigma(\bar\nu_e e \rightarrow \bar\nu_\mu \mu)}{dy} \, \frac{\Gamma(W \rightarrow \textrm{hadrons})}{\Gamma(W \rightarrow \bar\nu_\mu \mu)} ~, 
\label{eq:GRhad}
\end{equation}

\begin{equation}
\frac{A}{Z} \, \frac{d\sigma^{e}_{\nu_\ell, \ell}(E_\nu,y)}{dy} \equiv
\frac{d\sigma(\nu_\ell e \rightarrow \nu_e \ell)}{dy} = \frac{G_F^2 \, s}{\pi} \, \frac{ \left(1-(m_\ell^2 - m_e^2)/s\right)^2}{(1 + (1-y) \, s/M_W^2)^2} ~,
\label{eq:invelastic}
\end{equation}

\begin{equation}
\frac{A}{Z} \, \frac{d\sigma^{e}_{\nu_\ell, e}(E_\nu,y)}{dy} \equiv
\frac{d\sigma(\nu_\ell e \rightarrow \nu_\ell e)}{dy} = \frac{G_F^2 \, s}{\pi} \, \frac{g_R^2 \, (1-y)^2 + g_L^2}{(1 + y \, s/M_Z^2)^2} ~, 
\label{eq:nuelastic}
\end{equation}

\begin{equation}
\frac{A}{Z} \, \frac{d\sigma^{e}_{\bar\nu_\ell, e}(E_\nu,y)}{dy} \equiv
\frac{d\sigma(\bar\nu_\ell e \rightarrow \bar\nu_\ell e)}{dy} = \frac{G_F^2 \, s}{\pi} \, \frac{g_R^2 + g_L ^2 \, (1-y)^2}{(1 + y \, s/M_Z^2)^2} ~, 
\label{eq:anuelastic}
\end{equation}
where $\ell = \mu$ or $\tau$, $m_e$ and $m_\mu$ are the electron and muon masses, $s = 2 \, m_e \, E_\nu$, $G_F$ is the Fermi constant, $g_L = \sin^2\theta_W -1/2$ and $g_R = \sin^2 \theta_W$ are the chiral couplings of the $Z$ boson to the electron, $\theta_W$ is the weak mixing angle, $M_W$ and $M_Z$ are the gauge boson masses, $\Gamma_W = 2.085$~GeV is the $W$ boson width, $\Gamma(W \rightarrow \textrm{hadrons}) = 0.6760$ and $\Gamma(W \rightarrow \bar\nu_\mu \mu)=0.1057$~\cite{Agashe:2014kda}. The inelasticity $y$ is given by $E_\ell = y \, E_\nu$ for all flavors, $\ell = \{e, \mu, \tau\}$. Note that for the case of $\bar\nu_e$ interactions via the Glashow resonance with a pure hadronic final state, the true deposited energy is equal to the neutrino energy, so only the total cross section is relevant, see Eq.~(\ref{eq:eeh}).

\section{Full differential spectra formulas}
\label{app:totrates}

For an astrophysical flux $d \phi^a/d E_\nu$ (assuming the same spectrum and normalization for neutrinos and antineutrinos) with flavor combination at Earth $\{\alpha_e \, : \, \alpha_\mu \, : \, \alpha_\tau\}_\oplus$, the differential event spectra for topology $k = \{\textrm{sh, tr}\}$ is given by\footnote{Note that for the conventional and prompt atmospheric neutrino fluxes the relative contributions from each flavor, and for neutrinos and antineutrinos, are obtained by using the corresponding flux for each case.}
\begin{equation}
\frac{dN^{k, a}}{dE_{\textrm{dep}}} = \sum_{\ell=e,\mu,\tau} \alpha_\ell \, \frac{dN_{\ell}^{k, a}}{dE_{\textrm{dep}}}   ~.
\end{equation}

In this appendix we present the full expressions for the differential spectra, $dN_{\ell}^{k, f}/dE_{\textrm{dep}}$, for showers ($k=\textrm{sh}$) and tracks ($k=\textrm{tr}$) for each channel $\ell =\{e, \mu, \tau\}$, and for an incoming neutrino flux of type $f$ (astrophysical, conventional atmospheric or prompt atmospheric flux, $f=\{a, \nu, p\}$). The final spectra result from the sum of all the partial contributions from the different processes and are given by
\begin{eqnarray}
\frac{dN_{e}^{\textrm{sh}, f}}{dE_{\textrm{dep}}} & = & 
\frac{dN_{\nu_e}^{\textrm{sh,NC}, f}}{dE_{\textrm{dep}}} + 
\frac{dN_{\bar\nu_e}^{\textrm{sh,NC}, f}}{dE_{\textrm{dep}}} +
\frac{dN_{\nu_e}^{\textrm{sh,CC}, f}}{dE_{\textrm{dep}}} + 
\frac{dN_{\bar\nu_e}^{\textrm{sh,CC}, f}}{dE_{\textrm{dep}}} +
\frac{dN_{\nu_e}^{\textrm{sh}, e, f}}{dE_{\textrm{dep}}} + 
\frac{dN_{\bar\nu_e}^{\textrm{sh}, e, f}}{dE_{\textrm{dep}}} +
\frac{dN_{\bar\nu_e, \tau}^{\textrm{sh}, e, f}}{dE_{\textrm{dep}}} + 
\frac{dN_{\bar\nu_e, h}^{\textrm{sh}, e, f}}{dE_{\textrm{dep}}} ~, \\
\frac{dN_{\mu}^{\textrm{sh}, f}}{dE_{\textrm{dep}}} & = & 
\frac{dN_{\nu_\mu}^{\textrm{sh,NC}, f}}{dE_{\textrm{dep}}} + 
\frac{dN_{\bar\nu_\mu}^{\textrm{sh,NC}, f}}{dE_{\textrm{dep}}} +
\frac{dN_{\nu_\mu}^{\textrm{sh}, e, f}}{dE_{\textrm{dep}}} + 
\frac{dN_{\bar\nu_\mu}^{\textrm{sh}, e, f}}{dE_{\textrm{dep}}} ~, \\
\frac{dN_{\tau}^{\textrm{sh}, f}}{dE_{\textrm{dep}}} & = & 
\frac{dN_{\nu_\tau}^{\textrm{sh,NC}, f}}{dE_{\textrm{dep}}} + 
\frac{dN_{\bar\nu_\tau}^{\textrm{sh,NC}, f}}{dE_{\textrm{dep}}} +
\frac{dN_{\nu_\tau}^{\textrm{sh,CC}, f}}{dE_{\textrm{dep}}} + 
\frac{dN_{\bar\nu_\tau}^{\textrm{sh,CC}, f}}{dE_{\textrm{dep}}} +
\frac{dN_{\nu_\tau}^{\textrm{sh}, e, f}}{dE_{\textrm{dep}}} + 
\frac{dN_{\bar\nu_\tau}^{\textrm{sh}, e, f}}{dE_{\textrm{dep}}} +
\frac{dN_{\bar\nu_\tau, \tau}^{\textrm{sh}, e, f}}{dE_{\textrm{dep}}} ~, \\
\frac{dN_{e}^{\textrm{tr}, f}}{dE_{\textrm{dep}}} & = & 
\frac{dN_{\bar\nu_e, \mu}^{\textrm{tr}, e, f}}{dE_{\textrm{dep}}} + 
\frac{dN_{\bar\nu_e, \tau}^{\textrm{tr}, e, f}}{dE_{\textrm{dep}}}~, \\
\frac{dN_{\mu}^{\textrm{tr}, f}}{dE_{\textrm{dep}}} & = & 
\frac{dN_{\nu_\mu}^{\textrm{tr,CC}, f}}{dE_{\textrm{dep}}} + 
\frac{dN_{\bar\nu_\mu}^{\textrm{tr,CC}, f}}{dE_{\textrm{dep}}} +
\frac{dN_{\nu_\mu, \mu}^{\textrm{tr}, e, f}}{dE_{\textrm{dep}}} ~, \\
\frac{dN_{\tau}^{\textrm{tr}, f}}{dE_{\textrm{dep}}} & = & 
\frac{dN_{\nu_\tau}^{\textrm{tr,CC}, f}}{dE_{\textrm{dep}}} + 
\frac{dN_{\bar\nu_\tau}^{\textrm{tr,CC}, f}}{dE_{\textrm{dep}}} +
\frac{dN_{\nu_\tau, \tau}^{\textrm{tr}, e, f}}{dE_{\textrm{dep}}} ~,
\end{eqnarray}
and all the partial contributions from the different processes are detailed below in this appendix.

Showers are induced by both $\nu_e$ and $\nu_\tau$ CC interactions with nucleons, as well as by NC interactions of neutrinos of all three flavors with nucleons. The differential shower spectrum, in terms of the measured deposited energy, by NC interactions for neutrinos (and analogously for antineutrinos) with flavor $\ell$ reads
\begin{equation}
\label{eq:nc}
\frac{dN^{\textrm{sh,NC}, f}_{\nu_\ell}}{dE_{\textrm{dep}}}  =  T \, N_A \, \int_{0}^{\infty} dE_\nu \, Att^f_{\nu_\ell}(E_\nu) \, \frac{d \phi^f_{\nu_\ell}(E_\nu) }{d E_\nu} \, \int^{1}_{0} dy \, M_{\textrm{eff}}(E^{\textrm{NC}}) \,  R(E^{\textrm{NC}}, E_{\textrm{dep}}, \sigma(E^{\textrm{NC}})) \, \frac{d\sigma^{\textrm{NC}}_{\nu_\ell} (E_\nu,y)}{dy}  ~,
\end{equation}
where $T=988$~days is the time of data taking, $N_A = 6.022 \times 10^{-23}$~g$^{-1}$ and $E_\nu y = (E_\nu-E'_\nu)$ is the hadronic shower energy, with $E'_\nu$ the energy of the outgoing neutrino. The attenuation and regeneration factor due to the absorption of neutrinos when traversing the Earth is given by $Att^f_{\nu_\ell}(E_\nu)$ (see Sec.~\ref{sec:attenuation}) and the detector effective mass as a function of the true deposited energy by $M_{\textrm{eff}}(E_{\textrm{true}})$ (see Sec.~\ref{sec:ICmass}).  The incoming neutrino flux of type $f$ ($f=\{a, \nu, p\}$ for the astrophysical, atmospheric neutrino and atmospheric muon flux, respectively) is $d \phi^f_{\nu_\ell} (E_\nu)/d E_\nu$, and $d\sigma^{\textrm{NC}}_{\nu_\ell}/dy$ is the neutrino-nucleon NC differential cross section. The energy resolution function is given by $R(E_{\textrm{true}},E_{\textrm{dep}},\sigma(E_{\textrm{true}}))$. We assume the uncertainty on the true EM-equivalent deposited energy, $E_{\textrm{true}}$, to be given by the error on the measured EM-equivalent deposited energy, $E_{\textrm{dep}}$, and we perform two single-parameter ($\varepsilon$) fits, for $E_{\textrm{dep}} < E_{\textrm{true}}$ and $E_{\textrm{dep}} \geq E_{\textrm{true}}$, with the function $\sigma(E_{\textrm{true}}) = \varepsilon \, E_{\textrm{true}}$ within the observed energy range using the 36 (shower and track) events detected by IceCube after 988~days. Since upper and lower errors are different, we represent $R$ with two half-Gaussians,
\begin{equation}
R(E_{\textrm{true}},E_{\textrm{dep}},\sigma(E_{\textrm{true}})) = \frac{1}{\sqrt{2 \pi \sigma^2}} \,  e^{-\frac{(E_{\textrm{true}}-E_{\textrm{dep}})^2}{2 \, \sigma^2}} ~,
\end{equation}
where the best fits for the lower and upper dispersions are
\begin{equation}
\sigma(E_{\textrm{true}}) = \begin{cases}
0.121 \ E_{\textrm{true}} & \textrm{if } E_{\textrm{dep}} < E_{\textrm{true}} \\[2ex]
0.125 \ E_{\textrm{true}} & \textrm{if } E_{\textrm{dep}} \geq E_{\textrm{true}} ~.
\end{cases}
\label{eq:error}
\end{equation}

The differential shower spectrum produced after CC neutrino-nucleon interactions has contributions from  $\nu_e$ and $\nu_\tau$ (and similarly from $\bar\nu_e$ and $\bar\nu_\tau$). In the case of $\nu_e$ (and analogously $\bar\nu_e$) CC interactions, the differential shower spectrum in terms of the measured deposited energy is given by
\begin{equation}
\frac{dN^{\textrm{sh,CC}, f}_{\nu_e}}{dE_{\textrm{dep}}}  =  T \, N_A \, \int_{0}^{\infty} dE_\nu \, Att^f_{\nu_e}(E_\nu) \, \frac{d \phi^f_{\nu_e}(E_\nu) }{d E_\nu} \, \int^{1}_{0} dy \, M_{\textrm{eff}}(E^{\textrm{CC}}_{e}) \,  R(E^{\textrm{CC}}_{e}, E_{\textrm{dep}}, \sigma(E^{\textrm{CC}}_{e})) \, \frac{d\sigma^{\textrm{CC}}_{\nu_e} (E_\nu,y)}{dy}  ~,
\label{eq:cce}
\end{equation}

In the case of $\nu_\tau$ (and analogously $\bar\nu_\tau$) CC interactions, the differential shower spectrum in terms of the measured deposited energy reads
\begin{eqnarray}
\frac{dN^{\textrm{sh,CC}, f}_{\nu_\tau}}{dE_{\textrm{dep}}}  & = & T \, N_A \, \int_{0}^{\infty} dE_\nu \, Att^f_{\nu_\tau}(E_\nu) \, \frac{d \phi^f_{\nu_\tau}(E_\nu) }{d E_\nu} \, \int^{1}_{0} dy \, \frac{d\sigma^{\textrm{CC}}_{\nu_\tau} (E_\nu,y)}{dy}  \, \int_{0}^{1} dz \\
& \times & \sum_{c=h,e} \, \left( D_\tau(E_\tau) \,   M_{\textrm{eff}}(E^{\textrm{CC}}_{\tau, c}) \, R(E^{\textrm{CC}}_{\tau, c}, E_{\textrm{dep}}, \sigma(E^{\textrm{CC}}_{\tau, c})) +  (1-D_\tau (E_\tau)) \, M_{\textrm{eff}}(E_h) \, R(E_h, E_{\textrm{dep}}, \sigma(E_h)) \,\right) \frac{dn_c (z)}{dz} ~, \nonumber 
\label{eq:cct}
\end{eqnarray}
where $dn_c(z)/dz$ is the energy distribution of the daughter $\nu_\tau$ or $e$ with energy $E_{\nu_\tau, \tau}$ ($z = E_{\nu_\tau, \tau}/E_\tau$) or $E_{e, \tau}$ ($z = E_{e, \tau}/E_\tau$) from $\tau$ decay via the hadronic or electronic channel ($c = \{h, \, e\}$), respectively~\cite{Dutta:2000jv}.  The fraction of tau leptons decaying inside the detector, $D_\tau(E_\tau)$, is defined in Eq.~(\ref{eq:dtau}) and a fit in terms of a Pad\'e approximant, with an accuracy at the 0.1\% level, is given by
\begin{equation}
D_\tau(E_\tau) = \frac{1 + p_1 \, \left(E_\tau/10 \, \textrm{PeV}\right)}{1 + q_1 \, \left(E_\tau/10 \, \textrm{PeV}\right) + q_2 \, \left(E_\tau/10 \, \textrm{PeV}\right)^2} ~,
\label{eq:PadeD}
\end{equation}
with $p_1=0.883$, $q_1=1.66$ and $q_2=1.15$.

On the other hand, tracks are produced in CC $\nu_\mu$ and $\nu_\tau$  interactions (followed by the $\tau$ decay into $\nu_\tau \nu_\mu \mu$). For $\nu_\mu$ (and analogously $\bar\nu_\mu$)  CC interactions, the differential track spectrum in terms of the measured deposited energy reads
\begin{equation}
\frac{dN^{\textrm{tr,CC}, f}_{\nu_\mu}}{dE_{\textrm{dep}}}  =  T \, N_A \, \int_{0}^{\infty} dE_\nu \, Att^f_{\nu_\mu}(E_\nu) \, \frac{d \phi^f_{\nu_\mu}(E_\nu) }{d E_\nu} \, \int^{1}_{0} dy \, M_{\textrm{eff}}(E^{\textrm{CC}}_{\mu}) \,  R(E^{\textrm{CC}}_{\mu}, E_{\textrm{dep}}, \sigma(E^{\textrm{CC}}_{\mu})) \, \frac{d\sigma^{\textrm{CC}}_{\nu_\mu} (E_\nu,y)}{dy}  ~,
\label{eq:ccm}
\end{equation}

The differential track spectrum from $\nu_\tau$ (and analogously from $\bar\nu_\tau$) CC interactions is
\begin{eqnarray}
\frac{dN^{\textrm{tr,CC}, f}_{\nu_\tau}}{dE_{\textrm{dep}}}  & = & T \, N_A \, \int_{0}^{\infty} dE_\nu \, Att^f_{\nu_\tau}(E_\nu) \, \frac{d \phi^f_{\nu_\tau}(E_\nu) }{d E_\nu} \, \int^{1}_{0} dy \, \frac{d\sigma^{\textrm{CC}}_{\nu_\tau} (E_\nu,y)}{dy}  \\
& \times & \int_{0}^{1} dz  \, \left( D_\tau(E_\tau) \,   M_{\textrm{eff}}(E^{\textrm{CC}}_{\tau, \mu}) \, R(E^{\textrm{CC}}_{\tau, \mu}, E_{\textrm{dep}}, \sigma(E^{\textrm{CC}}_{\tau, \mu})) +  (1-D_\tau (E_\tau)) \, M_{\textrm{eff}}(E_h) R(E_h, E_{\textrm{dep}}, \sigma(E_h)) \,\right) \frac{dn_\mu (z)}{dz} ~, \nonumber
\label{eq:cctt}
\end{eqnarray}
where $dn_\mu(z)/dz$ is the energy distribution of the daughter $\mu$ with energy $E_{\mu, \tau}$ ($z = E_{\mu, \tau}/E_\tau$) from $\tau$ decay~\cite{Dutta:2000jv}. The true EM-equivalent deposited energies are given by
\begin{eqnarray}
E_h & = & F_h (E_\nu \, y)  \, E_\nu \, y  ~, \\
E_{\ell} & = & E_\nu \, (1-y)  ~, \hspace{5mm} \textrm{(with } \ell=\{e, \mu, \tau\}) ~,\\
E^{\textrm{NC}} & = & E_h  ~, \\
E^{\textrm{CC}}_{e} & = & E_h + E_e  ~, \\
E^{\textrm{CC}}_{\mu} & = & E_h + F_\mu \, (E_\mu + a/b) ~, \\
E^{\textrm{CC}}_{\tau, h} & = & E_h + F_h (E_\tau \, (1-z)) \, E_\tau \, (1-z)  ~, \\ 
E^{\textrm{CC}}_{\tau, e} & = & E_h + E_\tau \, z  ~, \\
E^{\textrm{CC}}_{\tau, \mu} & = &E_h + F_{\mu, \tau} (E_\tau) \, (E_\tau \, z  + a/b)~,
\end{eqnarray}
where $F_h(E_X)$ is defined in Eq.~(\ref{eq:Fhdef}), $a$ and $b$ in Eq.~(\ref{eq:dedx}), $F_\mu = 0.119$ in Eq.~(\ref{eq:avgloss}) and $F_{\mu, \tau}(E_\tau)$ in Eq.~(\ref{eq:avglossmutau}). 

The function $F_h(E_X)$ can be parametrized as~\cite{Gabriel:1993ai}
\begin{equation}
F_h (E_X) = 1- (1-f_0)\left(\frac{E_X}{E_0}\right)^{-m} ~, 
\label{eq:Fh}
\end{equation}
where the values of the parameters resulting from a fit to simulations of hadronic showers induced by NC neutrino-nucleon interactions are $f_0 = 0.467$, $E_0 = 0.399$~GeV and $m = 0.130$~\cite{Kowalski}. 

As for $a$ and $b$, we use tabulated data for the muon loss rate in ice~\cite{Agashe:2014kda} and perform a fit taking both parameters as constants.  The resulting values are: $a = 0.206$~GeV/m and $b = 3.21 \times 10^{-4}$~m$^{-1}$. This fit is accurate at the few percent level above 400~GeV and below the percent level above 40~TeV.

The average fraction of energy lost along a track of a muon produced in a tau decay after a $\nu_\tau$ or $\bar\nu_\tau$ CC interaction inside the detector, $F_{\mu, \tau}(E_\tau)$, is defined in Eq.~(\ref{eq:avglossmutau}) and a fit in terms of a Pad\'e approximant, with an accuracy at the 0.2\% level, is given by
\begin{equation}
F_{\mu, \tau}(E_\tau) = F_\mu \, \frac{1 + \tilde p_1 \, \left(E_\tau/10 \, \textrm{PeV}\right)}{1 + \tilde q_1 \, \left(E_\tau/10 \, \textrm{PeV}\right) + \tilde q_2 \, \left(E_\tau/10 \, \textrm{PeV}\right)^2} ~,
\label{eq:PadeF}
\end{equation}
with $\tilde p_1=0.984$, $\tilde q_1=1.01$ and $\tilde q_2=1.03$.

The contributions to the differential shower spectrum from neutrino or antineutrino interactions with electrons from different processes are given by
\begin{equation}
\frac{dN^{\textrm{sh}, e, f}_{\nu_\ell, e}}{dE_{\textrm{dep}}}  =  T \, N_A \, \int_{0}^{\infty} dE_\nu \, Att^f_{\nu_\ell}(E_\nu) \, \frac{d \phi^f_{\nu_\ell}(E_\nu) }{d E_\nu} \, \int^{1}_{0} dy \, M_{\textrm{eff}}(E^{e}_{e}) \,  R(E^{e}_{e}, E_{\textrm{dep}}, \sigma(E^{e}_{e})) \, \frac{d\sigma^{\textrm{e}}_{\nu_\ell, e} (E_\nu,y)}{dy}  ~,
\label{eq:ee}
\end{equation}
where $\nu_\ell = \{\nu_e, \nu_\mu, \nu_\tau, \bar\nu_e, \bar\nu_\mu, \bar\nu_\tau\}$, and

\begin{eqnarray}
\frac{dN^{\textrm{sh}, e, f}_{\bar\nu_e, \tau}}{dE_{\textrm{dep}}}  & =  & T \, N_A \, \int_{0}^{\infty} dE_\nu \, Att^f_{\bar\nu_e}(E_\nu) \, \frac{d \phi^f_{\bar\nu_e}(E_\nu) }{d E_\nu} \, \int^{1}_{0} dy \, \frac{d\sigma^{e}_{\bar\nu_e, \tau} (E_\nu,y)}{dy}  \nonumber \\
& & \times  \int_{0}^{1} dz  \, \sum_{c=h,e} D_\tau(E_\tau) \,   M_{\textrm{eff}}(E^{e}_{\tau, c}) \, R(E^{e}_{\tau, c}, E_{\textrm{dep}}, \sigma(E^{e}_{\tau, c})) \, \frac{dn_c (z)}{dz} ~,
\label{eq:eet}
\end{eqnarray}

\begin{equation}
\frac{dN^{\textrm{sh}, e, f}_{\bar\nu_e, h}}{dE_{\textrm{dep}}}  =  T \, N_A \, \int_{0}^{\infty} dE_\nu \, Att^f_{\bar\nu_e}(E_\nu) \, \frac{d \phi^f_{\bar\nu_e}(E_\nu) }{d E_\nu} \, M_{\textrm{eff}}(E^{e}_{e, h}) \,  R(E^{e}_{e, h}, E_{\textrm{dep}}, \sigma(E^{e}_{e, h})) \, \sigma^{e}_{\bar\nu_e, h} (E_\nu)  ~,
\label{eq:eeh}
\end{equation}

\begin{eqnarray}
\frac{dN^{\textrm{sh}, e, f}_{\nu_\tau, \tau}}{dE_{\textrm{dep}}}  & =  & T \, N_A \, \int_{0}^{\infty} dE_\nu \, Att^f_{\nu_\tau}(E_\nu) \, \frac{d \phi^f_{\nu_\tau}(E_\nu) }{d E_\nu} \, \int^{1}_{0} dy \, \frac{d\sigma^{e}_{\nu_\tau, \tau} (E_\nu,y)}{dy}  \nonumber \\
& & \times  \int_{0}^{1} dz  \, \sum_{c=h,e} D_\tau(E_\tau) \,   M_{\textrm{eff}}(E^{e}_{\tau, c}) \, R(E^{e}_{\tau, c}, E_{\textrm{dep}}, \sigma(E^{e}_{\tau, c})) \, \frac{dn_c (z)}{dz} ~,
\label{eq:ett}
\end{eqnarray}

\begin{figure}[t]
	\includegraphics[width=\textwidth]{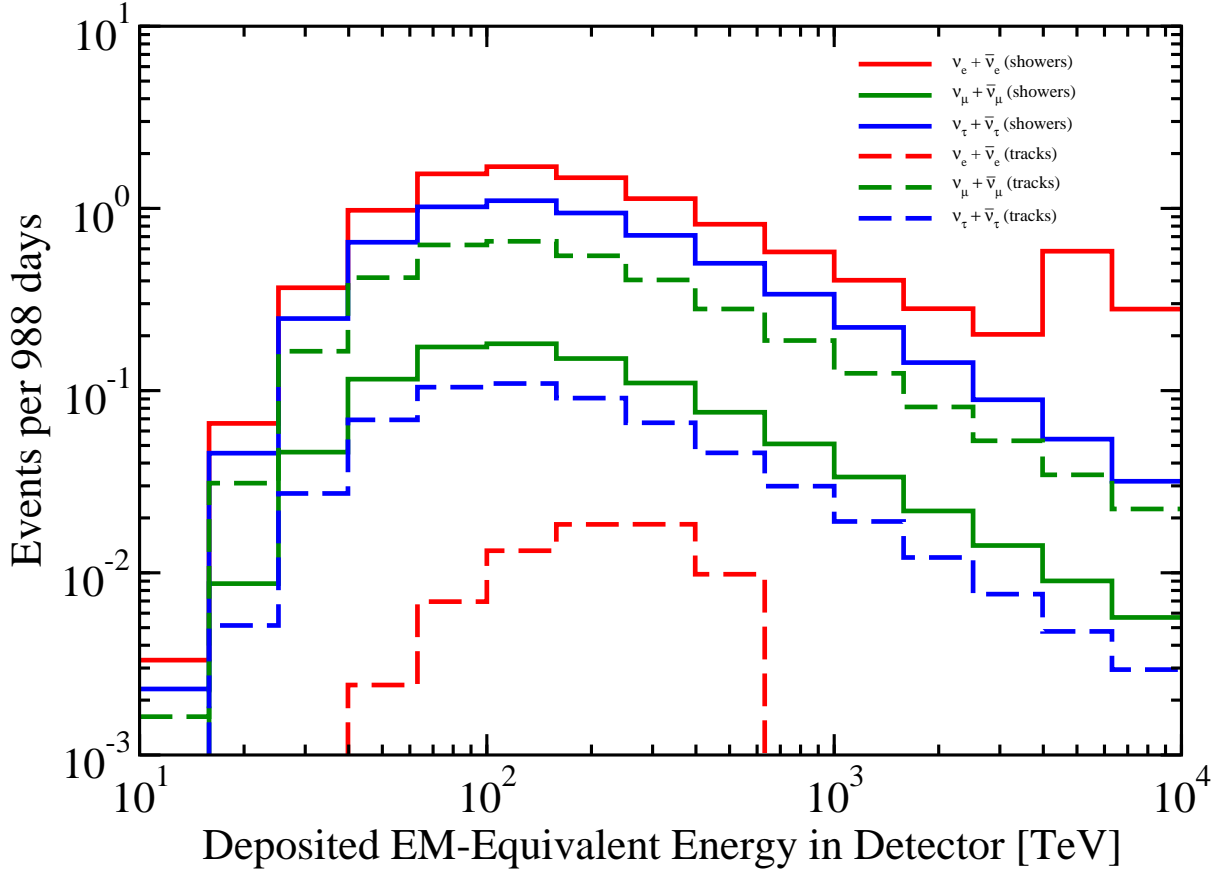}
	\caption{\textbf{\textit{Event spectra in IceCube, as a function of the EM-equivalent deposited energy, of tracks and showers for each flavor after 988~days}} (summing neutrinos and antineutrinos), for an isotropic power-law spectrum, per flavor, $E_\nu^2 \, d\phi^a/dE_\nu = 1.5 \times 10^{-8} \, (E_\nu/100 \, \textrm{TeV})^{-0.3} \, \textrm{GeV} \, \textrm{cm}^{-2} \, \textrm{s}^{-1} \, \textrm{sr}^{-1}$~\cite{Aartsen:2014gkd}.  Showers (tracks) are depicted as solid (dashed) histograms. The contributions from $\nu_e + \bar\nu_e$, $\nu_\mu + \bar\nu_\mu$	and $\nu_\tau + \bar\nu_\tau$ are represented by red, green and blue histograms, respectively.}
	\label{fig:app}
\end{figure}

The contributions to the differential track spectrum from neutrino or antineutrino interactions with electrons from different processes are given by
\begin{eqnarray}
\frac{dN^{\textrm{tr}, e, f}_{\nu_\ell, \tau}}{dE_{\textrm{dep}}}  & =  & T \, N_A \, \int_{0}^{\infty} dE_\nu \, Att^f_{\nu_\ell}(E_\nu) \, \frac{d \phi^f_{\nu_\ell}(E_\nu) }{d E_\nu} \, \int^{1}_{0} dy \, \frac{d\sigma^{e}_{\nu_\ell, \tau} (E_\nu,y)}{dy}  \nonumber \\
& & \times \int_{0}^{1} dz  \, D_\tau(E_\tau) \,   M_{\textrm{eff}}(E^{e}_{\tau, \mu}) \, R(E^{e}_{\tau, \mu}, E_{\textrm{dep}}, \sigma(E^{e}_{\tau, \mu}))  \, \frac{dn_\mu (z)}{dz} ~,
\label{eq:eettr}
\end{eqnarray}
where $\nu_\ell = \{\bar\nu_e, \nu_\tau\}$, and

\begin{equation}
\frac{dN^{\textrm{tr}, e, f}_{\nu_\ell, \mu}}{dE_{\textrm{dep},i}}  =  T \, N_A \, \int_{0}^{\infty} dE_\nu \, Att^f_{\nu_\ell}(E_\nu) \, \frac{d \phi^f_{\nu_\ell}(E_\nu) }{d E_\nu} \, \int^{1}_{0} dy \, M_{\textrm{eff}}(E^{e}_{\mu}) \,  R(E^{e}_{\mu}, E_{\textrm{dep},i}, \sigma(E^{e}_{\mu})) \, \frac{d\sigma^{e}_{\nu_\ell, \mu} (E_\nu,y)}{dy}  ~,
\label{eq:eaem}
\end{equation}
with $\nu_\ell = \{\bar\nu_e, \nu_\mu\}$.

The true EM-equivalent deposited energies for all the neutrino-electron processes are
\begin{eqnarray}
E^{e}_{e} & = & E_\nu \, y  ~, \\
E^{e}_{\tau, h} & = & F_h(E_\nu \, y \, (1-z)) \, E_\nu \, y \, (1-z) ~, \\
E^{e}_{\tau, e} & = & E_\nu \, y \, z   ~, \\
E^{e}_{e, h} & = & F_h (E_\nu) \, E_\nu  ~, \\	
E^{e}_{\tau, \mu} & = & F_{\mu, \tau} (E_\nu \, y) \, (E_\nu \, y \, z + a/b)  ~, \\
E^{e}_{\mu} & = & F_\mu \, (E_\nu \, y + a/b) ~.
\end{eqnarray}

In Fig.~\ref{fig:app} we show the event spectra of showers and tracks for each flavor (summing neutrinos and antineutrinos) for the best fit IceCube spectra~\cite{Aartsen:2014gkd}, \textit{i.e.}, $E_\nu^2 \, d\phi^a/dE_\nu = 1.5 \times 10^{-8} \, (E_\nu/100 \, \textrm{TeV})^{-0.3} \, \textrm{GeV} \, \textrm{cm}^{-2} \, \textrm{s}^{-1} \, \textrm{sr}^{-1}$, per flavor.  The effect of the Glashow resonance on the $\bar\nu_e$-induced event spectra is clearly visible in the red histograms. The shower spectrum for $\nu_e + \bar\nu_e$ (red solid histogram) shows a bump above a few PeV and the resonant interactions of $\bar\nu_e$ []but also the no-resonant interactions of $\nu_e$; see Eq.~(\ref{eq:invelastic})] with electrons also give rise to tracks (red dashed histogram), via the the leptonic decay of the produced $W$ bosons. We also note the similar shape of all the event distributions (except from the red dashed histogram), as a function of the EM-equivalent deposited energy, below a few PeV.

\pagebreak

\bibliographystyle{apsrev4-1}
\bibliography{flavors}

\end{document}